\documentclass[english,aps,prl,twocolumn,amsmath,amssymb,showpacs,superscriptaddress,notitlepage,longbibliography]{revtex4-1}
\usepackage[T1]{fontenc}
\usepackage[utf8]{inputenc}
\usepackage{graphicx}
\usepackage{amsfonts}
\usepackage{braket}
\usepackage{tikz}
\makeatletter
\usepackage{multirow}
\usepackage{amsmath}
\usepackage{amssymb}
\usepackage{bbm}
\usepackage{enumitem}
\usepackage{mathrsfs}
\usepackage[colorlinks=true,linkcolor=blue,anchorcolor=red,citecolor=blue,urlcolor=blue]{hyperref}
\usepackage[caption=false]{subfig}
\usepackage{lipsum}
\usepackage{balance}
\usepackage[normalem]{ulem}

\makeatother

\usepackage{babel}

\begin{document}

\title{Inter-orbital Cooper pairing at finite energies in Rashba surface states}

\author{Philipp R\"u{\ss}mann}
\email{philipp.ruessmann@uni-wuerzburg.de}
\affiliation{Institute for Theoretical Physics and Astrophysics, University of W\"urzburg, D-97074 W\"urzburg, Germany}
\affiliation{Peter Gr\"unberg Institut and Institute for Advanced Simulation, Forschungszentrum J\"ulich and JARA, D-52425 J\"ulich Germany}

\author{Masoud Bahari}
\email{masoud.bahari@physik.uni-wuerzburg.de}
\affiliation{Institute for Theoretical Physics and Astrophysics, University of W\"urzburg, D-97074 W\"urzburg, Germany}
\affiliation{W\"urzburg-Dresden Cluster of Excellence ct.qmat, Germany}

\author{Stefan Bl\"ugel}
\affiliation{Peter Gr\"unberg Institut and Institute for Advanced Simulation, Forschungszentrum J\"ulich and JARA, D-52425 J\"ulich Germany}

\author{Bj\"orn Trauzettel}
\affiliation{Institute for Theoretical Physics and Astrophysics, University of W\"urzburg, D-97074 W\"urzburg, Germany}
\affiliation{W\"urzburg-Dresden Cluster of Excellence ct.qmat, Germany}

\date{\today}
\begin{abstract}
    Multi-band effects in hybrid structures provide a rich playground for unconventional superconductivity. We combine two complementary approaches based on density-functional theory (DFT) and effective low-energy model
    theory in order to investigate the proximity effect in a Rashba surface state in contact to an $s$-wave superconductor. We discuss these synergistic approaches and combine the effective model and DFT analysis at the example of a Au/Al heterostructure. This allows to predict finite-energy superconducting pairing due to the interplay of the Rashba surface state of Au, and hybridization with the electronic structure of superconducting Al. We investigate the nature of the induced superconducting pairing and quantify its mixed singlet-triplet character. Our findings demonstrate general recipes to explore real material systems that exhibit inter-orbital pairing away from the Fermi energy.
\end{abstract}
\maketitle


\section{Introduction}

Materials that exhibit strong spin orbit coupling (SOC) build the foundation for a plethora of physical phenomena~\cite{Manchon2015, Bihlmayer2022} with applications ranging from non-collinear topological magnetic textures (e.g.\ skyrmions)~\cite{Fert2017} over spinorbitronics~\cite{Manchon2015} or topological insulators~\cite{Hasan2010} to quantum information processing~\cite{Alicea2011, Lutchyn2018, Frolov2020, Flensberg2021}.
Combining different materials in heterostructures not only gives rise to breaking of symmetries, which is essential to Rashba SOC~\cite{Rashba1960}, but it also allows us to tailor proximity effects, where the emergent physics of the heterostructure as a whole is richer than the sum of its constituents. In the past, this has attracted a lot of interest in the context of increasing SOC in graphene~\cite{Avsar2014, Gmitra2017, Island2019}.
Combining a strong-SOC material with a superconductor is, moreover, of particular use to realize topological superconductivity, that can host Majorana zero modes (MZMs). In turn, MZMs are building blocks of topological qubits~\cite{Nayak2008}. 

In this work, we study the inter-orbital physics inherent to heterostructures consisting of superconductors and Rashba materials. In a novel way, we combine theoretical modelling of two complementary approaches that have their roots in rather disjoint communities focusing on either microscopic or mesoscopic physics. We combine the predictive power of material-specific DFT simulations with the physical insights of an analytically solvable low-energy model. 
The Bogoliubov-de Gennes (BdG) formalism~\cite{deGennes1966, BdGbook} is the basis for both models, in particular, the DFT-based description of the superconducting state, commonly referred to as Kohn-Sham Bogoliubov-de Gennes (KS-BdG) approach~\cite{Oliveira1988, Lueders2005, Csire2015, Kawamura2020, Linscheid2015}.
While DFT naturally accounts for multi-band effects, the effective low-energy model with a simpler treatment of only a few bands allows us to identify the symmetry of the superconducting pairing.
{Crystal symmetries have profound effects. For example, they may or may not cause wavefunctions to overlap, which is visible in DFT calculations.
A group-theoretic analysis allows us to infer possible (unconventional) pairing channels from crystal symmetries~\cite{Scheurer2017}. However, group theory alone does not tell us which of the possible pairing channels really matters in a given material. Hence, only the combination of both approaches (DFT and group theory) is able to predict the emergence of experimentally relevant (unconventional) pairing channels in the laboratory.}

\begin{figure}[htbp]
    \centering
    \includegraphics[width=\linewidth]{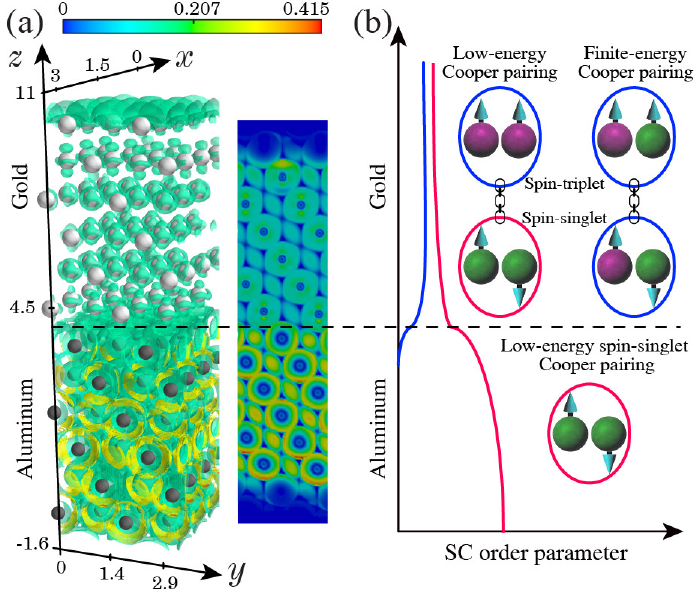}
    \caption{(a) Localization of the electron density (arb.\ units) around the Fermi energy throughout the Al/Au heterostructure. The background shows a cut through $x=0$. (b) Illustration of {three kinds of} Cooper pair tunneling and the formation of different singlet/triplet components due to the Rashba surface state. {Cooper pairs formed by electrons originating from different orbitals are denoted by different colors.}}
    \label{fig:density3D}
\end{figure}

Rashba SOC is intimately related to orbital mixing, often involving $p$ electrons~\cite{Bihlmayer2022}. Evidence for strong Rashba-SOC is found in a variety of materials ranging from heavy metal surfaces like Au or Ir and surface alloys (e.g.\ $\sqrt{3}\times\sqrt{3}$ Bi/Ag)~\cite{LaShell1996, Varykhalov2012, Ast2007}, over semiconductors like InSb~\cite{NadjPerge2010}, to topological insulators (e.g.\ Bi$_2$Se$_3$)~\cite{King2011}. We investigate the combination of such metals in hybrid structures with common superconductors{, where multi-band effects are essential. In general, multi-band effects have crucial implications. They are, for instance, relevant for transport across superconductor-semiconductor interfaces in presence of Fermi surface mismatch~\cite{Breunig2021}, and play a major role in the superconducting diode effect~\cite{Ando2020,Wu2022,Zhang2022}.}

As a {prime example} experiencing the multi-band physics of a proximitized Rashba state, we identify the interface between aluminium (Al) and gold (Au). This combination {allows us} to study the proximity effect with Rashba surface states. On the one hand, Al is a well-known and widely used $s$-wave superconductor whose valence band electrons are of $s-p$ orbital character.
On the other hand, Au is a simple heavy metal where effects of strong SOC are particularly pronounced. In fact, as a consequence of strong SOC, the (111) surface of Au hosts a set of two spin-momentum-locked Rashba surface states~\cite{LaShell1996, Nicolay2001, Henk2003, Henk2004, Hoesch2004, Wissing2013}.
Both Al and Au grow in the face centered crystal (fcc) structure and their lattice constants vary only marginally~\cite{Villars2016-1, Villars2016-2}.
Hence, epitaxial growth of this heterostructure is feasible.
{It} is ideally suited to gain insight into (i) hybridization of the electronic structure of Al- and Au-derived bands at the interface, (ii)  proximity effect of the SOC from Au into the superconductor Al, (iii) interplay of the superconducting proximity effect and SOC in this multi-band system, and (iv) mixed singlet-triplet nature of induced superconducting pairing.
{The hybridized electronic structure in the Al/Au heterostructure and the emerging superconducting pairing channels due to multi-band effects are depicted in Fig.~\ref{fig:density3D}.}

This {article} is structured as follows. In Sec.~\ref{seq:normal}, the normal state electronic structure of the Au/Al heterostructure is discussed with DFT and low-energy model {approaches}. In Sec.~\ref{seq:SC}, the DFT and model {access} to superconducting heterostructures are presented with emphasis on complementary insights. {This modelling} allows us to study the proximity effects of {SOC and superconductivity in multi-band systems at the example of Al/Au interfaces}. We conclude in Sec.~\ref{seq:concl}, where we also comment on the feasibility of experimental detection of our predictions.


\section{Normal state spectrum} \label{seq:normal}

The DFT and model-based approaches {described in this article} are complementary and uniquely distinct in their methodologies. The DFT-based numerical calculations provide an \textit{ab-initio} approach to the description of the electronic structure of the normal state, their scope encompasses \emph{all} electronic degrees of freedom, resulting in a precise and extensive representation applicable to a broad range of materials merely from the knowledge about the crystal structure. Consequently, the intricate band structure generated by this method can be complex, comprising several bands with diverse orbital and spin character. 

The effective low-energy model aims to simplify the complexity of the electronic structure by describing only a few bands, particularly those {close to} the $\Gamma$-point and the Fermi level. The model-based approach has the distinct advantage of deriving analytical expressions that can be applied to a wide range of material classes. Additionally, the model enables the analysis and inclusion of certain symmetries. For instance, only odd terms in $\mathbf{k}$ might appear in certain parts of the model Hamiltonian. To create a model that applies to a {real} material, it is, however, necessary to determine model parameters by fitting to experimental or DFT data.

\subsection{Density functional theory results}

\begin{figure}[t]
    \centering
    \includegraphics[width=\linewidth]{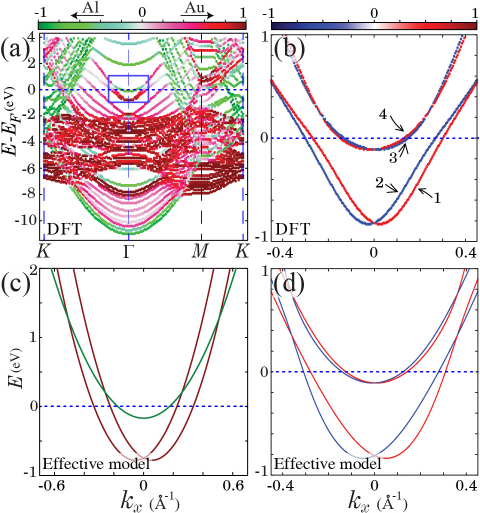}
    \caption{(a) DFT 
    {band structure} for the Al/Au hybrid structure {consisting of} 6 layers of each element. Colorbar shows the localization of the states. (b) Enlarged view of spectrum close to the Fermi energy denoted by the blue rectangle in panel (a). Colorbar shows  the spin polarization $\langle s_y \rangle$ (arb.\ units). (c) [(d)] Excitation spectrum of the  normal state obtained by low-energy model Hamiltonian, given in Eq.~\eqref{eq:modelNormalState}, close to the Fermi energy in absence [presence] of band hybridization and including third-order Rashba spin-orbit coupling, i.e., $F_0=g=0$ [$F_0=0.2$ and $g=-8.45$]. Other model parameters are given in Tab.~\ref{Table1}.}
    \label{Fig1.pdf}
\end{figure}

Our DFT calculations for heterostructures, {consisting of thin Al and Au films,} are summarized in Figs.~{\ref{fig:density3D} and} \ref{Fig1.pdf}(a,b). Both Al and Au have a face-centered cubic (fcc) crystal structure with lattice constants of $4.08$\AA\ and $4.05$\AA, respectively~\cite{Villars2016-1, Villars2016-2}. 
We investigate an ideal interface in the close-packed (111) surface of the fcc lattice.
To model the heterostructure, we use a unit cell that consists of 6 layers of Al and 6 layers of Au with the average experimental lattice constant of Al and Au, {differing only} by about $0.4\%$ from their respective bulk lattice constants.
For our DFT calculations, we employ the full-potential relativistic Korringa-Kohn-Rostoker Green function method, as implemented in the JuKKR code~\cite{jukkr}. This allows us to include the effect of superconductivity on the footings of the Bogoliubov-de Gennes formalism~\cite{Ruessmann2022a}. {Computational details are provided} in App.~\ref{app:ComputationDetails}.

The electronic structure of Au below the Fermi level is dominated by the fully occupied shell of $d$-electrons around $-2$ to $-8$\,eV (see App.~\ref{app:DFTnormal} for the corresponding DOS).
In thin-film heterostructures (called ``slabs''), the electrons are confined inside the slab, leading to finite-size quantization and the appearance of {two-dimensional} quantum{-}well states {manifested} as a series of discrete bands in the region where the bulk electronic structure is projected into the surface Brillouin zone. The presence of surfaces and interfaces, and the possible appearance of broken bonds, often leads to additional surface states or surface resonances in the electronic structure.
For the Au(111) surface, Rashba surface states appear {in surface projection of the bulk $L$-gap} around the $\Gamma$ point {of the surface Brillouin zone}.
{They are} are of $s$-$p_z$ orbital character \cite{LaShell1996,Henk2003}. 

The region around $\Gamma$, highlighted by the blue box in Fig.~\ref{Fig1.pdf}(a), is the focus of our study. {It is enlarged in} Fig.~\ref{Fig1.pdf}(b).
The in-plane component of the spin-polarization ($s_y$) perpendicular to the direction of the momentum ($k_x$) is shown by the color coding of the bands. Note that, due to crystal symmetries, $s_x$ is exactly zero in the plane through $k_y=0$, and $s_z$ is negligibly small. From the full band structure information {based on} DFT, we select {a regime} of interest for the analytical effective low-energy model. We restrict our analysis to the four states labeled 1-4, which (at small $|\mathbf{k}|$ close to $\Gamma$) are derived from the Rashba surface state of Au (states 1,2) or from Al (states 3,4), respectively. 
The Al states (3,4) have a quadratic dispersion and show much weaker spin-splitting.
{Importantly, our study reveals the existence of only a single pair of Au Rashba surface states localized at the interface of Au and vacuum. Notably, no second pair of states arises from the interface of Al and Au. This can be deduced from the localization of these states depicted in App.~\ref{app:DFTnormal}. The real-space distribution of the charge density at the Fermi energy is shown in Fig.~\ref{fig:density3D}a. We conclude that the scattering potential at the interface is weak enough to prevent the formation of a second state at the Al/Au interface.}

Aluminium is a light metal with negligible intrinsic SOC. The small SOC-induced spin-splitting seen for states 3,4 is merely a result of a proximity-induced SOC from Au to Al, hinting at sizable hybridization of the electronic structure of Al and Au.
In Sec.~\ref{seq:kpmodel}, we discuss in detail that, at higher momenta, the parabolas of the Al-derived states and the Rashba surface states intersect and hybridize, resulting in more delocalized states throughout the entire Al-Au heterostructure. This hybridization can be attributed to the compatible orbital character of the Al and Au bands, which both possess $s$-$p_z$ like orbital character. Ultimately, this hybridization leads to the proximity effect of the spin-orbit coupling (SOC) observed in the Al quantum well states.


\subsection{Effective low-energy model}
\label{seq:kpmodel}
Complementary to our DFT results, we develop an effective four-band model Hamiltonian to evaluate the spectral properties of the heterostructure in an analytical manner. Guided by the insights from our DFT calculation, we construct a model for the proximitized Rashba surface state. We note a hybridization of spin-split Au surface bands and {the} doubly degenerate Al band near the Fermi energy. Thus, we propose
the normal state model Hamiltonian to be 
\begin{equation}
    \!H_{N}\!=\!\!\sum_{\mathbf{k}}\!\left(c_{\mathbf{k},\text{Al}}^{\dagger},c_{\mathbf{k},\text{Au}}^{\dagger}\right)\!\left(\!\!\begin{array}{cc}
    \hat{h}_{\text{Al}}(\mathbf{k}) & F_{0}\hat{\sigma}_{0}\\
    F_{0}\hat{\sigma}_{0} & \hat{h}_{\text{Au}}(\mathbf{k})
    \end{array}\!\!\right)\!\left(\!\begin{array}{c}
    c_{\mathbf{k},\text{Al}}\\
    c_{\mathbf{k},\text{Au}}
    \end{array}\!\right)\!,\!\label{eq:modelNormalState}
\end{equation}
where the electron annihilation operator is denoted as $c_{\mathbf{k},\nu}=(c_{\mathbf{k},\nu,s},c_{\mathbf{k},\nu,-s})^{T}$
labeled by the 2D momentum vector $\mathbf{k}=(k_{x},k_{y})$
with orbital $(\nu\in\{\text{Al},\text{Au}\})$ and spin $(s\in\{\uparrow,\downarrow\})$
degrees of freedom. $F_{0}$ signifies the hybridization strength
between Al and Au bands. Furthermore, $\hat{h}_{\text{Al}(\text{Au})}(\mathbf{k})$
denotes the $2\times2$ sector for the Al (Au) segment given by 
\begin{align}
    \hat{h}_{\text{Al}}(\mathbf{k}) & =(\alpha_{\text{Al}}k^{2}-\mu_{\text{Al}})\hat{\sigma}_{0},\\
    \hat{h}_{\text{Au}}(\mathbf{k}) & =(\alpha_{\text{Au}}k^{2}-\mu_{\text{Au}})\hat{\sigma}_{0}+\lambda(\hat{\sigma}_{x}k_{y}-k_{x}\hat{\sigma}_{y})\nonumber \\
     & +g(\hat{\sigma}_{x}(k_{y}^{3}+k_{y}k_{x}^{2})-(k_{x}^{3}+k_{x}k_{y}^{2})\hat{\sigma}_{y}),\label{Eq: Au normals-state}
\end{align}
 where $k\equiv|\mathbf{k}|=\sqrt{k_{x}^{2}+k_{y}^{2}}$; $\alpha_{\text{Al}(\text{Au})}$ and $\mu_{\text{Al}(\text{Au})}$
{characterize} mass term and chemical potential for Al (Au) bands,
respectively. First (third) order spin-orbit coupling in the Au sector is parametrized
by $\lambda$ ($g$) leading to broken inversion symmetry, i.e.,
$\hat{h}_{\text{Au}}(-\mathbf{k})\neq\hat{h}_{\text{Au}}(\mathbf{k})$.
It is worth noting that even though the band spin-splitting of the Rashba surface state is isotropic in
$\text{Au}$~\cite{LaShell1996}, it is necessary
to consider higher order polynomials for the Rashba {SOC}
in the heterostructure to match the dispersion calculated from first-principles. {We attribute this {observation} to the reduced $C_{3v}$ point group symmetry of the interface built into the DFT model via the chosen crystal structure.} We obtain the third
order polynomial presented in the last term of Eq.~\eqref{Eq: Au normals-state} by taking the direct product of the irreducible representations of
$C_{3v}$~\cite{MSRef_ThirdOrder}. {Hence, this} normal-state model is constructed by
intuition {employing} the $\mathbf{k}\cdot\mathbf{p}$ approach.
This is evident in our formulation of the Hamiltonian, where we combine a Rashba model up to third order describing the Au layer with a quadratic dispersion for the Al layer and a band hybridization term $F_{0}$.

{For simplicity,} we focus on the 1D Brillouin zone, i.e., $\mathbf{k}=(k_{x},0)$, since our model is rotationally symmetric.
Therefore, the excitation spectra of the hybrid structure become
\begin{align}
    E_{\mathbf{k},s}^{s^{\prime}} & =\frac{1}{2}\left(\mathcal{E}_{\text{Al}}+\mathcal{E}_{\text{Au}}^{s}+s^{\prime}\sqrt{(\mathcal{E}_{\text{Al}}-\mathcal{E}_{\text{Au}}^{s})^{2}+4F_{0}^{2}}\right),\label{HybE}
\end{align}
where $s,s^{\prime}\in\{+,-\}$. The quadratic band in the $\text{Al}$
segment is denoted as $\mathcal{E}_{\text{Al}}=\alpha_{\text{Al}}k^{2}-\mu_{\text{Al}}$,
and the spin-split band in the $\text{Au}$ segment {as} $\mathcal{E}_{\text{Au}}^{\pm}=\alpha_{\text{Au}}k^{2}\pm(k\lambda+gk^{3})-\mu_{\text{Au}}$ [Fig.~\ref{Fig1.pdf}(c)].
{Due to} hybridization, an effective spin-orbit coupling
is induced in the doubly degenerate Al bands, ultimately leading
to the lifting of their degeneracy. After fitting to the DFT data, the analytical spectra given by Eq.~\eqref{HybE} are in excellent agreement with
the DFT calculation, compare Figs.~\ref{Fig1.pdf}(b)
and (d).
\begin{center}
    \begin{table}
        \begin{centering}
        \caption{Values for the parameters of the low-energy model
        given in Eq.~$\eqref{HybE}$.}\label{Table1}
            \begin{tabular}{|c|c|c|c|c|c|}
                \hline 
                 & $\alpha_{i}\,(\text{eV}\cdot\text{\AA}^{2})$ & $\mu_{i}\,(\text{eV})$ & $F_{0}\,(\text{eV})$ & $\lambda\,(\text{eV}\cdot\text{\AA})$ & $g\,(\text{eV}\cdot\text{\AA}^{3})$\tabularnewline
                \hline 
                $\text{Al}$ & 5.6 & 0.17 & \multirow{2}{*}{0.2} & \multirow{2}{*}{1.1} & \multirow{2}{*}{-8.45}\tabularnewline
                \cline{1-3} \cline{2-3} \cline{3-3} 
                $\text{Au}$ & 10 & 0.75 &  &  & \tabularnewline
                \hline 
            \end{tabular}
        \par\end{centering}
    \end{table}
\par\end{center}


\section{Superconducting excitation spectrum} \label{seq:SC}

In general, a microscopic theoretical description of the superconducting excitations can be achieved on the basis of the Bogoliubov-de Gennes (BdG) formalism \cite{deGennes1966, BdGbook}, a generalization of the BCS theory of superconductivity~\cite{BCS}.
The BdG formalism is based on the Hamiltonian 
\begin{equation}
    \hat{\mathcal{H}}_{\text{BdG}}=\left(
        \begin{array}{cc}
            \hat{H}_0 & \hat{\Delta}\\{}
            [\hat{\Delta}]^{\dagger} & -\hat{H}_0^{*}
        \end{array}
    \right), \label{eq:HBdGgeneral}
\end{equation}
where $\hat{H}_0$ {denotes} the normal state Hamiltonian and $\hat{\Delta}$ the superconducting pairing between particle and hole blocks.
The BdG method is also key to the extension of DFT for superconductors~\cite{Oliveira1988, Lueders2005, Csire2015, Kawamura2020, Linscheid2015}, commonly referred to as Kohn-Sham Bogoliubov-de Gennes (KS-BdG) formalism.
One major difference between DFT and model formulations is that Eq.~\eqref{eq:HBdGgeneral} is formulated in real-space (DFT) or momentum space (model){, if translation invariance is given}.

\subsection{Kohn-Sham Bogoliubov-de Gennes formalism}

\begin{figure}
    \centering
    \includegraphics[width=\linewidth]{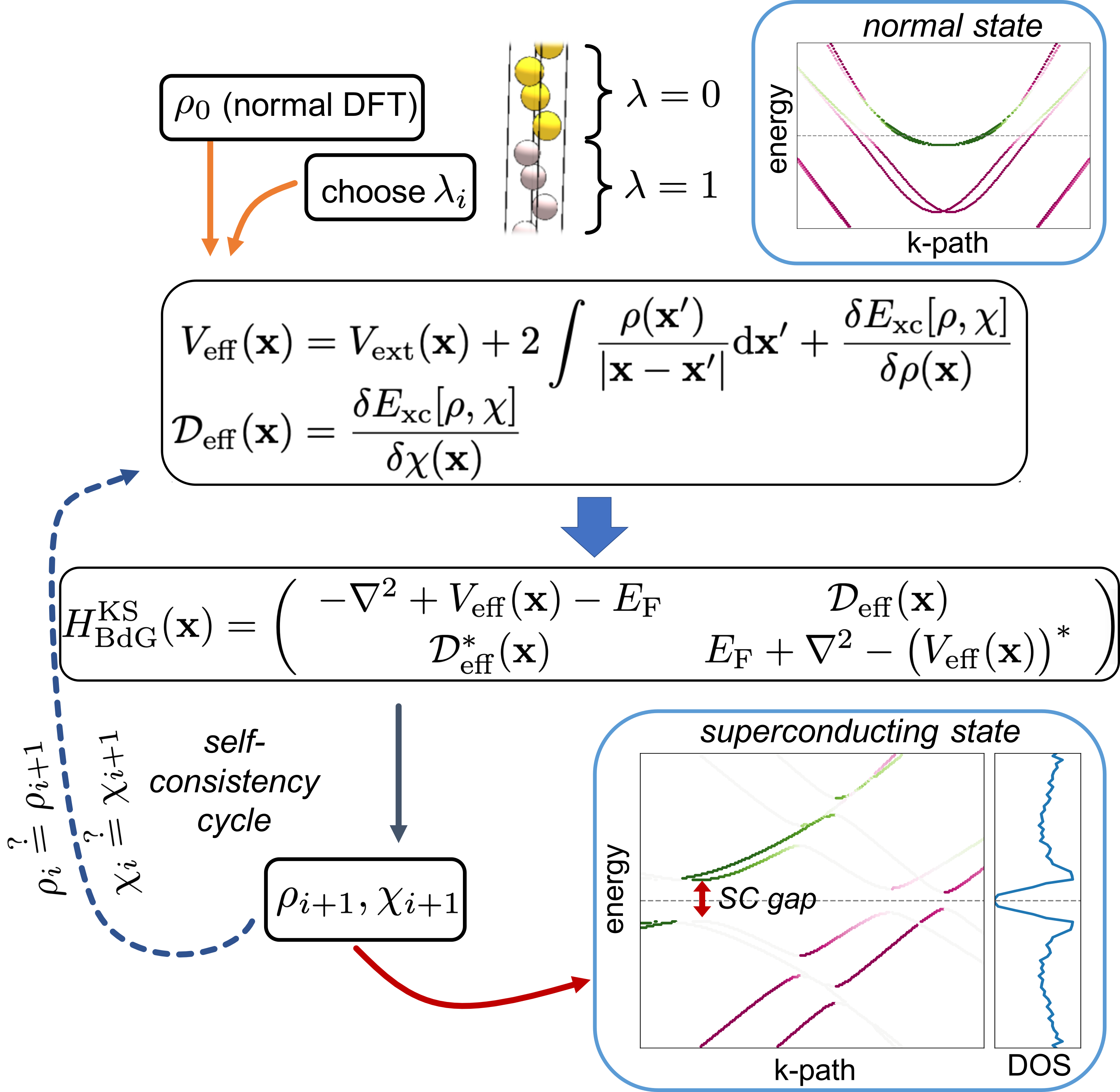}
    \caption{Schematic overview of a KS-BdG simulation that starts from the crystal structure which (in a standard DFT calculation) gives the ground state density $\rho_0$. For the superconducting state, the KS-BdG equations are then solved self-consistently to {obtain} charge and anomalous densities $\rho, \chi$ in the superconducting state which determines the superconducting band structure.
    }
    \label{fig:DFTscheme}
\end{figure}

The central task in the superconducting DFT approach (sketched in Fig.~\ref{fig:DFTscheme}) is to solve the Kohn-Sham BdG (KS-BdG) equation~\cite{Oliveira1988, Suvasini1993, Csire2015}
\begin{equation}
    H^\mathrm{KS}_\mathrm{BdG}(\mathbf{x}) \Psi^\mathrm{KS}_\nu(\mathbf{x}) = \varepsilon_\nu \Psi^\mathrm{KS}_\nu(\mathbf{x}),
\end{equation}
which is a reformulation of the Schr\"odinger equation (or Dirac equation if relativistic effects are taken into account) in terms of an effective single particle picture. The effective single-particle wavefunctions in Nambu space $\Psi^\mathrm{KS}_\nu(\mathbf{x}) = (u_\nu(\mathbf{x}), v_\nu(\mathbf{x}))^T$ describe, respectively, the particle and hole components at {excitation} energy $\varepsilon_\nu$ ($\nu$ is a band index labelling the electronic degrees of freedom).
The KS-BdG Hamiltonian can be written in matrix form as~\cite{Csire2015, Ruessmann2022a}
\begin{equation}
	H^\mathrm{KS}_{\mathrm{BdG}}(\mathbf{x}) = \left(
	\begin{array}{cc}
		H_0^\mathrm{KS}(\mathbf{x})-E_{\mathrm{F}} & \Delta_{\mathrm{eff}}(\mathbf{x}) \\
		\Delta^*_{\mathrm{eff}}(\mathbf{x}) & E_\mathrm{F}-\bigl(H_0^\mathrm{KS}(\mathbf{x})\bigr)^*,
		\label{eq:HKSBdG}
	\end{array}
	\right)
\end{equation}
where $E_\mathrm{F}$ is the Fermi energy. The normal state Hamiltonian
\begin{equation}
    H^\mathrm{KS}_0(\mathbf{x}) = -\nabla^2+V_{\mathrm{eff}}(\mathbf{x}),
\end{equation}
and the effective superconducting pairing potential $\Delta_\mathrm{eff}$ appear in {the} Kohn-Sham formulation (Rydberg atomic units are used where $\hbar=1$). For $\Delta_\mathrm{eff}=0$, the KS-BdG equation reduces to solving the conventional Kohn-Sham equation of DFT that describes the electronic structure of the normal state.

The effective single-particle potentials in Eq.~(\ref{eq:HKSBdG}) are functionals of the charge density $\rho(\mathbf{x})$ and the anomalous density $\chi(\mathbf{x})$ (the superconducting order parameter)\cite{Oliveira1988, Suvasini1993},
\begin{eqnarray}
	V_{\mathrm{eff}}(\mathbf{x}) &=& V_{\mathrm{ext}}(\mathbf{x}) + 2 \int \frac{\rho(\mathbf{x}')}{|\mathbf{x}-\mathbf{x}'|} \mathrm{d}\mathbf{x}' + \frac{\delta E_\mathrm{xc}[\rho,\chi]}{\delta \rho(\mathbf{x})}, \label{eq:Veff} \\ 
	\Delta_{\mathrm{eff}}(\mathbf{x}) &=& \frac{\delta E_\mathrm{xc}[\rho,\chi]}{\delta \chi(\mathbf{x})}, \label{eq:Deff}
\end{eqnarray}
where functional derivatives of the exchange correlation functional $E_\mathrm{xc}$ appear {requiring} a self-consistent solution of the non-linear KS-BdG equations.
The exchange correlation functional can be expressed as~\cite{Suvasini1993}
\begin{equation}
	E_\mathrm{xc}[\rho,\chi] = E_\mathrm{xc}^0[\rho] - \int\chi^*(\mathbf{x})\,\lambda\,\chi(\mathbf{x})\,\mathrm{d} \mathbf{x},
	\label{eq:Exc}
\end{equation}
where the conventional exchange-correlation functional  $E^0_\mathrm{xc}$ {is the} standard DFT {term (in the normal state).}

It is important to note that the above formulation of the KS-BdG equations assume a simplified form of the superconducting pairing kernel~\cite{Suvasini1993} (i.e.\ the second term in {Eq.~\eqref{eq:Exc}}) which reduces $\lambda$ to simple {constants within the cells surrounding the atoms} that are however allowed to take different values throughout the computational unit cell. This assumes that the pairing interaction is local in space. {This} approximation was successfully used to study conventional $s$-wave superconductors~\cite{Csire2015, CsireSchoenecker2016, Tombulk, Ruessmann2022a}, heterostructures of $s$-wave superconductors and non-superconductors~\cite{Csire2016,  Csireheterostruc2016, Ruessmann2022b}, or impurities embedded into superconductors~\cite{Tomimp, Nyari2021}.
{Hence,} the effective pairing interaction takes the simple form~\cite{Suvasini1993}
\begin{equation}
  \Delta_\mathrm{eff}(\mathbf{x}) = \lambda_i \chi(\mathbf{x})
  \label{eq:Deff2}
\end{equation}
where $\lambda_i$ is a set of effective coupling constants describing the intrinsic superconducting coupling that is allowed to depend on the position $i$ in the unit cell.

Finally, the {charge} density $\rho$ and {the anomalous density} $\chi$ are calculated from the particle {($u_\nu$)} and hole components {($v_\nu$)} of the wavefunction
\begin{eqnarray}
	\!\!\!\!\rho(\mathbf{x}) \!\!&=& \!\!2\sum_\nu f(\varepsilon_\nu)|u_\nu(\mathbf{x})|^2 \!+\! [1-\!f(\varepsilon_\nu)]|v_\nu(\mathbf{x})|^2, \\
	\chi(\mathbf{x}) &=& \sum_\nu [1-2f(\varepsilon_\nu)]u_\nu(\mathbf{x})v_\nu^*(\mathbf{x}),
\end{eqnarray}
where $f(\varepsilon)$ is the Fermi-Dirac distribution function and the summation over $\nu$ includes the full spectrum of the KS-BdG Hamiltonian.


\subsection{DFT results for superconducting Al/Au}

\begin{figure}
    \centering
    \includegraphics[width=\linewidth]{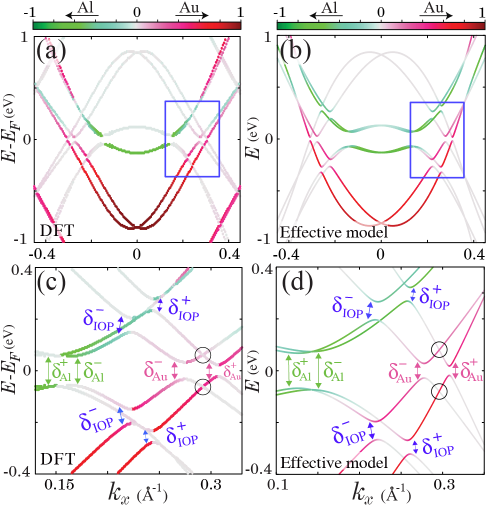}
    \caption{Superconducting band structure of the Al/Au heterostructure obtained by (a) DFT and (b) low-energy model. The red/green and grey bands indicate the particle and hole character of the BdG bands, respectively. The red/green color of the particle bands indicate the localization of the states. Panels (c) and (d) show enlarged view of the region marked by the blue box in (a) where five different superconducting avoided crossings emerge (labeled $\delta_\mathrm{Al}^\pm$, $\delta_\mathrm{Au}^\pm$, and $\delta_\mathrm{IOP}^\pm$). The absence of avoided crossings marked by black circles in (c) and (d) is {due to} pseudo-spin-rotational symmetry. For illustration purposes, we show results for scaled-up values of the superconducting pairing. The model parameters for the analytical model are those given in Table~\ref{Table1} and $\Delta=0.4F_0$.}
    \label{fig:DFT_SC}
\end{figure}

For the superconducting state, we assume that only Al has an intrinsic
superconducting coupling and set the layer-dependent coupling constant
in the {KS-BdG} calculation to 
\begin{equation}
    \lambda_{i}=\left\{ \begin{array}{l}
    \lambda_{\mathrm{Al}},\quad\mathrm{if}\ i\in\mathrm{Al},\\
    0,\quad\mathrm{else},
    \end{array}\right.
\end{equation}
where $\lambda_{\mathrm{Al}}$ is a positive real-valued constant
and $i$ is an index counting the atomic layers in the Al/Au heterostructure.
{While the value of $\lambda_\mathrm{Al}$ can be regarded as a fitting parameter in this approach, we stress that only an integral quantity, leading to the overall superconducting gap size in Al, is fitted. {Other spectral properties} like avoided crossings and proximity effects are in fact predictions {of} this theory.}
The results of our {KS-BdG} simulations and analytical model for the Al/Au heterostructure
are summarized in Fig.~\ref{fig:DFT_SC}.
For {better visibility,} we show results for scaled-up values of the superconducting {pairing}. The general trends we discuss here are, however, transferable
from large to small {pairing strengths} {with} only {quantitative changes.} We find
superconducting {gaps and avoided crossings} at low and finite excitation energies{, labelled with $\delta$ in Fig.~\ref{fig:DFT_SC}(c)}.
These avoided crossings are rooted in the $s$-wave superconductivity {induced} from the Al segment included in the DFT-based simulations by $\lambda_{i}$ (the \emph{only}
adjustable parameter in our description of the superconducting state).
The hybridization between Al and Au bands {enables} Cooper pair tunneling
from the superconductor into the metal {(see Fig.~\ref{fig:density3D}b)}. This results
in a superconducting proximity effect in the Rashba surface state {of Au}.
The large spin-splitting of the Rashba surface state allows for the
{pairing} to have triplet character because the superconducting hybridization
happens between quasiparticle bands with identical pseudo-spin degree of freedom.
{This will be further {explained} in the effective model analysis of Sec.~\ref{seq:SCmodel}.}

{The} DFT calculations {disclose} the anisotropy of the {pairing gap} (see Fig.~\ref{fig:DFT_SC}), which is
stronger for the Rashba state at smaller momentum with {$\delta^{-}_{\text{Au}}\approx0.51\delta^{\pm}_{\text{Al}}$}
and {decreases} to {$\delta^{+}_{\text{Au}}\approx0.38\delta^{\pm}_{\text{Al}}$}
for the state at larger momentum. Furthermore, we also observe
that inter-orbital {pairings appear} away from the Fermi energy, as indicated by $\delta_\mathrm{IOP}^{\pm}$, where the states with dominant Au orbital character and pseudo-spin-up intersects with the hole states with dominant Al orbital character having pseudo-spin-down degrees of freedom. This phenomenon has been referred to as inter-band pairing \cite{Interband-2009,Interband-2015,Interband-2017,Interband-2018,Interband-2019}, mirage gap \cite{Mirage}, and finite-energy Cooper pairing \cite{FECooper-2022-1,FECooper-2022-2,FECooper-2023}. However, conclusive experimental evidence supporting it is still elusive. {The} Al/Au hybrid structure {presented here} provides a simple system in which such {finite-energy pairing} can be observed.

Similar to the two {pairing gaps} $\delta_\mathrm{Au}^\pm$ in the Rashba surface state, the DFT calculation
shows that the inter-orbital {pairings $\delta_\mathrm{IOP}$} also {decrease at} larger momentum, i.e., $\delta_{\mathrm{IOP}}^-/\delta_{\mathrm{Al}}=0.60$
to {$\delta_{\mathrm{IOP2}}^+/\delta_{\mathrm{Al}}=0.44$}. 
Based on these observations, we pose {four} questions:
\begin{enumerate}[label=(Q.\arabic*), topsep=2pt, itemsep=-2pt]
    \item Is inter-orbital pairing exclusively the result of superconducting order or other mechanisms?\label{Q1}
    \item What determines the magnitude of the {finite-energy pairing}?
    \label{Q2}
    \item What is the magnitude of the induced spin-singlet and triplet components of the effective pairing?\label{Q3}
    \item What specific symmetries are responsible for protecting certain electron-hole band crossings that occur away from the Fermi energy?\label{Q4}
\end{enumerate}
{These questions will be answered in the following sections.}


\subsection{Effective low-energy model for the superconducting heterostructure}\label{seq:SCmodel}

Based on an effective low-energy, we can achieve a deeper understanding of the KS-BdG results.
The results of our low-energy model are illustrated in Figs.~\ref{fig:DFT_SC}(b) and (d). They are obtained by {the} model introduced in Sec.~\ref{seq:kpmodel}. 
In order to obtain an analytical characterization
of the {superconducting} pairing in the heterostructure, it is necessary to construct
a BdG formalism for our minimal model, \textit{cf.}\ Eq.~$\eqref{eq:modelNormalState}$.
Assuming that the superconducting {pairing} arises
from the Al layer, we model the single-particle pairing operator
as
\begin{equation}
    H_{\Delta}=\sum_{\mathbf{k}}\left(c_{\mathbf{k},\text{Al}}^{\dagger},c_{\mathbf{k},\text{Au}}^{\dagger}\right)\left(\begin{array}{cc}
    \Delta i\hat{\sigma}_{y} & 0\\
    0 & 0
    \end{array}\right)\left(\begin{array}{c}
    c_{-\mathbf{k},\text{Al}}^{\dagger}\\
    c_{-\mathbf{k},\text{Au}}^{\dagger}
    \end{array}\right),
\end{equation}
where $\Delta$ denotes the superconducting pairing strength, and
the nonvanishing diagonal entry corresponds to $s$-wave spin singlet
pairing {in} the Al {layer}. Since pure Au does not become
a superconductor at  {experimentally relevant} temperatures, the pairing {strength} in the
Au {layer} is {put to} zero.

It is {illuminating} to represent
the BdG Hamiltonian in the eigenbasis of the normal state, given in
Eq.~$\eqref{eq:modelNormalState}$, as defined by the $8\times8$ matrix
in Nambu space
\begin{equation}
    \!\!\hat{H}_{\text{\text{BdG}}}\!=\!\!\left(\!\begin{array}{cccc}
    \hat{N}_{\mathbf{k}}^{++} & 0 & \hat{\varDelta}_{\mathbf{k}}^{++} & \hat{\varDelta}_{\mathbf{k}}^{+-}\\
    0 & \hat{N}_{\mathbf{k}}^{--} & \hat{\varDelta}_{\mathbf{k}}^{-+} & \hat{\varDelta}_{\mathbf{k}}^{--}\\{}
    [\hat{\varDelta}_{\mathbf{k}}^{++}]^{\dagger} & [\hat{\varDelta}_{\mathbf{k}}^{-+}]^{\dagger} & -\hat{N}_{-\mathbf{k}}^{++} & 0\\{}
    [\hat{\varDelta}_{\mathbf{k}}^{+-}]^{\dagger} & [\hat{\varDelta}_{\mathbf{k}}^{--}]^{\dagger} & 0 & -\hat{N}_{-\mathbf{k}}^{--}
    \end{array}\!\right)\!,\label{Masoud: BdG band basis}
\end{equation}
where the diagonal entries are the normal state dispersion relations
$\hat{N}_{\mathbf{k}}^{\nu\nu}=\text{diag}(E_{\mathbf{k},+}^{\nu},E_{\mathbf{k},-}^{\nu})$
with $\nu=\pm$. Note that $E_{\mathbf{k},\pm}^{+}$ ($E_{\mathbf{k},\pm}^{-}$)
refer to the upper (lower) spin-split bands which predominantly exhibit
Al (Au) orbital character for small momenta, as can be seen
in Fig.~$\ref{Fig1.pdf}$. Furthermore, the off-diagonal block in
$\hat{H}_{\text{\text{BdG}}}$ is the pairing matrix projected onto
the band basis {(\textit{cf.}\ App.~\ref{Masoud Appendix: Band basis})} as obtained by
\begin{equation}
    \hat{\mathcal{V}}_{\mathbf{k}}^{\dagger}\text{diag}(\Delta i\hat{\sigma}_{y},0)\hat{\mathcal{V}}_{-\mathbf{k}}^{\dagger T}=\left(\begin{array}{cc}
    \hat{\varDelta}_{\mathbf{k}}^{++} & \hat{\varDelta}_{\mathbf{k}}^{+-}\\
    \hat{\varDelta}_{\mathbf{k}}^{-+} & \hat{\varDelta}_{\mathbf{k}}^{--}
    \end{array}\right),\label{ProjectedPairing}
\end{equation}
where $\hat{\mathcal{V}}_{\mathbf{k}}$ is the matrix of eigenvectors
associated to the eigenvalue $\hat{\mathscr{E}}_{\mathbf{k}}$ of the normal state Hamiltonian. $\hat{\varDelta}_{\mathbf{k}}^{++}$
($\hat{\varDelta}_{\mathbf{k}}^{--}$) {correspond to} the intra-band
pairing matrices, specifically pairing between $E_{\mathbf{k},+}^{+}$
and $E_{\mathbf{k},-}^{+}$ ($E_{\mathbf{k},+}^{-}$ and $E_{\mathbf{k},-}^{-}$)
with their hole counterparts {leading to the superconducting gap for Al, i.e., $\delta_{\mathrm{Al}}$, and the proximity-induced pairing gaps labeled by ($\delta_{\mathrm{Au}}^{\pm}$) in Fig.~\ref{fig:DFT_SC}.} Such matrices are explicitly given by
the relation
\begin{equation}
    \hat{\varDelta}_{\mathbf{k}}^{\nu\nu}=\frac{i\Delta}{2}\left(\begin{array}{cc}
    0 & 1+\nu G_{\mathbf{k}}^{-}\\
    -1-\nu G_{\mathbf{k}}^{+} & 0
    \end{array}\right),\label{Intra-BandPairingMatrix}
\end{equation}
where $\nu=+(-)$ and 
\begin{equation}
    G_{\mathbf{k}}^{\pm}=\frac{\mathcal{E}_{\text{Al}}-\mathcal{E}_{\text{Au}}^{\pm}}{\sqrt{[\mathcal{E}_{\text{Al}}-\mathcal{E}_{\text{Au}}^{\pm}]^{2}+4F_{0}^{2}}}.\label{GPM factor}
\end{equation}
In Eq.~$\eqref{ProjectedPairing}$, $\hat{\varDelta}_{\mathbf{k}}^{+-}$
($\hat{\varDelta}_{\mathbf{k}}^{-+}$) indicates the inter-orbital
pairing, i.e., pairing between electron bands $E_{\mathbf{k},+}^{+}$
and $E_{\mathbf{k},-}^{+}$ with hole band bands $-E_{-\mathbf{k},+}^{-}$
and $-E_{-\mathbf{k},-}^{-}$. This gives rise to the emergence of
finite-energy Cooper pairing {resulting} in avoided crossings at finite excitation energy ($\delta_{\mathrm{IOP}}^\pm$) in Fig.~\ref{fig:DFT_SC} (c) and (d).
The explicit form for the inter-band pairing matrix is given by
\begin{equation}
    \hat{\varDelta}_{\mathbf{k}}^{+-}=\Delta F_{0}^{2}\left(\begin{array}{cc}
    0 & \frac{-4i}{\Lambda_{\mathbf{k},1}^{-}\Lambda_{\mathbf{k},2}^{-}}\\
    \frac{4i}{\Lambda_{\mathbf{k},1}^{+}\Lambda_{\mathbf{k},2}^{+}} & 0
    \end{array}\right),\label{Inter-orbitaPairing}
\end{equation}
with 
\begin{equation}
    \Lambda_{\mathbf{k},l}^{\pm}\!=\!\sqrt{\left(\mathcal{E}_{\text{Al}}-\mathcal{E}_{\text{Au}}^{\pm}\right)^{2}\left(1+(-1)^{l}/G_{\mathbf{k}}^{\pm}\right)^{2}+4F_{0}^{2}},\!\label{GamPMfactor}
\end{equation}
where $l=\{1,2\}$. Importantly, the interplay between band
hybridization and superconductivity, manifested by $\Delta F_{0}^{2}$
in Eq.~$\eqref{Inter-orbitaPairing}$, intrinsically allows for the
emergence of finite-energy pairing. Therefore, the inter-orbital pairing is not induced solely by superconducting order but also by band hybridization in the normal state. This is the answer to question~\ref{Q1}. 


\subsection{Pairing symmetry analysis}

\begin{figure}
    \centering
    \includegraphics[width=\linewidth]{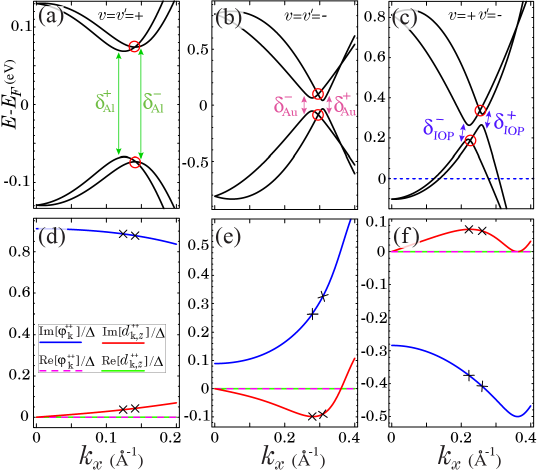}
    \caption{
    Effective superconducting excitation spectra for (a) $\hat{H}_{\mathbf{k},\text{eff}}^{++}$
    with $\Delta=0.4F_{0}$ (b) $\hat{H}_{\mathbf{k},\text{eff}}^{--}$
    with $\Delta=0.8F_{0}$, and (c) $\hat{H}_{\mathbf{k},\text{eff}}^{+-}$
    with $\Delta=0.4F_{0}$. (d-f) Real and imaginary part of the pseudo-spin-singlet
    and pseudo-spin-triplet components of the effective pairing matrix
    associated {with} the dispersion {relation} illustrated {in} the top panels. The model
    parameters are the same as those given in Table~\ref{Table1}.}
    \label{fig:EffSpectra}
\end{figure}

In order to determine the pairing symmetry in the hybrid structure, it is essential to establish an effective formalism that {concentrates on} either low \emph{or} finite excitation energies. {With this respect}, it is necessary to derive a $4\times4$ matrix formalism from the {full} $8\times8$  BdG Hamiltonian $\hat{H}_{\text{BdG}}$. This can be done by utilizing the downfolding method
{specified} in App.~$\ref{APP:FoldingDown}$. The downfolding method yields the effective model that enables us to investigate the superconducting properties within a {given} set of energy bands. As mentioned {above}, there are three distinct sets of spin-split bands where pairing occurs. These bands are characterized by $\nu=\nu^{\prime}=+(-)$, indicating that the pairing takes place at the Fermi energy, where the energy bands possess predominant Al (Au) orbital character. Another set of bands corresponds to the inter-orbital bands, where Al-dominated states intersect with Au-dominated hole states (and vice versa). Thus, the general form for the $4\times4$ effective superconducting Hamiltonian becomes 
\begin{equation}
    \hat{H}_{\mathbf{k},\text{eff}}^{+\iota}=\left(\begin{array}{cc}
    \hat{N}_{\mathbf{k}}^{++}+\hat{\xi}_{1} & \hat{\varDelta}_{\mathbf{k},\text{eff}}^{+\iota}\\{}
    [\hat{\varDelta}_{\mathbf{k},\text{eff}}^{+\iota}]^{\dagger} & -\hat{N}_{-\mathbf{k}}^{\iota\iota}+\hat{\xi}_{2}
    \end{array}\right),\label{EffectiveSupHam}
\end{equation}
where the diagonal entries $\hat{\xi}_{1(2)}$ are the energy shifts arsing
from multiband effects given by
\begin{align}
    \hat{\xi}_{1} & =\hat{\varDelta}_{\mathbf{k}}^{+\nu}\frac{1}{\omega+\hat{N}_{-\mathbf{k}}^{\nu\nu}}[\hat{\varDelta}_{\mathbf{k}}^{+\nu}]^{\dagger},\\
    \hat{\xi}_{2} & =\big[\hat{\varDelta}_{\mathbf{k}}^{-\iota}\big]^{\dagger}\frac{1}{\omega-\hat{N}_{\mathbf{k}}^{--}}\hat{\varDelta}_{\mathbf{k}}^{-\iota},
\end{align}
and $\omega$ is a constant. In addition, the effective pairing matrix in Eq.~$\eqref{EffectiveSupHam}$ becomes
\begin{equation}
    \!\hat{\varDelta}_{\mathbf{k},\text{eff}}^{+\iota}\!=\!\hat{\varDelta}_{\mathbf{k}}^{+\iota}+\hat{\varDelta}_{\mathbf{k}}^{+\nu}\frac{1}{\omega+\hat{N}_{-\mathbf{k}}^{\nu\nu}}[\hat{\varDelta}_{\mathbf{k}}^{-\nu}]^{\dagger}\frac{1}{\omega-\hat{N}_{\mathbf{k}}^{--}}\hat{\varDelta}_{\mathbf{k}}^{-\iota}.\!\!\label{EffPairing}
\end{equation}
 The effective intra-(inter-)orbital superconducting Hamiltonian, i.e.,
$\hat{H}_{\mathbf{k},\text{eff}}^{++(+-)}$, can be obtained by setting
$\iota=+(-)$ and $\nu=-(+)$.
Note that $\hat{H}_{\mathbf{k},\text{eff}}^{--}$ can also be derived
by substituting $(+)\leftrightarrow(-)$, and setting $\iota=-$ and
$\nu=+$ in Eqs.~($\ref{EffectiveSupHam}$-$\ref{EffPairing}$). 
The spectra of the effective superconducting Hamiltonians $\hat{H}_{\mathbf{k},\text{eff}}^{++}$, $\hat{H}_{\mathbf{k},\text{eff}}^{--}$, and $\hat{H}_{\mathbf{k},\text{eff}}^{+-}$ are obtained numerically and depicted in Fig.~\ref{fig:EffSpectra}(a-c).
Additionally, the magnitudes of the pseudo-spin-singlet and triplet components corresponding to these spectra are illustrated in Fig.~\ref{fig:EffSpectra}(d-f).

Importantly, the proximity-induced intra- and inter-orbital pairing states 
are mixtures of singlet {and} triplet states due to broken inversion symmetry in the Au layer. Based on our model, only the $z$-component of the $\mathbf{d}$ vector, {i.e.,  $\hat{\varDelta}^{+\iota}_{\mathbf{k},\text{eff}}(i\hat{\sigma}_{y})^{-1}=\varphi_{\mathbf{k}}^{+\iota}\hat{\sigma}_{0}+\mathbf{d}^{+\iota}_\mathbf{k}\cdot\hat{\mathbf{\sigma}}$}, either at the Fermi energy or finite excitation energies is present. According
to Eqs.~\eqref{Intra-BandPairingMatrix} and \eqref{Inter-orbitaPairing},
the pairing matrices are off-diagonal. Therefore, $\hat{\varDelta}_{\mathbf{k},\text{eff}}^{+\iota}$
becomes an off-diagonal matrix reflecting an effective mixed-pairing
state having nonvanishing pseudo-spin-singlet $\varphi_{\mathbf{k}}^{\nu\nu}$
and pseudo-spin-triplet $d_{\mathbf{k},z}^{\nu\nu}$ character obtained
as
\begin{align}
    \varphi_{\mathbf{k}}^{\nu\nu} & =\frac{i\Delta}{4}\left[2+\nu\left(G_{\mathbf{k}}^{-}+G_{\mathbf{k}}^{+}\right)\right],\label{Intra-Singlet}\\
    d_{\mathbf{k},z}^{\nu\nu} & =\frac{i\Delta}{4}\nu\left[G_{\mathbf{k}}^{-}-G_{\mathbf{k}}^{+}\right].\label{Intra_triplet}
\end{align}
where $\nu\in\{+,-\}$. Note that we have excluded terms proportional to the third order of $\Delta$ in Eqs.~\eqref{Intra-Singlet} and \eqref{Intra_triplet} as they are negligibly small in the weak pairing limit. {It is worth mentioning that the property $G_{-\mathbf{k}}^{\pm}=G_{\mathbf{k}}^{\mp}$ leads
to even (odd) parity for the pseudo-spin-singlet (triplet) state,
i.e., $\varphi_{-\mathbf{k}}^{\nu\nu}=\varphi_{\mathbf{k}}^{\nu\nu}$
$(d_{-\mathbf{k},z}^{\nu\nu}=-d_{\mathbf{k},z}^{\nu\nu})$.} The inter-orbital pairing components
take the form
\begin{align}
    \varphi_{\mathbf{k}}^{+-} & =\Delta F_{0}^{2}\left(\frac{-2i}{\Lambda_{\mathbf{k},1}^{-}\Lambda_{\mathbf{k},2}^{-}}-\frac{2i}{\Lambda_{\mathbf{k},1}^{+}\Lambda_{\mathbf{k},2}^{+}}\right),\label{Inter_singlet}\\
    d_{\mathbf{k},z}^{+-} & =\Delta F_{0}^{2}\left(\frac{-2i}{\Lambda_{\mathbf{k},1}^{-}\Lambda_{\mathbf{k},2}^{-}}+\frac{2i}{\Lambda_{\mathbf{k},1}^{+}\Lambda_{\mathbf{k},2}^{+}}\right).\label{Inter_Triplet}
\end{align}
Overall, we observe that the pseudo-spin-singlet component is consistently larger in magnitude than the triplet component, see Fig.~\ref{fig:EffSpectra}(d-f).
Note that the pairing state becomes purely pseudo-spin-singlet in the absence
of either band hybridization or Rashba spin-orbit coupling, i.e., when
$F_{0}=0$ or $\lambda=g=0$. Therefore, the pseudo-spin-triplet component
originates from the interplay between Rashba surface states and band hybridization.

\begin{figure}
    \centering
    \includegraphics[width=\linewidth]{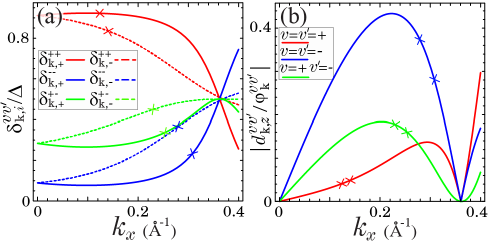}
    \caption{(a) Magnitude of the effective superconducting avoided crossings for {different} pairing potentials. (b) Strength of pseudo-spin-triplet {$d^{\nu\nu'}_{\mathbf{k},z}$} compared to pseudo-spin-singlet {$\varphi^{\nu\nu'}_{\mathbf{k},z}$}
    for the effective intra-orbital pairing potential, namely $\hat{\varDelta}_{\mathbf{k},\text{eff}}^{++}$
    and $\hat{\varDelta}_{\mathbf{k},\text{eff}}^{--}$, as well as inter-orbital
    pairing potential $\hat{\varDelta}_{\mathbf{k},\text{eff}}^{+-}$.
    }
    \label{fig:PairingAmplitudes}
\end{figure}

The size of the {avoided crossing in the spectrum of the} effective pairing Hamiltonian, as expressed in Eq.~\eqref{EffectiveSupHam}, is given by
\begin{equation}
    \!\!\!\delta_{\mathbf{k},\pm}^{\nu\nu^{\prime}}\!\!=\!\!\sqrt{|\varphi_{\mathbf{k}}^{\nu\nu^{\prime}}|^{2}\!+\!|d_{\mathbf{k},z}^{\nu\nu^{\prime}}|^{2}\!\pm\!|(\varphi_{\mathbf{k}}^{\nu\nu^{\prime}})^{*}d_{\mathbf{k},z}^{\nu\nu^{\prime}}\!+\!\varphi_{\mathbf{k}}^{\nu\nu^{\prime}}(d_{\mathbf{k},z}^{\nu\nu^{\prime}})^{*}|},\!
\end{equation}
where the third term effectively accounts for the anisotropy observed in the {{magnitude} of the avoided crossing}, as initially demonstrated in the KS-BdG simulation in Figs.~\ref{fig:DFT_SC} and \ref{fig:EffSpectra}. This {point} addresses question~\ref{Q2}.
{Note that the Fermi surface of the hybrid structure consists of
four circular rings. The inner rings are primarily composed of spin-split
Al states, while they are surrounded by predominantly spin-split
Au states. The superconducting hybridization
happens at four Fermi momenta, i.e., 
\begin{align}
\!\!\!|\mathbf{k}^{F}| & \in\{k_{1}^{\text{Al}},k_{2}^{\text{Al}},k_{1}^{\text{Au}},k_{2}^{\text{Au}}\}\!\\
 & =\!\pm\!\{0.124,0.141,0.278,0.308\}\!\,\text{Å}^{-1}.\!\!\!
\end{align}
At the above momenta, we have defined the following quantities
\begin{equation}
\!\!\!\delta_{\mathrm{Al}}\!\equiv\!\delta_{k_{1}^{\text{Al}},+}^{++}\!\approx\!\delta_{k_{2}^{\text{Al}},-}^{++},\ \delta_{\mathrm{Au}}^{-}\!\equiv\!\delta_{k_{1}^{\text{Au}},-}^{--},\ \delta_{\mathrm{Au}}^{+}\!\equiv\!\delta_{k_{2}^{\text{Au}},+}^{--}.\!\!
\end{equation}
Therefore the full pairing gap for the hybrid structure at the Fermi
energy can be determined by $\delta_{\mathrm{Au}}^{+}=\text{min}(\delta_{\mathrm{Al}},\delta_{\mathrm{Au}}^{-},\delta_{\mathrm{Au}}^{+})$. The inter-orbital Cooper pairing away from the Fermi energy
happens at momenta $k_{1}^{\text{IOP}}=0.221\,\text{Å}^{-1}$ and
$k_{2}^{\text{IOP}}=0.26\,\text{Å}^{-1}$. Accordingly, the magnitude of
finite-energy Cooper pairing is defined by $\delta_{\text{IOP}}^{-}\equiv\delta_{k_{1}^{\text{IOP}},-}^{+-}$
and $\delta_{\text{IOP}}^{+}\equiv\delta_{k_{2}^{\text{IOP}},+}^{+-}$.

}
The magnitudes of both intra- and inter-orbital {pairings} are plotted in Fig.~\ref{fig:PairingAmplitudes}(b).
Apparently, the intra-orbital bands labeled by $\nu=\nu^{\prime}=+(-)$ exhibit the largest (smallest) {pairing gap} at low momenta, indicating a dominant Al (Au) orbital character. Interestingly, the inter-orbital pairing {leads to larger avoided crossings}  compared to the intra-orbital pairing of predominantly Au electrons. The Fermi momenta for the intra-orbital energy bands are marked in blue and red crosses at {${k}=0.124\,\text{\AA}^{-1}$, ${k}=0.141\,\text{\AA}^{-1}$, ${k}=0.278\,\text{\AA}^{-1}$, and ${k}=0.308\,\text{\AA}^{-1}$}, respectively. 
At these momenta, the {pairing} anisotropy for Al-dominated states is slightly larger than the energy bands with dominant Au orbital character. Importantly, we observe that the {pairing} anisotropy disappears at critical momenta {${k}_{c}=0.368\,${\AA}$^{-1}$}, resulting in identical sizes for the pairing potentials. This occurs because the induced intra- and inter-orbital pairing {becomes} a purely pseudo-spin-singlet state by eliminating the spin-split nature of the bands, specifically, $d_{\mathbf{k}_{c},z}^{++}=d_{\mathbf{k}_{c},z}^{--}=d_{\mathbf{k}_{c},z}^{+-}=0$. The critical momenta can be determined by setting $\mathcal{E}_{\text{Al}}-\mathcal{E}_{\text{Au}}^{\pm}=0$ according to Eqs.~\eqref{GPM factor} and \eqref{GamPMfactor}. In general, the proximity-induced pairing exhibits a stronger presence of the pseudo-spin-singlet component over the triplet component, i.e., $d_{\mathbf{k},z}^{\nu\nu^{\prime}}/\varphi_{\mathbf{k}}^{\nu\nu^{\prime}}<1$, as  illustrated in Fig.~\ref{fig:PairingAmplitudes}(a). {This} answers question \ref{Q3}. {Notably}, among the various pairing potentials, the Au-dominated states, labeled by $\nu=\nu^\prime=-$, display the largest contribution from the triplet component.

\subsection{Finite-energy inter-orbital {avoided crossing} with external magnetic fields}

Note that we do not observe the occurrence
of an inter-band {pairing} between the two dominant Rashba states
displaying opposite spin-polarization marked by black
and red circles in Figures~\ref{fig:DFT_SC}
and \ref{fig:EffSpectra}, respectively. These
crossings {are} protected by time-reversal and spin-rotational symmetries.
They can, however, be lifted if an external Zeeman  field is
applied to the heterostructure. This point answers question \ref{Q4}.

\begin{figure}
    \centering
    \includegraphics[width=\linewidth]{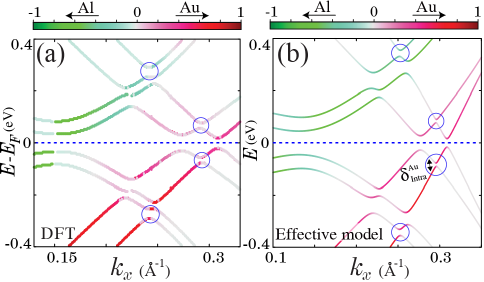}
    \caption{Superconducting band structure of Al/Au obtained by {(a)} DFT calculations and (b) analytical model in the presence of Zeeman magnetic field of size $B=2$\,mRy. Finite-energy Cooper pairings, highlighted by blue circles, emerge due to the interplay between superconductivity and magnetic field. The colorbar indicates particle (red/green) and hole (grey) components of the BdG spectra. The model parameters for the analytical model are those given in Table~\ref{Table1} and $\Delta=0.4F_0$.}
    \label{fig:DFT_SC_bfield}
\end{figure}

The effect of an external  {magnetic} field
on the electronic structure is shown in Fig.~\ref{fig:DFT_SC_bfield}, both from DFT and low-energy {model perspective}.
As the Zeeman field strength increases, the Rashba spin-split bands
undergo further splitting.This shift of the bands
leads to a decreasing superconducting energy gap in predominant Al states
because spin up and {spin} down {states} are shifted away from each other. For
large external magnetic fields, the gap closes completely and superconductivity
is destroyed at the critical field of the superconductor. Note that
the inter-band pairing between particle-hole Rashba states at finite
excitation energy is clearly visible before the superconducting gap
for Al states closes.

\section{Discussion and Conclusion} \label{seq:concl}

Our results show the existence of finite-energy pairing due to the complex multi-band effects arising in the proximity effect of heterostructures between $s$-wave superconductors and heavy metals hosting Rashba surface states. The main ingredients are: 
\begin{itemize}[itemsep=-2pt, topsep=2pt]
    \item[(i)] $s$-wave superconductivity,
    \item[(ii)] surface states originating from the normal metal,
    \item[(iii)] Rashba SOC in the normal metal,
    \item[(iv)] significant hybridization between Rashba surface states and electronic structure of the $s$-wave superconductor.
\end{itemize}
If all these requirements are met finite-energy pairing emerges between discrete states of the superconductor \emph{and} the Rashba surface states. This unconventional pairing leads to avoided crossings in the BdG band structures. In our case, discrete states in the superconductor are pronounced due to finite-size effects of the thin Al films. Their location relative to the position of the Au surface states can be fine-tuned by appropriate doping or film thickness. This allows us to control at which finite energy the inter-orbital pairing between Al and Au Rashba states occurs.

The size of the observable avoided crossings for the Al/Au heterostructure crucially depends on the superconducting gap of the superconductor (summarized in Tabs.~\ref{tab:expDeltaTcHc} and \ref{tab:DFTDeltaHc} of the Appendix).
Aluminium has a critical temperature of $T_c\approx1\,\mathrm{T}$ and a critical magnetic field of $H_c\approx10\,\mathrm{mT}$~\cite{Eisenstein1954}. In the thin film limit, both $T_c$ and $H_c$ increase substantially~\cite{Meservey1971,Court2008,VanWeerdenburg2023}, together with an increased size of the superconducting gap of $\delta_\mathrm{Al}\approx300\,\mu\mathrm{eV}$ ~\cite{Court2008}.
The proximity-induced {pairings} at zero (within the Au Rashba bands) and finite excitation energy (due to Al-Au inter-orbital pairing) are of size {$\delta_\mathrm{IOP}^\pm\approx100-200\,\mu\mathrm{eV}$}. The Au-Au inter-orbital avoided crossing that only opens up under a finite magnetic field is of size {$\delta_\mathrm{Intra}^\mathrm{Au}\approx30-50\,\mu\mathrm{eV}$} for values of the magnetic field well below the critical field of Al. 
These energy scales are rather small but within experimental reach. Note that energy resolutions below $10\,\mu\mathrm{eV}$ can be achieved at low temperatures~\cite{Schwenk2020}, see also App.~\ref{APP:experimental detection} for further details.

Suitable materials engineering might further enhance the chances to detect and eventually exploit finite-energy pairings. A strong Rashba effect is typically seen in $p$-electron materials. Superconductors whose electronic structure close to the Fermi level is dominated by $sp$-electrons, as it is the case for Al, are therefore well suited to achieve strong hybridization with Rashba materials. Consequently, other superconductors with larger superconducting gaps (e.g.\ Pb with $T_c\approx7.2\,\mathrm{K}$), that nevertheless have dominating $p$-electron character {responsible for} superconductivity, are promising to increase the observable size of the finite-energy pairing. Furthermore, replacing Au by the Bi/Ag(111) surface alloy, which shows a gigantic Rashba effect~\cite{Ast2007}, is another option for optimization. 
Apart from Rashba-type SOC, also bulk-inversion asymmetric crystals (e.g.\ BiTeI or IrBiSe~\cite{Ishizaka2011, ZhonghaoLiu2020}) where additionally Dresselhaus-type SOC-induced spin-momentum locking can be present, could be explored in this context. 
Observing {finite-energy pairing} under broken pseudo-spin-rotational symmetry benefits from a material with larger $g$-factor to increase the response to the magnetic field. InSb nanowires could be interesting systems for this purpose~\cite{Quantum2021}. Moreover, van der Waals heterostructures are rich material combinations where proximity effects and inter-orbital pairing can be explored~\cite{Geim2013}. {In these systems,} the possibility of engineering the band structures via Moir\'e superlattices provides additional knobs to tune their physical properties~\cite{Andrei2021}.

Albeit the abundance of heterostructures currently under investigation in the context of the search for MZMs or superconducting spintronics, multi-band physics in heterostructures remains largely unexplored. 
A variety of emergent phenomena can be explored in materials which show strong multi-band effects. For instance, multi-band superconductors can lead to exotic odd-frequency superconductivity~\cite{Triola2020}. Suitable materials engineering might further promote control over the mixed singlet-triplet character of the finite-energy pairing, that we demonstrate for Al/Au. This could be useful to control spin-triplet superconductivity, that in turn plays a pivotal role in superconducting spintronics~\cite{Eschrig2011, Linder2015}.
{Moreover, spin-3/2 superconductors (e.g.\ YPtBi) or superconductors that show local inversion symmetry breaking in their crystal structures (e.g.\ CeRh$_2$As$_2$) are other examples where multi-band physics and broken symmetries inherently leads to unconventional pairing~\cite{Kim2018,Khim2021}.}
Finally, novel topological superconducting pairing at finite energies~\cite{FECooper-2023} is another exciting direction for future research in real materials beyond model-based calculations.

In summary, in our combined DFT and low-energy model approach we study the proximity effect in a heterostructure of Au with strong Rashba SOC and the $s$-wave superconductor Al. We show the existence of finite-energy {pairing} in the superconducting state and analyze the mixed singlet-triplet character of the proximity-induced pairing. Combining the strengths of predictive DFT simulations with the insights from model calculations, our results pave the way towards a deeper understanding and experimental detection of multi-band effects in superconducting heterostructures.


\begin{acknowledgements}
    \section*{Acknowledgements}
    We acknowledge stimulating discussions with {Juba Bouaziz,} Julia Link, {and} Carsten Timm.
    We thank the Bavarian Ministry of Economic Affairs, Regional Development and Energy for financial support within the High-Tech Agenda Project ``Bausteine f\"ur das Quantencomputing auf Basis topologischer Materialien mit experimentellen und theoretischen Ansätzen''.
	The work was also supported by the SFB1170 ToCoTronics and the W\"urzburg-Dresden Cluster of Excellence ct.qmat, EXC2147, Project Id 390858490.
	Furthermore, this work was funded by the Deutsche Forschungsgemeinschaft (DFG, German Research Foundation) under Germany's Excellence Strategy -- Cluster of Excellence Matter and Light for Quantum Computing (ML4Q) EXC 2004/1 -- 390534769. We are also grateful for computing time granted by the JARA Vergabegremium and provided on the JARA Partition part of the supercomputer CLAIX at RWTH Aachen University (project number jara0191). 
\end{acknowledgements}

\bibliographystyle{apsrev4-1}
\bibliography{references}

\begin{thebibliography}{84}%
\makeatletter
\providecommand \@ifxundefined [1]{%
 \@ifx{#1\undefined}
}%
\providecommand \@ifnum [1]{%
 \ifnum #1\expandafter \@firstoftwo
 \else \expandafter \@secondoftwo
 \fi
}%
\providecommand \@ifx [1]{%
 \ifx #1\expandafter \@firstoftwo
 \else \expandafter \@secondoftwo
 \fi
}%
\providecommand \natexlab [1]{#1}%
\providecommand \enquote  [1]{``#1''}%
\providecommand \bibnamefont  [1]{#1}%
\providecommand \bibfnamefont [1]{#1}%
\providecommand \citenamefont [1]{#1}%
\providecommand \href@noop [0]{\@secondoftwo}%
\providecommand \href [0]{\begingroup \@sanitize@url \@href}%
\providecommand \@href[1]{\@@startlink{#1}\@@href}%
\providecommand \@@href[1]{\endgroup#1\@@endlink}%
\providecommand \@sanitize@url [0]{\catcode `\\12\catcode `\$12\catcode
  `\&12\catcode `\#12\catcode `\^12\catcode `\_12\catcode `\%12\relax}%
\providecommand \@@startlink[1]{}%
\providecommand \@@endlink[0]{}%
\providecommand \url  [0]{\begingroup\@sanitize@url \@url }%
\providecommand \@url [1]{\endgroup\@href {#1}{\urlprefix }}%
\providecommand \urlprefix  [0]{URL }%
\providecommand \Eprint [0]{\href }%
\providecommand \doibase [0]{http://dx.doi.org/}%
\providecommand \selectlanguage [0]{\@gobble}%
\providecommand \bibinfo  [0]{\@secondoftwo}%
\providecommand \bibfield  [0]{\@secondoftwo}%
\providecommand \translation [1]{[#1]}%
\providecommand \BibitemOpen [0]{}%
\providecommand \bibitemStop [0]{}%
\providecommand \bibitemNoStop [0]{.\EOS\space}%
\providecommand \EOS [0]{\spacefactor3000\relax}%
\providecommand \BibitemShut  [1]{\csname bibitem#1\endcsname}%
\let\auto@bib@innerbib\@empty
\bibitem [{\citenamefont {Manchon}\ \emph {et~al.}(2015)\citenamefont
  {Manchon}, \citenamefont {Koo}, \citenamefont {Nitta}, \citenamefont
  {Frolov},\ and\ \citenamefont {Duine}}]{Manchon2015}%
  \BibitemOpen
  \bibfield  {author} {\bibinfo {author} {\bibfnamefont {A.}~\bibnamefont
  {Manchon}}, \bibinfo {author} {\bibfnamefont {H.~C.}\ \bibnamefont {Koo}},
  \bibinfo {author} {\bibfnamefont {J.}~\bibnamefont {Nitta}}, \bibinfo
  {author} {\bibfnamefont {S.~M.}\ \bibnamefont {Frolov}}, \ and\ \bibinfo
  {author} {\bibfnamefont {R.~A.}\ \bibnamefont {Duine}},\ }\href {\doibase
  10.1038/nmat4360} {\bibfield  {journal} {\bibinfo  {journal} {Nature
  Materials}\ }\textbf {\bibinfo {volume} {14}},\ \bibinfo {pages} {871}
  (\bibinfo {year} {2015})}\BibitemShut {NoStop}%
\bibitem [{\citenamefont {Bihlmayer}\ \emph {et~al.}(2022)\citenamefont
  {Bihlmayer}, \citenamefont {No{\"{e}}l}, \citenamefont {Vyalikh},
  \citenamefont {Chulkov},\ and\ \citenamefont {Manchon}}]{Bihlmayer2022}%
  \BibitemOpen
  \bibfield  {author} {\bibinfo {author} {\bibfnamefont {G.}~\bibnamefont
  {Bihlmayer}}, \bibinfo {author} {\bibfnamefont {P.}~\bibnamefont
  {No{\"{e}}l}}, \bibinfo {author} {\bibfnamefont {D.~V.}\ \bibnamefont
  {Vyalikh}}, \bibinfo {author} {\bibfnamefont {E.~V.}\ \bibnamefont
  {Chulkov}}, \ and\ \bibinfo {author} {\bibfnamefont {A.}~\bibnamefont
  {Manchon}},\ }\href {\doibase 10.1038/s42254-022-00490-y} {\bibfield
  {journal} {\bibinfo  {journal} {Nature Reviews Physics}\ }\textbf {\bibinfo
  {volume} {4}},\ \bibinfo {pages} {642} (\bibinfo {year} {2022})}\BibitemShut
  {NoStop}%
\bibitem [{\citenamefont {Fert}\ \emph {et~al.}(2017)\citenamefont {Fert},
  \citenamefont {Reyren},\ and\ \citenamefont {Cros}}]{Fert2017}%
  \BibitemOpen
  \bibfield  {author} {\bibinfo {author} {\bibfnamefont {A.}~\bibnamefont
  {Fert}}, \bibinfo {author} {\bibfnamefont {N.}~\bibnamefont {Reyren}}, \ and\
  \bibinfo {author} {\bibfnamefont {V.}~\bibnamefont {Cros}},\ }\href {\doibase
  10.1038/natrevmats.2017.31} {\bibfield  {journal} {\bibinfo  {journal} {Nat
  Rev Mater}\ }\textbf {\bibinfo {volume} {2}},\ \bibinfo {pages} {17031}
  (\bibinfo {year} {2017})}\BibitemShut {NoStop}%
\bibitem [{\citenamefont {Hasan}\ and\ \citenamefont {Kane}(2010)}]{Hasan2010}%
  \BibitemOpen
  \bibfield  {author} {\bibinfo {author} {\bibfnamefont {M.~Z.}\ \bibnamefont
  {Hasan}}\ and\ \bibinfo {author} {\bibfnamefont {C.~L.}\ \bibnamefont
  {Kane}},\ }\href {\doibase 10.1103/RevModPhys.82.3045} {\bibfield  {journal}
  {\bibinfo  {journal} {Rev. Mod. Phys.}\ }\textbf {\bibinfo {volume} {82}},\
  \bibinfo {pages} {3045} (\bibinfo {year} {2010})}\BibitemShut {NoStop}%
\bibitem [{\citenamefont {Alicea}\ \emph {et~al.}(2011)\citenamefont {Alicea},
  \citenamefont {Oreg}, \citenamefont {Refael}, \citenamefont {von Oppen},\
  and\ \citenamefont {Fisher}}]{Alicea2011}%
  \BibitemOpen
  \bibfield  {author} {\bibinfo {author} {\bibfnamefont {J.}~\bibnamefont
  {Alicea}}, \bibinfo {author} {\bibfnamefont {Y.}~\bibnamefont {Oreg}},
  \bibinfo {author} {\bibfnamefont {G.}~\bibnamefont {Refael}}, \bibinfo
  {author} {\bibfnamefont {F.}~\bibnamefont {von Oppen}}, \ and\ \bibinfo
  {author} {\bibfnamefont {M.~P.~A.}\ \bibnamefont {Fisher}},\ }\href {\doibase
  10.1038/nphys1915} {\bibfield  {journal} {\bibinfo  {journal} {Nature
  Physics}\ }\textbf {\bibinfo {volume} {7}},\ \bibinfo {pages} {412} (\bibinfo
  {year} {2011})}\BibitemShut {NoStop}%
\bibitem [{\citenamefont {Lutchyn}\ \emph {et~al.}(2018)\citenamefont
  {Lutchyn}, \citenamefont {Bakkers}, \citenamefont {Kouwenhoven},
  \citenamefont {Krogstrup}, \citenamefont {Marcus},\ and\ \citenamefont
  {Oreg}}]{Lutchyn2018}%
  \BibitemOpen
  \bibfield  {author} {\bibinfo {author} {\bibfnamefont {R.~M.}\ \bibnamefont
  {Lutchyn}}, \bibinfo {author} {\bibfnamefont {E.~P. A.~M.}\ \bibnamefont
  {Bakkers}}, \bibinfo {author} {\bibfnamefont {L.~P.}\ \bibnamefont
  {Kouwenhoven}}, \bibinfo {author} {\bibfnamefont {P.}~\bibnamefont
  {Krogstrup}}, \bibinfo {author} {\bibfnamefont {C.~M.}\ \bibnamefont
  {Marcus}}, \ and\ \bibinfo {author} {\bibfnamefont {Y.}~\bibnamefont
  {Oreg}},\ }\href {\doibase 10.1038/s41578-018-0003-1} {\bibfield  {journal}
  {\bibinfo  {journal} {Nature Reviews Materials}\ }\textbf {\bibinfo {volume}
  {3}},\ \bibinfo {pages} {52} (\bibinfo {year} {2018})}\BibitemShut {NoStop}%
\bibitem [{\citenamefont {Frolov}\ \emph {et~al.}(2020)\citenamefont {Frolov},
  \citenamefont {Manfra},\ and\ \citenamefont {Sau}}]{Frolov2020}%
  \BibitemOpen
  \bibfield  {author} {\bibinfo {author} {\bibfnamefont {S.~M.}\ \bibnamefont
  {Frolov}}, \bibinfo {author} {\bibfnamefont {M.~J.}\ \bibnamefont {Manfra}},
  \ and\ \bibinfo {author} {\bibfnamefont {J.~D.}\ \bibnamefont {Sau}},\ }\href
  {\doibase 10.1038/s41567-020-0925-6} {\bibfield  {journal} {\bibinfo
  {journal} {Nature Physics}\ }\textbf {\bibinfo {volume} {16}},\ \bibinfo
  {pages} {718} (\bibinfo {year} {2020})}\BibitemShut {NoStop}%
\bibitem [{\citenamefont {Flensberg}\ \emph {et~al.}(2021)\citenamefont
  {Flensberg}, \citenamefont {von Oppen},\ and\ \citenamefont
  {Stern}}]{Flensberg2021}%
  \BibitemOpen
  \bibfield  {author} {\bibinfo {author} {\bibfnamefont {K.}~\bibnamefont
  {Flensberg}}, \bibinfo {author} {\bibfnamefont {F.}~\bibnamefont {von
  Oppen}}, \ and\ \bibinfo {author} {\bibfnamefont {A.}~\bibnamefont {Stern}},\
  }\href {\doibase 10.1038/s41578-021-00336-6} {\bibfield  {journal} {\bibinfo
  {journal} {Nature Reviews Materials}\ }\textbf {\bibinfo {volume} {6}},\
  \bibinfo {pages} {944–958} (\bibinfo {year} {2021})}\BibitemShut {NoStop}%
\bibitem [{\citenamefont {Rashba}(1960)}]{Rashba1960}%
  \BibitemOpen
  \bibfield  {author} {\bibinfo {author} {\bibfnamefont {E.~I.}\ \bibnamefont
  {Rashba}},\ }\href@noop {} {\bibfield  {journal} {\bibinfo  {journal} {Sov.
  Phys. Solid State}\ }\textbf {\bibinfo {volume} {2}},\ \bibinfo {pages}
  {1109} (\bibinfo {year} {1960})}\BibitemShut {NoStop}%
\bibitem [{\citenamefont {Avsar}\ \emph {et~al.}(2014)\citenamefont {Avsar},
  \citenamefont {Tan}, \citenamefont {Taychatanapat}, \citenamefont
  {Balakrishnan}, \citenamefont {Koon}, \citenamefont {Yeo}, \citenamefont
  {Lahiri}, \citenamefont {Carvalho}, \citenamefont {Rodin}, \citenamefont
  {{O’Farrell}}, \citenamefont {G.~Eda},\ and\ \citenamefont
  {\"Ozyilmaz}}]{Avsar2014}%
  \BibitemOpen
  \bibfield  {author} {\bibinfo {author} {\bibfnamefont {A.}~\bibnamefont
  {Avsar}}, \bibinfo {author} {\bibfnamefont {J.}~\bibnamefont {Tan}}, \bibinfo
  {author} {\bibfnamefont {T.}~\bibnamefont {Taychatanapat}}, \bibinfo {author}
  {\bibfnamefont {J.}~\bibnamefont {Balakrishnan}}, \bibinfo {author}
  {\bibfnamefont {G.}~\bibnamefont {Koon}}, \bibinfo {author} {\bibfnamefont
  {Y.}~\bibnamefont {Yeo}}, \bibinfo {author} {\bibfnamefont {J.}~\bibnamefont
  {Lahiri}}, \bibinfo {author} {\bibfnamefont {A.}~\bibnamefont {Carvalho}},
  \bibinfo {author} {\bibfnamefont {A.}~\bibnamefont {Rodin}}, \bibinfo
  {author} {\bibfnamefont {E.}~\bibnamefont {{O’Farrell}}}, \bibinfo {author}
  {\bibfnamefont {A.~C.~N.}\ \bibnamefont {G.~Eda}}, \ and\ \bibinfo {author}
  {\bibfnamefont {B.}~\bibnamefont {\"Ozyilmaz}},\ }\href {\doibase
  10.1038/ncomms5875} {\bibfield  {journal} {\bibinfo  {journal} {Nature
  Communications}\ }\textbf {\bibinfo {volume} {5}},\ \bibinfo {pages} {4875}
  (\bibinfo {year} {2014})}\BibitemShut {NoStop}%
\bibitem [{\citenamefont {Gmitra}\ and\ \citenamefont
  {Fabian}(2017)}]{Gmitra2017}%
  \BibitemOpen
  \bibfield  {author} {\bibinfo {author} {\bibfnamefont {M.}~\bibnamefont
  {Gmitra}}\ and\ \bibinfo {author} {\bibfnamefont {J.}~\bibnamefont
  {Fabian}},\ }\href {\doibase 10.1103/PhysRevLett.119.146401} {\bibfield
  {journal} {\bibinfo  {journal} {Phys. Rev. Lett.}\ }\textbf {\bibinfo
  {volume} {119}},\ \bibinfo {pages} {146401} (\bibinfo {year}
  {2017})}\BibitemShut {NoStop}%
\bibitem [{\citenamefont {Island}\ \emph {et~al.}(2019)\citenamefont {Island},
  \citenamefont {Cui}, \citenamefont {Lewandowski}, \citenamefont {Khoo},
  \citenamefont {{E. M.}}, \citenamefont {Zhou}, \citenamefont {Rhodes},
  \citenamefont {Hone}, \citenamefont {Taniguchi}, \citenamefont {Watanabe},
  \citenamefont {Levitov}, \citenamefont {Zaletel},\ and\ \citenamefont
  {Young}}]{Island2019}%
  \BibitemOpen
  \bibfield  {author} {\bibinfo {author} {\bibfnamefont {J.}~\bibnamefont
  {Island}}, \bibinfo {author} {\bibfnamefont {X.}~\bibnamefont {Cui}},
  \bibinfo {author} {\bibfnamefont {C.}~\bibnamefont {Lewandowski}}, \bibinfo
  {author} {\bibfnamefont {J.}~\bibnamefont {Khoo}}, \bibinfo {author}
  {\bibfnamefont {S.}~\bibnamefont {{E. M.}}}, \bibinfo {author} {\bibfnamefont
  {H.}~\bibnamefont {Zhou}}, \bibinfo {author} {\bibfnamefont {D.}~\bibnamefont
  {Rhodes}}, \bibinfo {author} {\bibfnamefont {J.}~\bibnamefont {Hone}},
  \bibinfo {author} {\bibfnamefont {T.}~\bibnamefont {Taniguchi}}, \bibinfo
  {author} {\bibfnamefont {K.}~\bibnamefont {Watanabe}}, \bibinfo {author}
  {\bibfnamefont {L.}~\bibnamefont {Levitov}}, \bibinfo {author} {\bibfnamefont
  {M.}~\bibnamefont {Zaletel}}, \ and\ \bibinfo {author} {\bibfnamefont
  {A.}~\bibnamefont {Young}},\ }\href {\doibase 10.1038/s41586-019-1304-2}
  {\bibfield  {journal} {\bibinfo  {journal} {Nature}\ }\textbf {\bibinfo
  {volume} {571}},\ \bibinfo {pages} {85–89} (\bibinfo {year}
  {2019})}\BibitemShut {NoStop}%
\bibitem [{\citenamefont {Nayak}\ \emph {et~al.}(2008)\citenamefont {Nayak},
  \citenamefont {Simon}, \citenamefont {Stern}, \citenamefont {Freedman},\ and\
  \citenamefont {Das~Sarma}}]{Nayak2008}%
  \BibitemOpen
  \bibfield  {author} {\bibinfo {author} {\bibfnamefont {C.}~\bibnamefont
  {Nayak}}, \bibinfo {author} {\bibfnamefont {S.~H.}\ \bibnamefont {Simon}},
  \bibinfo {author} {\bibfnamefont {A.}~\bibnamefont {Stern}}, \bibinfo
  {author} {\bibfnamefont {M.}~\bibnamefont {Freedman}}, \ and\ \bibinfo
  {author} {\bibfnamefont {S.}~\bibnamefont {Das~Sarma}},\ }\href {\doibase
  10.1103/RevModPhys.80.1083} {\bibfield  {journal} {\bibinfo  {journal} {Rev.
  Mod. Phys.}\ }\textbf {\bibinfo {volume} {80}},\ \bibinfo {pages} {1083}
  (\bibinfo {year} {2008})}\BibitemShut {NoStop}%
\bibitem [{\citenamefont {De~Gennes}(1966)}]{deGennes1966}%
  \BibitemOpen
  \bibfield  {author} {\bibinfo {author} {\bibfnamefont {P.~G.}\ \bibnamefont
  {De~Gennes}},\ }\href@noop {} {\emph {\bibinfo {title} {{Superconductivity of
  metals and alloys}}}}\ (\bibinfo  {publisher} {W. A. Benjamin},\ \bibinfo
  {address} {New York, NY},\ \bibinfo {year} {1966})\BibitemShut {NoStop}%
\bibitem [{\citenamefont {Zhu}(2016)}]{BdGbook}%
  \BibitemOpen
  \bibfield  {author} {\bibinfo {author} {\bibfnamefont {J.-X.}\ \bibnamefont
  {Zhu}},\ }\href {\doibase 10.1007/978-3-319-31314-6} {\emph {\bibinfo {title}
  {{Bogoliubov-de Gennes Method and Its Applications}}}}\ (\bibinfo
  {publisher} {Springer International Publishing},\ \bibinfo {year}
  {2016})\BibitemShut {NoStop}%
\bibitem [{\citenamefont {Oliveira}\ \emph {et~al.}(1988)\citenamefont
  {Oliveira}, \citenamefont {Gross},\ and\ \citenamefont
  {Kohn}}]{Oliveira1988}%
  \BibitemOpen
  \bibfield  {author} {\bibinfo {author} {\bibfnamefont {L.~N.}\ \bibnamefont
  {Oliveira}}, \bibinfo {author} {\bibfnamefont {E.~K.~U.}\ \bibnamefont
  {Gross}}, \ and\ \bibinfo {author} {\bibfnamefont {W.}~\bibnamefont {Kohn}},\
  }\href {\doibase 10.1103/PhysRevLett.60.2430} {\bibfield  {journal} {\bibinfo
   {journal} {Phys. Rev. Lett.}\ }\textbf {\bibinfo {volume} {60}},\ \bibinfo
  {pages} {2430} (\bibinfo {year} {1988})}\BibitemShut {NoStop}%
\bibitem [{\citenamefont {L\"uders}\ \emph {et~al.}(2005)\citenamefont
  {L\"uders}, \citenamefont {Marques}, \citenamefont {Lathiotakis},
  \citenamefont {Floris}, \citenamefont {Profeta}, \citenamefont {Fast},
  \citenamefont {Continenza}, \citenamefont {Massidda},\ and\ \citenamefont
  {Gross}}]{Lueders2005}%
  \BibitemOpen
  \bibfield  {author} {\bibinfo {author} {\bibfnamefont {M.}~\bibnamefont
  {L\"uders}}, \bibinfo {author} {\bibfnamefont {M.~A.~L.}\ \bibnamefont
  {Marques}}, \bibinfo {author} {\bibfnamefont {N.~N.}\ \bibnamefont
  {Lathiotakis}}, \bibinfo {author} {\bibfnamefont {A.}~\bibnamefont {Floris}},
  \bibinfo {author} {\bibfnamefont {G.}~\bibnamefont {Profeta}}, \bibinfo
  {author} {\bibfnamefont {L.}~\bibnamefont {Fast}}, \bibinfo {author}
  {\bibfnamefont {A.}~\bibnamefont {Continenza}}, \bibinfo {author}
  {\bibfnamefont {S.}~\bibnamefont {Massidda}}, \ and\ \bibinfo {author}
  {\bibfnamefont {E.~K.~U.}\ \bibnamefont {Gross}},\ }\href {\doibase
  10.1103/PhysRevB.72.024545} {\bibfield  {journal} {\bibinfo  {journal} {Phys.
  Rev. B}\ }\textbf {\bibinfo {volume} {72}},\ \bibinfo {pages} {024545}
  (\bibinfo {year} {2005})}\BibitemShut {NoStop}%
\bibitem [{\citenamefont {Csire}\ \emph {et~al.}(2015)\citenamefont {Csire},
  \citenamefont {\'Ujfalussy}, \citenamefont {Cserti},\ and\ \citenamefont
  {Gy\ifmmode~\mbox{\H{o}}\else \H{o}\fi{}rffy}}]{Csire2015}%
  \BibitemOpen
  \bibfield  {author} {\bibinfo {author} {\bibfnamefont {G.}~\bibnamefont
  {Csire}}, \bibinfo {author} {\bibfnamefont {B.}~\bibnamefont {\'Ujfalussy}},
  \bibinfo {author} {\bibfnamefont {J.}~\bibnamefont {Cserti}}, \ and\ \bibinfo
  {author} {\bibfnamefont {B.}~\bibnamefont {Gy\ifmmode~\mbox{\H{o}}\else
  \H{o}\fi{}rffy}},\ }\href {\doibase 10.1103/PhysRevB.91.165142} {\bibfield
  {journal} {\bibinfo  {journal} {Phys. Rev. B}\ }\textbf {\bibinfo {volume}
  {91}},\ \bibinfo {pages} {165142} (\bibinfo {year} {2015})}\BibitemShut
  {NoStop}%
\bibitem [{\citenamefont {Kawamura}\ \emph {et~al.}(2020)\citenamefont
  {Kawamura}, \citenamefont {Hizume},\ and\ \citenamefont
  {Ozaki}}]{Kawamura2020}%
  \BibitemOpen
  \bibfield  {author} {\bibinfo {author} {\bibfnamefont {M.}~\bibnamefont
  {Kawamura}}, \bibinfo {author} {\bibfnamefont {Y.}~\bibnamefont {Hizume}}, \
  and\ \bibinfo {author} {\bibfnamefont {T.}~\bibnamefont {Ozaki}},\ }\href
  {\doibase 10.1103/PhysRevB.101.134511} {\bibfield  {journal} {\bibinfo
  {journal} {Phys. Rev. B}\ }\textbf {\bibinfo {volume} {101}},\ \bibinfo
  {pages} {134511} (\bibinfo {year} {2020})}\BibitemShut {NoStop}%
\bibitem [{\citenamefont {Linscheid}\ \emph {et~al.}(2015)\citenamefont
  {Linscheid}, \citenamefont {Sanna}, \citenamefont {Essenberger},\ and\
  \citenamefont {Gross}}]{Linscheid2015}%
  \BibitemOpen
  \bibfield  {author} {\bibinfo {author} {\bibfnamefont {A.}~\bibnamefont
  {Linscheid}}, \bibinfo {author} {\bibfnamefont {A.}~\bibnamefont {Sanna}},
  \bibinfo {author} {\bibfnamefont {F.}~\bibnamefont {Essenberger}}, \ and\
  \bibinfo {author} {\bibfnamefont {E.~K.~U.}\ \bibnamefont {Gross}},\ }\href
  {\doibase 10.1103/PhysRevB.92.024505} {\bibfield  {journal} {\bibinfo
  {journal} {Phys. Rev. B}\ }\textbf {\bibinfo {volume} {92}},\ \bibinfo
  {pages} {024505} (\bibinfo {year} {2015})}\BibitemShut {NoStop}%
\bibitem [{\citenamefont {Scheurer}\ \emph {et~al.}(2017)\citenamefont
  {Scheurer}, \citenamefont {Agterberg},\ and\ \citenamefont
  {Schmalian}}]{Scheurer2017}%
  \BibitemOpen
  \bibfield  {author} {\bibinfo {author} {\bibfnamefont {M.~S.}\ \bibnamefont
  {Scheurer}}, \bibinfo {author} {\bibfnamefont {D.~F.}\ \bibnamefont
  {Agterberg}}, \ and\ \bibinfo {author} {\bibfnamefont {J.}~\bibnamefont
  {Schmalian}},\ }\href {\doibase 10.1038/s41535-016-0008-1} {\bibfield
  {journal} {\bibinfo  {journal} {npj Quantum Materials}\ }\textbf {\bibinfo
  {volume} {2}},\ \bibinfo {pages} {9} (\bibinfo {year} {2017})}\BibitemShut
  {NoStop}%
\bibitem [{\citenamefont {{LaShell}}\ \emph {et~al.}(1996)\citenamefont
  {{LaShell}}, \citenamefont {{McDougall}},\ and\ \citenamefont
  {Jensen}}]{LaShell1996}%
  \BibitemOpen
  \bibfield  {author} {\bibinfo {author} {\bibfnamefont {S.}~\bibnamefont
  {{LaShell}}}, \bibinfo {author} {\bibfnamefont {B.~A.}\ \bibnamefont
  {{McDougall}}}, \ and\ \bibinfo {author} {\bibfnamefont {E.}~\bibnamefont
  {Jensen}},\ }\href {\doibase 10.1103/PhysRevLett.77.3419} {\bibfield
  {journal} {\bibinfo  {journal} {Phys. Rev. Lett.}\ }\textbf {\bibinfo
  {volume} {77}},\ \bibinfo {pages} {3419} (\bibinfo {year}
  {1996})}\BibitemShut {NoStop}%
\bibitem [{\citenamefont {Varykhalov}\ \emph {et~al.}(2012)\citenamefont
  {Varykhalov}, \citenamefont {Marchenko}, \citenamefont {Scholz},
  \citenamefont {Rienks}, \citenamefont {Kim}, \citenamefont {Bihlmayer},
  \citenamefont {S\'anchez-Barriga},\ and\ \citenamefont
  {Rader}}]{Varykhalov2012}%
  \BibitemOpen
  \bibfield  {author} {\bibinfo {author} {\bibfnamefont {A.}~\bibnamefont
  {Varykhalov}}, \bibinfo {author} {\bibfnamefont {D.}~\bibnamefont
  {Marchenko}}, \bibinfo {author} {\bibfnamefont {M.~R.}\ \bibnamefont
  {Scholz}}, \bibinfo {author} {\bibfnamefont {E.~D.~L.}\ \bibnamefont
  {Rienks}}, \bibinfo {author} {\bibfnamefont {T.~K.}\ \bibnamefont {Kim}},
  \bibinfo {author} {\bibfnamefont {G.}~\bibnamefont {Bihlmayer}}, \bibinfo
  {author} {\bibfnamefont {J.}~\bibnamefont {S\'anchez-Barriga}}, \ and\
  \bibinfo {author} {\bibfnamefont {O.}~\bibnamefont {Rader}},\ }\href
  {\doibase 10.1103/PhysRevLett.108.066804} {\bibfield  {journal} {\bibinfo
  {journal} {Phys. Rev. Lett.}\ }\textbf {\bibinfo {volume} {108}},\ \bibinfo
  {pages} {066804} (\bibinfo {year} {2012})}\BibitemShut {NoStop}%
\bibitem [{\citenamefont {Ast}\ \emph {et~al.}(2007)\citenamefont {Ast},
  \citenamefont {Henk}, \citenamefont {Ernst}, \citenamefont {Moreschini},
  \citenamefont {Falub}, \citenamefont {Pacil\'e}, \citenamefont {Bruno},
  \citenamefont {Kern},\ and\ \citenamefont {Grioni}}]{Ast2007}%
  \BibitemOpen
  \bibfield  {author} {\bibinfo {author} {\bibfnamefont {C.~R.}\ \bibnamefont
  {Ast}}, \bibinfo {author} {\bibfnamefont {J.}~\bibnamefont {Henk}}, \bibinfo
  {author} {\bibfnamefont {A.}~\bibnamefont {Ernst}}, \bibinfo {author}
  {\bibfnamefont {L.}~\bibnamefont {Moreschini}}, \bibinfo {author}
  {\bibfnamefont {M.~C.}\ \bibnamefont {Falub}}, \bibinfo {author}
  {\bibfnamefont {D.}~\bibnamefont {Pacil\'e}}, \bibinfo {author}
  {\bibfnamefont {P.}~\bibnamefont {Bruno}}, \bibinfo {author} {\bibfnamefont
  {K.}~\bibnamefont {Kern}}, \ and\ \bibinfo {author} {\bibfnamefont
  {M.}~\bibnamefont {Grioni}},\ }\href {\doibase 0.1103/PhysRevLett.98.186807}
  {\bibfield  {journal} {\bibinfo  {journal} {Phys. Rev. Lett.}\ }\textbf
  {\bibinfo {volume} {98}},\ \bibinfo {pages} {186807} (\bibinfo {year}
  {2007})}\BibitemShut {NoStop}%
\bibitem [{\citenamefont {Nadj-Perge}\ \emph {et~al.}(2010)\citenamefont
  {Nadj-Perge}, \citenamefont {Frolov}, \citenamefont {Bakkers},\ and\
  \citenamefont {Kouwenhoven}}]{NadjPerge2010}%
  \BibitemOpen
  \bibfield  {author} {\bibinfo {author} {\bibfnamefont {S.}~\bibnamefont
  {Nadj-Perge}}, \bibinfo {author} {\bibfnamefont {S.~M.}\ \bibnamefont
  {Frolov}}, \bibinfo {author} {\bibfnamefont {E.~P. A.~M.}\ \bibnamefont
  {Bakkers}}, \ and\ \bibinfo {author} {\bibfnamefont {L.~P.}\ \bibnamefont
  {Kouwenhoven}},\ }\href {\doibase 10.1038/nature09682} {\bibfield  {journal}
  {\bibinfo  {journal} {Nature}\ }\textbf {\bibinfo {volume} {468}},\ \bibinfo
  {pages} {1084–1087} (\bibinfo {year} {2010})}\BibitemShut {NoStop}%
\bibitem [{\citenamefont {King}\ \emph {et~al.}(2011)\citenamefont {King},
  \citenamefont {Hatch}, \citenamefont {Bianchi}, \citenamefont {Ovsyannikov},
  \citenamefont {Lupulescu}, \citenamefont {Landolt}, \citenamefont {Slomski},
  \citenamefont {Dil}, \citenamefont {Guan}, \citenamefont {Mi}, \citenamefont
  {Rienks}, \citenamefont {Fink}, \citenamefont {Lindblad}, \citenamefont
  {Svensson}, \citenamefont {Bao}, \citenamefont {Balakrishnan}, \citenamefont
  {Iversen}, \citenamefont {Osterwalder}, \citenamefont {Eberhardt},
  \citenamefont {Baumberger},\ and\ \citenamefont {Hofmann}}]{King2011}%
  \BibitemOpen
  \bibfield  {author} {\bibinfo {author} {\bibfnamefont {P.~D.~C.}\
  \bibnamefont {King}}, \bibinfo {author} {\bibfnamefont {R.~C.}\ \bibnamefont
  {Hatch}}, \bibinfo {author} {\bibfnamefont {M.}~\bibnamefont {Bianchi}},
  \bibinfo {author} {\bibfnamefont {R.}~\bibnamefont {Ovsyannikov}}, \bibinfo
  {author} {\bibfnamefont {C.}~\bibnamefont {Lupulescu}}, \bibinfo {author}
  {\bibfnamefont {G.}~\bibnamefont {Landolt}}, \bibinfo {author} {\bibfnamefont
  {B.}~\bibnamefont {Slomski}}, \bibinfo {author} {\bibfnamefont {J.~H.}\
  \bibnamefont {Dil}}, \bibinfo {author} {\bibfnamefont {D.}~\bibnamefont
  {Guan}}, \bibinfo {author} {\bibfnamefont {J.~L.}\ \bibnamefont {Mi}},
  \bibinfo {author} {\bibfnamefont {E.~D.~L.}\ \bibnamefont {Rienks}}, \bibinfo
  {author} {\bibfnamefont {J.}~\bibnamefont {Fink}}, \bibinfo {author}
  {\bibfnamefont {A.}~\bibnamefont {Lindblad}}, \bibinfo {author}
  {\bibfnamefont {S.}~\bibnamefont {Svensson}}, \bibinfo {author}
  {\bibfnamefont {S.}~\bibnamefont {Bao}}, \bibinfo {author} {\bibfnamefont
  {G.}~\bibnamefont {Balakrishnan}}, \bibinfo {author} {\bibfnamefont {B.~B.}\
  \bibnamefont {Iversen}}, \bibinfo {author} {\bibfnamefont {J.}~\bibnamefont
  {Osterwalder}}, \bibinfo {author} {\bibfnamefont {W.}~\bibnamefont
  {Eberhardt}}, \bibinfo {author} {\bibfnamefont {F.}~\bibnamefont
  {Baumberger}}, \ and\ \bibinfo {author} {\bibfnamefont {P.}~\bibnamefont
  {Hofmann}},\ }\href {\doibase 10.1103/PhysRevLett.107.096802} {\bibfield
  {journal} {\bibinfo  {journal} {Phys. Rev. Lett.}\ }\textbf {\bibinfo
  {volume} {107}},\ \bibinfo {pages} {096802} (\bibinfo {year}
  {2011})}\BibitemShut {NoStop}%
\bibitem [{\citenamefont {Breunig}\ \emph {et~al.}(2021)\citenamefont
  {Breunig}, \citenamefont {Zhang}, \citenamefont {Trauzettel},\ and\
  \citenamefont {Klapwijk}}]{Breunig2021}%
  \BibitemOpen
  \bibfield  {author} {\bibinfo {author} {\bibfnamefont {D.}~\bibnamefont
  {Breunig}}, \bibinfo {author} {\bibfnamefont {S.-B.}\ \bibnamefont {Zhang}},
  \bibinfo {author} {\bibfnamefont {B.}~\bibnamefont {Trauzettel}}, \ and\
  \bibinfo {author} {\bibfnamefont {T.~M.}\ \bibnamefont {Klapwijk}},\ }\href
  {\doibase 10.1103/PhysRevB.103.165414} {\bibfield  {journal} {\bibinfo
  {journal} {Phys. Rev. B}\ }\textbf {\bibinfo {volume} {103}},\ \bibinfo
  {pages} {165414} (\bibinfo {year} {2021})}\BibitemShut {NoStop}%
\bibitem [{\citenamefont {Ando}\ \emph {et~al.}(2020)\citenamefont {Ando},
  \citenamefont {Miyasaka}, \citenamefont {Li}, \citenamefont {Ishizuka},
  \citenamefont {Arakawa}, \citenamefont {Shiota}, \citenamefont {Moriyama},
  \citenamefont {Yanase},\ and\ \citenamefont {Ono}}]{Ando2020}%
  \BibitemOpen
  \bibfield  {author} {\bibinfo {author} {\bibfnamefont {F.}~\bibnamefont
  {Ando}}, \bibinfo {author} {\bibfnamefont {Y.}~\bibnamefont {Miyasaka}},
  \bibinfo {author} {\bibfnamefont {T.}~\bibnamefont {Li}}, \bibinfo {author}
  {\bibfnamefont {J.}~\bibnamefont {Ishizuka}}, \bibinfo {author}
  {\bibfnamefont {T.}~\bibnamefont {Arakawa}}, \bibinfo {author} {\bibfnamefont
  {Y.}~\bibnamefont {Shiota}}, \bibinfo {author} {\bibfnamefont
  {T.}~\bibnamefont {Moriyama}}, \bibinfo {author} {\bibfnamefont
  {Y.}~\bibnamefont {Yanase}}, \ and\ \bibinfo {author} {\bibfnamefont
  {T.}~\bibnamefont {Ono}},\ }\href {\doibase 10.1038/s41586-020-2590-4}
  {\bibfield  {journal} {\bibinfo  {journal} {Nature}\ }\textbf {\bibinfo
  {volume} {584}},\ \bibinfo {pages} {373–376} (\bibinfo {year}
  {2020})}\BibitemShut {NoStop}%
\bibitem [{\citenamefont {Wu}\ \emph {et~al.}(2022)\citenamefont {Wu},
  \citenamefont {Wang}, \citenamefont {Xu}, \citenamefont {Sivakumar},
  \citenamefont {Pasco}, \citenamefont {Filippozzi}, \citenamefont {amd
  Yu-Jia~Zeng}, \citenamefont {McQueen},\ and\ \citenamefont {Ali}}]{Wu2022}%
  \BibitemOpen
  \bibfield  {author} {\bibinfo {author} {\bibfnamefont {H.}~\bibnamefont
  {Wu}}, \bibinfo {author} {\bibfnamefont {Y.}~\bibnamefont {Wang}}, \bibinfo
  {author} {\bibfnamefont {Y.}~\bibnamefont {Xu}}, \bibinfo {author}
  {\bibfnamefont {P.~K.}\ \bibnamefont {Sivakumar}}, \bibinfo {author}
  {\bibfnamefont {C.}~\bibnamefont {Pasco}}, \bibinfo {author} {\bibfnamefont
  {U.}~\bibnamefont {Filippozzi}}, \bibinfo {author} {\bibfnamefont {S.~S.
  P.~P.}\ \bibnamefont {amd Yu-Jia~Zeng}}, \bibinfo {author} {\bibfnamefont
  {T.}~\bibnamefont {McQueen}}, \ and\ \bibinfo {author} {\bibfnamefont
  {M.~N.}\ \bibnamefont {Ali}},\ }\href {\doibase 10.1038/s41586-022-04504-8}
  {\bibfield  {journal} {\bibinfo  {journal} {Nature}\ }\textbf {\bibinfo
  {volume} {604}},\ \bibinfo {pages} {653–656} (\bibinfo {year}
  {2022})}\BibitemShut {NoStop}%
\bibitem [{\citenamefont {Zhang}\ \emph {et~al.}(2022)\citenamefont {Zhang},
  \citenamefont {Gu}, \citenamefont {Li}, \citenamefont {Hu},\ and\
  \citenamefont {Jiang}}]{Zhang2022}%
  \BibitemOpen
  \bibfield  {author} {\bibinfo {author} {\bibfnamefont {Y.}~\bibnamefont
  {Zhang}}, \bibinfo {author} {\bibfnamefont {Y.}~\bibnamefont {Gu}}, \bibinfo
  {author} {\bibfnamefont {P.}~\bibnamefont {Li}}, \bibinfo {author}
  {\bibfnamefont {J.}~\bibnamefont {Hu}}, \ and\ \bibinfo {author}
  {\bibfnamefont {K.}~\bibnamefont {Jiang}},\ }\href {\doibase
  10.1103/PhysRevX.12.041013} {\bibfield  {journal} {\bibinfo  {journal} {Phys.
  Rev. X}\ }\textbf {\bibinfo {volume} {12}},\ \bibinfo {pages} {041013}
  (\bibinfo {year} {2022})}\BibitemShut {NoStop}%
\bibitem [{\citenamefont {Nicolay}\ \emph {et~al.}(2001)\citenamefont
  {Nicolay}, \citenamefont {Reinert}, \citenamefont {H\"ufner},\ and\
  \citenamefont {Blaha}}]{Nicolay2001}%
  \BibitemOpen
  \bibfield  {author} {\bibinfo {author} {\bibfnamefont {G.}~\bibnamefont
  {Nicolay}}, \bibinfo {author} {\bibfnamefont {F.}~\bibnamefont {Reinert}},
  \bibinfo {author} {\bibfnamefont {S.}~\bibnamefont {H\"ufner}}, \ and\
  \bibinfo {author} {\bibfnamefont {P.}~\bibnamefont {Blaha}},\ }\href
  {\doibase 10.1103/PhysRevB.65.033407} {\bibfield  {journal} {\bibinfo
  {journal} {Phys. Rev. B}\ }\textbf {\bibinfo {volume} {65}},\ \bibinfo
  {pages} {033407} (\bibinfo {year} {2001})}\BibitemShut {NoStop}%
\bibitem [{\citenamefont {Henk}\ \emph {et~al.}(2003)\citenamefont {Henk},
  \citenamefont {Ernst},\ and\ \citenamefont {Bruno}}]{Henk2003}%
  \BibitemOpen
  \bibfield  {author} {\bibinfo {author} {\bibfnamefont {J.}~\bibnamefont
  {Henk}}, \bibinfo {author} {\bibfnamefont {A.}~\bibnamefont {Ernst}}, \ and\
  \bibinfo {author} {\bibfnamefont {P.}~\bibnamefont {Bruno}},\ }\href
  {\doibase 10.1103/PhysRevB.68.165416} {\bibfield  {journal} {\bibinfo
  {journal} {Phys. Rev. B}\ }\textbf {\bibinfo {volume} {68}},\ \bibinfo
  {pages} {165416} (\bibinfo {year} {2003})}\BibitemShut {NoStop}%
\bibitem [{\citenamefont {Henk}\ \emph {et~al.}(2004)\citenamefont {Henk},
  \citenamefont {Hoesch}, \citenamefont {Osterwalder}, \citenamefont {Ernst},\
  and\ \citenamefont {Bruno}}]{Henk2004}%
  \BibitemOpen
  \bibfield  {author} {\bibinfo {author} {\bibfnamefont {J.}~\bibnamefont
  {Henk}}, \bibinfo {author} {\bibfnamefont {M.}~\bibnamefont {Hoesch}},
  \bibinfo {author} {\bibfnamefont {J.}~\bibnamefont {Osterwalder}}, \bibinfo
  {author} {\bibfnamefont {A.}~\bibnamefont {Ernst}}, \ and\ \bibinfo {author}
  {\bibfnamefont {P.}~\bibnamefont {Bruno}},\ }\href
  {http://stacks.iop.org/0953-8984/16/i=43/a=002} {\bibfield  {journal}
  {\bibinfo  {journal} {Journal of Physics: Condensed Matter}\ }\textbf
  {\bibinfo {volume} {16}},\ \bibinfo {pages} {7581} (\bibinfo {year}
  {2004})}\BibitemShut {NoStop}%
\bibitem [{\citenamefont {Hoesch}\ \emph {et~al.}(2004)\citenamefont {Hoesch},
  \citenamefont {Muntwiler}, \citenamefont {Petrov}, \citenamefont
  {Hengsberger}, \citenamefont {Patthey}, \citenamefont {Shi}, \citenamefont
  {Falub}, \citenamefont {Greber},\ and\ \citenamefont
  {Osterwalder}}]{Hoesch2004}%
  \BibitemOpen
  \bibfield  {author} {\bibinfo {author} {\bibfnamefont {M.}~\bibnamefont
  {Hoesch}}, \bibinfo {author} {\bibfnamefont {M.}~\bibnamefont {Muntwiler}},
  \bibinfo {author} {\bibfnamefont {V.~N.}\ \bibnamefont {Petrov}}, \bibinfo
  {author} {\bibfnamefont {M.}~\bibnamefont {Hengsberger}}, \bibinfo {author}
  {\bibfnamefont {L.}~\bibnamefont {Patthey}}, \bibinfo {author} {\bibfnamefont
  {M.}~\bibnamefont {Shi}}, \bibinfo {author} {\bibfnamefont {M.}~\bibnamefont
  {Falub}}, \bibinfo {author} {\bibfnamefont {T.}~\bibnamefont {Greber}}, \
  and\ \bibinfo {author} {\bibfnamefont {J.}~\bibnamefont {Osterwalder}},\
  }\href {\doibase 10.1103/PhysRevB.69.241401} {\bibfield  {journal} {\bibinfo
  {journal} {Phys. Rev. B}\ }\textbf {\bibinfo {volume} {69}},\ \bibinfo
  {pages} {241401(R)} (\bibinfo {year} {2004})}\BibitemShut {NoStop}%
\bibitem [{\citenamefont {Wissing}\ \emph {et~al.}(2013)\citenamefont
  {Wissing}, \citenamefont {Eibl}, \citenamefont {Zumb\"ulte}, \citenamefont
  {Schmidt}, \citenamefont {Braun}, \citenamefont {Minár}, \citenamefont
  {Ebert},\ and\ \citenamefont {Donath}}]{Wissing2013}%
  \BibitemOpen
  \bibfield  {author} {\bibinfo {author} {\bibfnamefont {S.~N.~P.}\
  \bibnamefont {Wissing}}, \bibinfo {author} {\bibfnamefont {C.}~\bibnamefont
  {Eibl}}, \bibinfo {author} {\bibfnamefont {A.}~\bibnamefont {Zumb\"ulte}},
  \bibinfo {author} {\bibfnamefont {A.~B.}\ \bibnamefont {Schmidt}}, \bibinfo
  {author} {\bibfnamefont {J.}~\bibnamefont {Braun}}, \bibinfo {author}
  {\bibfnamefont {J.}~\bibnamefont {Minár}}, \bibinfo {author} {\bibfnamefont
  {H.}~\bibnamefont {Ebert}}, \ and\ \bibinfo {author} {\bibfnamefont
  {M.}~\bibnamefont {Donath}},\ }\href
  {http://stacks.iop.org/1367-2630/15/i=10/a=105001} {\bibfield  {journal}
  {\bibinfo  {journal} {New Journal of Physics}\ }\textbf {\bibinfo {volume}
  {15}},\ \bibinfo {pages} {105001} (\bibinfo {year} {2013})}\BibitemShut
  {NoStop}%
\bibitem [{\citenamefont {Villars}\ and\ \citenamefont
  {Cenzual}(2016{\natexlab{a}})}]{Villars2016-1}%
  \BibitemOpen
  \bibinfo {editor} {\bibfnamefont {P.}~\bibnamefont {Villars}}\ and\ \bibinfo
  {editor} {\bibfnamefont {K.}~\bibnamefont {Cenzual}},\ eds.,\ \href
  {https://materials.springer.com/isp/crystallographic/docs/sd_1250064} {\emph
  {\bibinfo {title} {{Al Crystal Structure: Datasheet from ``PAULING FILE
  Multinaries Edition -- 2022'' in SpringerMaterials}}}}\ (\bibinfo
  {publisher} {Springer-Verlag Berlin Heidelberg {\&} Material Phases Data
  System (MPDS), Switzerland {\&} National Institute for Materials Science
  (NIMS), Japan},\ \bibinfo {year} {2016})\ \bibinfo {note} {accessed
  2023-05-23}\BibitemShut {NoStop}%
\bibitem [{\citenamefont {Villars}\ and\ \citenamefont
  {Cenzual}(2016{\natexlab{b}})}]{Villars2016-2}%
  \BibitemOpen
  \bibinfo {editor} {\bibfnamefont {P.}~\bibnamefont {Villars}}\ and\ \bibinfo
  {editor} {\bibfnamefont {K.}~\bibnamefont {Cenzual}},\ eds.,\ \href
  {https://materials.springer.com/isp/crystallographic/docs/sd_0261045} {\emph
  {\bibinfo {title} {{Au Crystal Structure: Datasheet from ``PAULING FILE
  Multinaries Edition -- 2022'' in SpringerMaterials}}}}\ (\bibinfo
  {publisher} {Springer-Verlag Berlin Heidelberg {\&} Material Phases Data
  System (MPDS), Switzerland {\&} National Institute for Materials Science
  (NIMS), Japan},\ \bibinfo {year} {2016})\ \bibinfo {note} {accessed
  2023-05-23}\BibitemShut {NoStop}%
\bibitem [{\citenamefont {{The JuKKR developers}}(2022)}]{jukkr}%
  \BibitemOpen
  \bibfield  {author} {\bibinfo {author} {\bibnamefont {{The JuKKR
  developers}}},\ }\href {\doibase 10.5281/zenodo.7284738} {\enquote {\bibinfo
  {title} {{The J\"ulich KKR Codes}},}\ } (\bibinfo {year} {2022}),\ \bibinfo
  {note} {\url{https://jukkr.fz-juelich.de}}\BibitemShut {NoStop}%
\bibitem [{\citenamefont {R\"u{\ss}mann}\ and\ \citenamefont
  {Bl\"ugel}(2022{\natexlab{a}})}]{Ruessmann2022a}%
  \BibitemOpen
  \bibfield  {author} {\bibinfo {author} {\bibfnamefont {P.}~\bibnamefont
  {R\"u{\ss}mann}}\ and\ \bibinfo {author} {\bibfnamefont {S.}~\bibnamefont
  {Bl\"ugel}},\ }\href {\doibase 10.1103/PhysRevB.105.125143} {\bibfield
  {journal} {\bibinfo  {journal} {Phys. Rev. B}\ }\textbf {\bibinfo {volume}
  {105}},\ \bibinfo {pages} {125143} (\bibinfo {year}
  {2022}{\natexlab{a}})}\BibitemShut {NoStop}%
\bibitem [{\citenamefont {Vajna}\ \emph {et~al.}(2012)\citenamefont {Vajna},
  \citenamefont {Simon}, \citenamefont {Szilva}, \citenamefont {Palotas},
  \citenamefont {Ujfalussy},\ and\ \citenamefont
  {Szunyogh}}]{MSRef_ThirdOrder}%
  \BibitemOpen
  \bibfield  {author} {\bibinfo {author} {\bibfnamefont {S.}~\bibnamefont
  {Vajna}}, \bibinfo {author} {\bibfnamefont {E.}~\bibnamefont {Simon}},
  \bibinfo {author} {\bibfnamefont {A.}~\bibnamefont {Szilva}}, \bibinfo
  {author} {\bibfnamefont {K.}~\bibnamefont {Palotas}}, \bibinfo {author}
  {\bibfnamefont {B.}~\bibnamefont {Ujfalussy}}, \ and\ \bibinfo {author}
  {\bibfnamefont {L.}~\bibnamefont {Szunyogh}},\ }\href {\doibase
  10.1103/PhysRevB.85.075404} {\bibfield  {journal} {\bibinfo  {journal} {Phys.
  Rev. B}\ }\textbf {\bibinfo {volume} {85}},\ \bibinfo {pages} {075404}
  (\bibinfo {year} {2012})}\BibitemShut {NoStop}%
\bibitem [{\citenamefont {Bardeen}\ \emph {et~al.}(1957)\citenamefont
  {Bardeen}, \citenamefont {Cooper},\ and\ \citenamefont {Schrieffer}}]{BCS}%
  \BibitemOpen
  \bibfield  {author} {\bibinfo {author} {\bibfnamefont {J.}~\bibnamefont
  {Bardeen}}, \bibinfo {author} {\bibfnamefont {L.~N.}\ \bibnamefont {Cooper}},
  \ and\ \bibinfo {author} {\bibfnamefont {J.~R.}\ \bibnamefont {Schrieffer}},\
  }\href@noop {} {\bibfield  {journal} {\bibinfo  {journal} {Phys. Rev.}\
  }\textbf {\bibinfo {volume} {108}},\ \bibinfo {pages} {1175} (\bibinfo {year}
  {1957})}\BibitemShut {NoStop}%
\bibitem [{\citenamefont {Suvasini}\ \emph {et~al.}(1993)\citenamefont
  {Suvasini}, \citenamefont {Temmerman},\ and\ \citenamefont
  {Gy\ifmmode~\mbox{\H{o}}\else \H{o}\fi{}rffy}}]{Suvasini1993}%
  \BibitemOpen
  \bibfield  {author} {\bibinfo {author} {\bibfnamefont {M.~B.}\ \bibnamefont
  {Suvasini}}, \bibinfo {author} {\bibfnamefont {W.~M.}\ \bibnamefont
  {Temmerman}}, \ and\ \bibinfo {author} {\bibfnamefont {B.~L.}\ \bibnamefont
  {Gy\ifmmode~\mbox{\H{o}}\else \H{o}\fi{}rffy}},\ }\href {\doibase
  10.1103/PhysRevB.48.1202} {\bibfield  {journal} {\bibinfo  {journal} {Phys.
  Rev. B}\ }\textbf {\bibinfo {volume} {48}},\ \bibinfo {pages} {1202}
  (\bibinfo {year} {1993})}\BibitemShut {NoStop}%
\bibitem [{\citenamefont {Csire}\ \emph
  {et~al.}(2016{\natexlab{a}})\citenamefont {Csire}, \citenamefont
  {Sch\"onecker},\ and\ \citenamefont
  {{\'{U}}jfalussy}}]{CsireSchoenecker2016}%
  \BibitemOpen
  \bibfield  {author} {\bibinfo {author} {\bibfnamefont {G.}~\bibnamefont
  {Csire}}, \bibinfo {author} {\bibfnamefont {S.}~\bibnamefont {Sch\"onecker}},
  \ and\ \bibinfo {author} {\bibfnamefont {B.}~\bibnamefont
  {{\'{U}}jfalussy}},\ }\href {\doibase 10.1103/PhysRevB.94.140502} {\bibfield
  {journal} {\bibinfo  {journal} {Phys. Rev. B}\ }\textbf {\bibinfo {volume}
  {94}},\ \bibinfo {pages} {140502(R)} (\bibinfo {year}
  {2016}{\natexlab{a}})}\BibitemShut {NoStop}%
\bibitem [{\citenamefont {Saunderson}\ \emph
  {et~al.}(2020{\natexlab{a}})\citenamefont {Saunderson}, \citenamefont
  {Annett}, \citenamefont {\'Ujfalussy}, \citenamefont {Csire},\ and\
  \citenamefont {Gradhand}}]{Tombulk}%
  \BibitemOpen
  \bibfield  {author} {\bibinfo {author} {\bibfnamefont {T.~G.}\ \bibnamefont
  {Saunderson}}, \bibinfo {author} {\bibfnamefont {J.~F.}\ \bibnamefont
  {Annett}}, \bibinfo {author} {\bibfnamefont {B.}~\bibnamefont {\'Ujfalussy}},
  \bibinfo {author} {\bibfnamefont {G.}~\bibnamefont {Csire}}, \ and\ \bibinfo
  {author} {\bibfnamefont {M.}~\bibnamefont {Gradhand}},\ }\href {\doibase
  10.1103/PhysRevB.101.064510} {\bibfield  {journal} {\bibinfo  {journal}
  {Phys. Rev. B}\ }\textbf {\bibinfo {volume} {101}},\ \bibinfo {pages}
  {064510} (\bibinfo {year} {2020}{\natexlab{a}})}\BibitemShut {NoStop}%
\bibitem [{\citenamefont {Csire}\ \emph
  {et~al.}(2016{\natexlab{b}})\citenamefont {Csire}, \citenamefont {Cserti},
  \citenamefont {T\"utt\ifmmode~\mbox{\H{o}}\else \H{o}\fi{}},\ and\
  \citenamefont {\'Ujfalussy}}]{Csire2016}%
  \BibitemOpen
  \bibfield  {author} {\bibinfo {author} {\bibfnamefont {G.}~\bibnamefont
  {Csire}}, \bibinfo {author} {\bibfnamefont {J.}~\bibnamefont {Cserti}},
  \bibinfo {author} {\bibfnamefont {I.}~\bibnamefont
  {T\"utt\ifmmode~\mbox{\H{o}}\else \H{o}\fi{}}}, \ and\ \bibinfo {author}
  {\bibfnamefont {B.}~\bibnamefont {\'Ujfalussy}},\ }\href {\doibase
  10.1103/PhysRevB.94.104511} {\bibfield  {journal} {\bibinfo  {journal} {Phys.
  Rev. B}\ }\textbf {\bibinfo {volume} {94}},\ \bibinfo {pages} {104511}
  (\bibinfo {year} {2016}{\natexlab{b}})}\BibitemShut {NoStop}%
\bibitem [{\citenamefont {{G. Csire and J. Cserti and B.
  \'{U}}jfalussy}(2016)}]{Csireheterostruc2016}%
  \BibitemOpen
  \bibfield  {author} {\bibinfo {author} {\bibnamefont {{G. Csire and J. Cserti
  and B. \'{U}}jfalussy}},\ }\href
  {http://stacks.iop.org/0953-8984/28/i=49/a=495701} {\bibfield  {journal}
  {\bibinfo  {journal} {Journal of Physics: Condensed Matter}\ }\textbf
  {\bibinfo {volume} {28}},\ \bibinfo {pages} {495701} (\bibinfo {year}
  {2016})}\BibitemShut {NoStop}%
\bibitem [{\citenamefont {R\"u{\ss}mann}\ and\ \citenamefont
  {Bl\"ugel}(2022{\natexlab{b}})}]{Ruessmann2022b}%
  \BibitemOpen
  \bibfield  {author} {\bibinfo {author} {\bibfnamefont {P.}~\bibnamefont
  {R\"u{\ss}mann}}\ and\ \bibinfo {author} {\bibfnamefont {S.}~\bibnamefont
  {Bl\"ugel}},\ }\href {\doibase 10.48550/arXiv.2208.14289} {\bibfield
  {journal} {\bibinfo  {journal} {arXiv:2208.14289}\ } (\bibinfo {year}
  {2022}{\natexlab{b}}),\ 10.48550/arXiv.2208.14289}\BibitemShut {NoStop}%
\bibitem [{\citenamefont {Saunderson}\ \emph
  {et~al.}(2020{\natexlab{b}})\citenamefont {Saunderson}, \citenamefont
  {Gyorgypal}, \citenamefont {Annett}, \citenamefont {Csire}, \citenamefont
  {\'Ujfalussy},\ and\ \citenamefont {Gradhand}}]{Tomimp}%
  \BibitemOpen
  \bibfield  {author} {\bibinfo {author} {\bibfnamefont {T.~G.}\ \bibnamefont
  {Saunderson}}, \bibinfo {author} {\bibfnamefont {Z.}~\bibnamefont
  {Gyorgypal}}, \bibinfo {author} {\bibfnamefont {J.~F.}\ \bibnamefont
  {Annett}}, \bibinfo {author} {\bibfnamefont {G.}~\bibnamefont {Csire}},
  \bibinfo {author} {\bibfnamefont {B.}~\bibnamefont {\'Ujfalussy}}, \ and\
  \bibinfo {author} {\bibfnamefont {M.}~\bibnamefont {Gradhand}},\ }\href
  {\doibase 10.1103/PhysRevB.102.245106} {\bibfield  {journal} {\bibinfo
  {journal} {Phys. Rev. B}\ }\textbf {\bibinfo {volume} {102}},\ \bibinfo
  {pages} {245106} (\bibinfo {year} {2020}{\natexlab{b}})}\BibitemShut
  {NoStop}%
\bibitem [{\citenamefont {Ny\'ari}\ \emph {et~al.}(2021)\citenamefont
  {Ny\'ari}, \citenamefont {L\'aszl\'offy}, \citenamefont {Szunyogh},
  \citenamefont {Csire}, \citenamefont {Park},\ and\ \citenamefont
  {Ujfalussy}}]{Nyari2021}%
  \BibitemOpen
  \bibfield  {author} {\bibinfo {author} {\bibfnamefont {B.}~\bibnamefont
  {Ny\'ari}}, \bibinfo {author} {\bibfnamefont {A.}~\bibnamefont
  {L\'aszl\'offy}}, \bibinfo {author} {\bibfnamefont {L.}~\bibnamefont
  {Szunyogh}}, \bibinfo {author} {\bibfnamefont {G.}~\bibnamefont {Csire}},
  \bibinfo {author} {\bibfnamefont {K.}~\bibnamefont {Park}}, \ and\ \bibinfo
  {author} {\bibfnamefont {B.}~\bibnamefont {Ujfalussy}},\ }\href {\doibase
  10.1103/PhysRevB.104.235426} {\bibfield  {journal} {\bibinfo  {journal}
  {Phys. Rev. B}\ }\textbf {\bibinfo {volume} {104}},\ \bibinfo {pages}
  {235426} (\bibinfo {year} {2021})}\BibitemShut {NoStop}%
\bibitem [{\citenamefont {Moreo}\ \emph {et~al.}(2009)\citenamefont {Moreo},
  \citenamefont {Daghofer}, \citenamefont {Nicholson},\ and\ \citenamefont
  {Dagotto}}]{Interband-2009}%
  \BibitemOpen
  \bibfield  {author} {\bibinfo {author} {\bibfnamefont {A.}~\bibnamefont
  {Moreo}}, \bibinfo {author} {\bibfnamefont {M.}~\bibnamefont {Daghofer}},
  \bibinfo {author} {\bibfnamefont {A.}~\bibnamefont {Nicholson}}, \ and\
  \bibinfo {author} {\bibfnamefont {E.}~\bibnamefont {Dagotto}},\ }\href
  {\doibase 10.1103/PhysRevB.80.104507} {\bibfield  {journal} {\bibinfo
  {journal} {Phys. Rev. B}\ }\textbf {\bibinfo {volume} {80}},\ \bibinfo
  {pages} {104507} (\bibinfo {year} {2009})}\BibitemShut {NoStop}%
\bibitem [{\citenamefont {Komendov\'a}\ \emph {et~al.}(2015)\citenamefont
  {Komendov\'a}, \citenamefont {Balatsky},\ and\ \citenamefont
  {Black-Schaffer}}]{Interband-2015}%
  \BibitemOpen
  \bibfield  {author} {\bibinfo {author} {\bibfnamefont {L.}~\bibnamefont
  {Komendov\'a}}, \bibinfo {author} {\bibfnamefont {A.~V.}\ \bibnamefont
  {Balatsky}}, \ and\ \bibinfo {author} {\bibfnamefont {A.~M.}\ \bibnamefont
  {Black-Schaffer}},\ }\href {\doibase 10.1103/PhysRevB.92.094517} {\bibfield
  {journal} {\bibinfo  {journal} {Phys. Rev. B}\ }\textbf {\bibinfo {volume}
  {92}},\ \bibinfo {pages} {094517} (\bibinfo {year} {2015})}\BibitemShut
  {NoStop}%
\bibitem [{\citenamefont {Triola}\ and\ \citenamefont
  {Balatsky}(2017)}]{Interband-2017}%
  \BibitemOpen
  \bibfield  {author} {\bibinfo {author} {\bibfnamefont {C.}~\bibnamefont
  {Triola}}\ and\ \bibinfo {author} {\bibfnamefont {A.~V.}\ \bibnamefont
  {Balatsky}},\ }\href {\doibase 10.1103/PhysRevB.95.224518} {\bibfield
  {journal} {\bibinfo  {journal} {Phys. Rev. B}\ }\textbf {\bibinfo {volume}
  {95}},\ \bibinfo {pages} {224518} (\bibinfo {year} {2017})}\BibitemShut
  {NoStop}%
\bibitem [{\citenamefont {Banerjee}\ \emph {et~al.}(2018)\citenamefont
  {Banerjee}, \citenamefont {Sundaresh}, \citenamefont {Ganesan},\ and\
  \citenamefont {Kumar}}]{Interband-2018}%
  \BibitemOpen
  \bibfield  {author} {\bibinfo {author} {\bibfnamefont {A.}~\bibnamefont
  {Banerjee}}, \bibinfo {author} {\bibfnamefont {A.}~\bibnamefont {Sundaresh}},
  \bibinfo {author} {\bibfnamefont {R.}~\bibnamefont {Ganesan}}, \ and\
  \bibinfo {author} {\bibfnamefont {P.~S.~A.}\ \bibnamefont {Kumar}},\ }\href
  {\doibase 10.1021/acsnano.8b07550} {\bibfield  {journal} {\bibinfo  {journal}
  {ACS Nano}\ }\textbf {\bibinfo {volume} {12}},\ \bibinfo {pages} {12665}
  (\bibinfo {year} {2018})}\BibitemShut {NoStop}%
\bibitem [{\citenamefont {Linder}\ and\ \citenamefont
  {Balatsky}(2019)}]{Interband-2019}%
  \BibitemOpen
  \bibfield  {author} {\bibinfo {author} {\bibfnamefont {J.}~\bibnamefont
  {Linder}}\ and\ \bibinfo {author} {\bibfnamefont {A.~V.}\ \bibnamefont
  {Balatsky}},\ }\href {\doibase 10.1103/RevModPhys.91.045005} {\bibfield
  {journal} {\bibinfo  {journal} {Rev. Mod. Phys.}\ }\textbf {\bibinfo {volume}
  {91}},\ \bibinfo {pages} {045005} (\bibinfo {year} {2019})}\BibitemShut
  {NoStop}%
\bibitem [{\citenamefont {Tang}\ \emph {et~al.}(2021)\citenamefont {Tang},
  \citenamefont {Bruder},\ and\ \citenamefont {Belzig}}]{Mirage}%
  \BibitemOpen
  \bibfield  {author} {\bibinfo {author} {\bibfnamefont {G.}~\bibnamefont
  {Tang}}, \bibinfo {author} {\bibfnamefont {C.}~\bibnamefont {Bruder}}, \ and\
  \bibinfo {author} {\bibfnamefont {W.}~\bibnamefont {Belzig}},\ }\href
  {\doibase 10.1103/PhysRevLett.126.237001} {\bibfield  {journal} {\bibinfo
  {journal} {Phys. Rev. Lett.}\ }\textbf {\bibinfo {volume} {126}},\ \bibinfo
  {pages} {237001} (\bibinfo {year} {2021})}\BibitemShut {NoStop}%
\bibitem [{\citenamefont {Bahari}\ \emph {et~al.}(2022)\citenamefont {Bahari},
  \citenamefont {Zhang},\ and\ \citenamefont {Trauzettel}}]{FECooper-2022-1}%
  \BibitemOpen
  \bibfield  {author} {\bibinfo {author} {\bibfnamefont {M.}~\bibnamefont
  {Bahari}}, \bibinfo {author} {\bibfnamefont {S.-B.}\ \bibnamefont {Zhang}}, \
  and\ \bibinfo {author} {\bibfnamefont {B.}~\bibnamefont {Trauzettel}},\
  }\href {\doibase 10.1103/PhysRevResearch.4.L012017} {\bibfield  {journal}
  {\bibinfo  {journal} {Phys. Rev. Res.}\ }\textbf {\bibinfo {volume} {4}},\
  \bibinfo {pages} {L012017} (\bibinfo {year} {2022})}\BibitemShut {NoStop}%
\bibitem [{\citenamefont {Chakraborty}\ and\ \citenamefont
  {Black-Schaffer}(2022)}]{FECooper-2022-2}%
  \BibitemOpen
  \bibfield  {author} {\bibinfo {author} {\bibfnamefont {D.}~\bibnamefont
  {Chakraborty}}\ and\ \bibinfo {author} {\bibfnamefont {A.~M.}\ \bibnamefont
  {Black-Schaffer}},\ }\href {\doibase 10.1103/PhysRevB.106.024511} {\bibfield
  {journal} {\bibinfo  {journal} {Phys. Rev. B}\ }\textbf {\bibinfo {volume}
  {106}},\ \bibinfo {pages} {024511} (\bibinfo {year} {2022})}\BibitemShut
  {NoStop}%
\bibitem [{\citenamefont {Bahari}\ \emph {et~al.}(2023)\citenamefont {Bahari},
  \citenamefont {Zhang}, \citenamefont {Li}, \citenamefont {Choi},
  \citenamefont {Timm},\ and\ \citenamefont {Trauzettel}}]{FECooper-2023}%
  \BibitemOpen
  \bibfield  {author} {\bibinfo {author} {\bibfnamefont {M.}~\bibnamefont
  {Bahari}}, \bibinfo {author} {\bibfnamefont {S.-B.}\ \bibnamefont {Zhang}},
  \bibinfo {author} {\bibfnamefont {C.-A.}\ \bibnamefont {Li}}, \bibinfo
  {author} {\bibfnamefont {S.-J.}\ \bibnamefont {Choi}}, \bibinfo {author}
  {\bibfnamefont {C.}~\bibnamefont {Timm}}, \ and\ \bibinfo {author}
  {\bibfnamefont {B.}~\bibnamefont {Trauzettel}},\ }\href@noop {} {\enquote
  {\bibinfo {title} {New type of helical topological superconducting pairing at
  finite excitation energies},}\ } (\bibinfo {year} {2023}),\ \Eprint
  {http://arxiv.org/abs/2210.11955} {arXiv:2210.11955 [cond-mat.mes-hall]}
  \BibitemShut {NoStop}%
\bibitem [{\citenamefont {Eisenstein}(1954)}]{Eisenstein1954}%
  \BibitemOpen
  \bibfield  {author} {\bibinfo {author} {\bibfnamefont {J.}~\bibnamefont
  {Eisenstein}},\ }\href {\doibase 10.1103/RevModPhys.26.277} {\bibfield
  {journal} {\bibinfo  {journal} {Rev. Mod. Phys.}\ }\textbf {\bibinfo {volume}
  {26}},\ \bibinfo {pages} {277} (\bibinfo {year} {1954})}\BibitemShut
  {NoStop}%
\bibitem [{\citenamefont {Meservey}\ and\ \citenamefont
  {Tedrow}(1971)}]{Meservey1971}%
  \BibitemOpen
  \bibfield  {author} {\bibinfo {author} {\bibfnamefont {R.}~\bibnamefont
  {Meservey}}\ and\ \bibinfo {author} {\bibfnamefont {P.}~\bibnamefont
  {Tedrow}},\ }\href {\doibase 10.1063/1.1659648} {\bibfield  {journal}
  {\bibinfo  {journal} {{Journal of Applied Physics}}\ }\textbf {\bibinfo
  {volume} {42}},\ \bibinfo {pages} {51} (\bibinfo {year} {1971})}\BibitemShut
  {NoStop}%
\bibitem [{\citenamefont {Court}\ \emph {et~al.}(2008)\citenamefont {Court},
  \citenamefont {Ferguson},\ and\ \citenamefont {Clark}}]{Court2008}%
  \BibitemOpen
  \bibfield  {author} {\bibinfo {author} {\bibfnamefont {N.}~\bibnamefont
  {Court}}, \bibinfo {author} {\bibfnamefont {A.}~\bibnamefont {Ferguson}}, \
  and\ \bibinfo {author} {\bibfnamefont {R.}~\bibnamefont {Clark}},\ }\href
  {\doibase 10.1088/0953-2048/21/01/015013} {\bibfield  {journal} {\bibinfo
  {journal} {Supercond. Sci. Technol.}\ }\textbf {\bibinfo {volume} {21}},\
  \bibinfo {pages} {015013} (\bibinfo {year} {2008})}\BibitemShut {NoStop}%
\bibitem [{\citenamefont {van Weerdenburg}\ \emph {et~al.}(2023)\citenamefont
  {van Weerdenburg}, \citenamefont {Kamlapure}, \citenamefont {Fyhn},
  \citenamefont {Huang}, \citenamefont {van Mullekom}, \citenamefont
  {Steinbrecher}, \citenamefont {Krogstrup}, \citenamefont {Linder},\ and\
  \citenamefont {Khajetoorians}}]{VanWeerdenburg2023}%
  \BibitemOpen
  \bibfield  {author} {\bibinfo {author} {\bibfnamefont {W.~M.}\ \bibnamefont
  {van Weerdenburg}}, \bibinfo {author} {\bibfnamefont {A.}~\bibnamefont
  {Kamlapure}}, \bibinfo {author} {\bibfnamefont {E.~H.}\ \bibnamefont {Fyhn}},
  \bibinfo {author} {\bibfnamefont {X.}~\bibnamefont {Huang}}, \bibinfo
  {author} {\bibfnamefont {N.~P.}\ \bibnamefont {van Mullekom}}, \bibinfo
  {author} {\bibfnamefont {M.}~\bibnamefont {Steinbrecher}}, \bibinfo {author}
  {\bibfnamefont {P.}~\bibnamefont {Krogstrup}}, \bibinfo {author}
  {\bibfnamefont {J.}~\bibnamefont {Linder}}, \ and\ \bibinfo {author}
  {\bibfnamefont {A.~A.}\ \bibnamefont {Khajetoorians}},\ }\href {\doibase
  10.1126/sciadv.adf5500} {\bibfield  {journal} {\bibinfo  {journal} {Science
  Advances}\ }\textbf {\bibinfo {volume} {9}},\ \bibinfo {pages} {1} (\bibinfo
  {year} {2023})}\BibitemShut {NoStop}%
\bibitem [{\citenamefont {Schwenk}\ \emph {et~al.}(2020)\citenamefont
  {Schwenk}, \citenamefont {Kim}, \citenamefont {Berwanger}, \citenamefont
  {Ghahari}, \citenamefont {Walkup}, \citenamefont {Slot}, \citenamefont {Le},
  \citenamefont {Cullen}, \citenamefont {Blankenship}, \citenamefont
  {Vranjkovic}, \citenamefont {Hug}, \citenamefont {Kuk}, \citenamefont
  {Giessibl},\ and\ \citenamefont {Stroscio}}]{Schwenk2020}%
  \BibitemOpen
  \bibfield  {author} {\bibinfo {author} {\bibfnamefont {J.}~\bibnamefont
  {Schwenk}}, \bibinfo {author} {\bibfnamefont {S.}~\bibnamefont {Kim}},
  \bibinfo {author} {\bibfnamefont {J.}~\bibnamefont {Berwanger}}, \bibinfo
  {author} {\bibfnamefont {F.}~\bibnamefont {Ghahari}}, \bibinfo {author}
  {\bibfnamefont {D.}~\bibnamefont {Walkup}}, \bibinfo {author} {\bibfnamefont
  {M.}~\bibnamefont {Slot}}, \bibinfo {author} {\bibfnamefont {S.}~\bibnamefont
  {Le}}, \bibinfo {author} {\bibfnamefont {W.}~\bibnamefont {Cullen}}, \bibinfo
  {author} {\bibfnamefont {S.}~\bibnamefont {Blankenship}}, \bibinfo {author}
  {\bibfnamefont {S.}~\bibnamefont {Vranjkovic}}, \bibinfo {author}
  {\bibfnamefont {H.}~\bibnamefont {Hug}}, \bibinfo {author} {\bibfnamefont
  {Y.}~\bibnamefont {Kuk}}, \bibinfo {author} {\bibfnamefont {F.}~\bibnamefont
  {Giessibl}}, \ and\ \bibinfo {author} {\bibfnamefont {J.}~\bibnamefont
  {Stroscio}},\ }\href {\doibase 10.1063/5.0005320} {\bibfield  {journal}
  {\bibinfo  {journal} {Rev Sci Instrum}\ }\textbf {\bibinfo {volume} {91}},\
  \bibinfo {pages} {071101} (\bibinfo {year} {2020})}\BibitemShut {NoStop}%
\bibitem [{\citenamefont {Ishizaka}\ \emph {et~al.}(2011)\citenamefont
  {Ishizaka}, \citenamefont {Bahramy}, \citenamefont {Murakawa}, \citenamefont
  {Sakano}, \citenamefont {Shimojima}, \citenamefont {Sonobe}, \citenamefont
  {Koizumi}, \citenamefont {Shin}, \citenamefont {Miyahara}, \citenamefont
  {Kimura}, \citenamefont {Miyamoto}, \citenamefont {Okuda}, \citenamefont
  {Namatame}, \citenamefont {Taniguchi}, \citenamefont {Arita}, \citenamefont
  {Nagaosa}, \citenamefont {Kobayashi}, \citenamefont {Murakami}, \citenamefont
  {Kumai}, \citenamefont {Kaneko}, \citenamefont {Onose},\ and\ \citenamefont
  {Tokura}}]{Ishizaka2011}%
  \BibitemOpen
  \bibfield  {author} {\bibinfo {author} {\bibfnamefont {K.}~\bibnamefont
  {Ishizaka}}, \bibinfo {author} {\bibfnamefont {M.~S.}\ \bibnamefont
  {Bahramy}}, \bibinfo {author} {\bibfnamefont {H.}~\bibnamefont {Murakawa}},
  \bibinfo {author} {\bibfnamefont {M.}~\bibnamefont {Sakano}}, \bibinfo
  {author} {\bibfnamefont {T.}~\bibnamefont {Shimojima}}, \bibinfo {author}
  {\bibfnamefont {T.}~\bibnamefont {Sonobe}}, \bibinfo {author} {\bibfnamefont
  {K.}~\bibnamefont {Koizumi}}, \bibinfo {author} {\bibfnamefont
  {S.}~\bibnamefont {Shin}}, \bibinfo {author} {\bibfnamefont {H.}~\bibnamefont
  {Miyahara}}, \bibinfo {author} {\bibfnamefont {A.}~\bibnamefont {Kimura}},
  \bibinfo {author} {\bibfnamefont {K.}~\bibnamefont {Miyamoto}}, \bibinfo
  {author} {\bibfnamefont {T.}~\bibnamefont {Okuda}}, \bibinfo {author}
  {\bibfnamefont {H.}~\bibnamefont {Namatame}}, \bibinfo {author}
  {\bibfnamefont {M.}~\bibnamefont {Taniguchi}}, \bibinfo {author}
  {\bibfnamefont {R.}~\bibnamefont {Arita}}, \bibinfo {author} {\bibfnamefont
  {N.}~\bibnamefont {Nagaosa}}, \bibinfo {author} {\bibfnamefont
  {K.}~\bibnamefont {Kobayashi}}, \bibinfo {author} {\bibfnamefont
  {Y.}~\bibnamefont {Murakami}}, \bibinfo {author} {\bibfnamefont
  {R.}~\bibnamefont {Kumai}}, \bibinfo {author} {\bibfnamefont
  {Y.}~\bibnamefont {Kaneko}}, \bibinfo {author} {\bibfnamefont
  {Y.}~\bibnamefont {Onose}}, \ and\ \bibinfo {author} {\bibfnamefont
  {Y.}~\bibnamefont {Tokura}},\ }\href {\doibase 10.1038/nmat3051} {\bibfield
  {journal} {\bibinfo  {journal} {Nature Materials}\ }\textbf {\bibinfo
  {volume} {10}},\ \bibinfo {pages} {521} (\bibinfo {year} {2011})}\BibitemShut
  {NoStop}%
\bibitem [{\citenamefont {Liu}\ \emph {et~al.}(2020)\citenamefont {Liu},
  \citenamefont {Thirupathaiah}, \citenamefont {Yaresko}, \citenamefont
  {Kushwaha}, \citenamefont {Gibson}, \citenamefont {Xia}, \citenamefont {Guo},
  \citenamefont {Shen}, \citenamefont {Cava},\ and\ \citenamefont
  {Borisenko}}]{ZhonghaoLiu2020}%
  \BibitemOpen
  \bibfield  {author} {\bibinfo {author} {\bibfnamefont {Z.}~\bibnamefont
  {Liu}}, \bibinfo {author} {\bibfnamefont {S.}~\bibnamefont {Thirupathaiah}},
  \bibinfo {author} {\bibfnamefont {A.~N.}\ \bibnamefont {Yaresko}}, \bibinfo
  {author} {\bibfnamefont {S.}~\bibnamefont {Kushwaha}}, \bibinfo {author}
  {\bibfnamefont {Q.}~\bibnamefont {Gibson}}, \bibinfo {author} {\bibfnamefont
  {W.}~\bibnamefont {Xia}}, \bibinfo {author} {\bibfnamefont {Y.}~\bibnamefont
  {Guo}}, \bibinfo {author} {\bibfnamefont {D.}~\bibnamefont {Shen}}, \bibinfo
  {author} {\bibfnamefont {R.~J.}\ \bibnamefont {Cava}}, \ and\ \bibinfo
  {author} {\bibfnamefont {S.~V.}\ \bibnamefont {Borisenko}},\ }\href {\doibase
  https://doi.org/10.1002/pssr.201900684} {\bibfield  {journal} {\bibinfo
  {journal} {physica status solidi (RRL) – Rapid Research Letters}\ }\textbf
  {\bibinfo {volume} {14}},\ \bibinfo {pages} {1900684} (\bibinfo {year}
  {2020})}\BibitemShut {NoStop}%
\bibitem [{\citenamefont {{Aghaee \textit{et al.} (Microsoft
  Quantum)}}(2023)}]{Quantum2021}%
  \BibitemOpen
  \bibfield  {author} {\bibinfo {author} {\bibfnamefont {M.}~\bibnamefont
  {{Aghaee \textit{et al.} (Microsoft Quantum)}}},\ }\href {\doibase
  10.1103/PhysRevB.107.245423} {\bibfield  {journal} {\bibinfo  {journal}
  {Phys. Rev. B}\ }\textbf {\bibinfo {volume} {107}},\ \bibinfo {pages}
  {245423} (\bibinfo {year} {2023})},\ \Eprint
  {http://arxiv.org/abs/2103.12217} {arXiv:2103.12217} \BibitemShut {NoStop}%
\bibitem [{\citenamefont {Geim}\ and\ \citenamefont
  {Grigorieva}(2013)}]{Geim2013}%
  \BibitemOpen
  \bibfield  {author} {\bibinfo {author} {\bibfnamefont {A.}~\bibnamefont
  {Geim}}\ and\ \bibinfo {author} {\bibfnamefont {I.}~\bibnamefont
  {Grigorieva}},\ }\href {\doibase 10.1038/nature12385} {\bibfield  {journal}
  {\bibinfo  {journal} {Nature}\ }\textbf {\bibinfo {volume} {499}},\ \bibinfo
  {pages} {419–425} (\bibinfo {year} {2013})}\BibitemShut {NoStop}%
\bibitem [{\citenamefont {Andrei}\ \emph {et~al.}(2021)\citenamefont {Andrei},
  \citenamefont {Efetov}, \citenamefont {Jarillo-Herrero}, \citenamefont
  {MacDonald}, \citenamefont {Mak}, \citenamefont {Senthil}, \citenamefont
  {Tutuc}, \citenamefont {Yazdani},\ and\ \citenamefont {Young}}]{Andrei2021}%
  \BibitemOpen
  \bibfield  {author} {\bibinfo {author} {\bibfnamefont {E.~Y.}\ \bibnamefont
  {Andrei}}, \bibinfo {author} {\bibfnamefont {D.~K.}\ \bibnamefont {Efetov}},
  \bibinfo {author} {\bibfnamefont {P.}~\bibnamefont {Jarillo-Herrero}},
  \bibinfo {author} {\bibfnamefont {A.~H.}\ \bibnamefont {MacDonald}}, \bibinfo
  {author} {\bibfnamefont {K.~F.}\ \bibnamefont {Mak}}, \bibinfo {author}
  {\bibfnamefont {T.}~\bibnamefont {Senthil}}, \bibinfo {author} {\bibfnamefont
  {E.}~\bibnamefont {Tutuc}}, \bibinfo {author} {\bibfnamefont
  {A.}~\bibnamefont {Yazdani}}, \ and\ \bibinfo {author} {\bibfnamefont
  {A.~F.}\ \bibnamefont {Young}},\ }\href {\doibase 10.1038/s41578-021-00284-1}
  {\bibfield  {journal} {\bibinfo  {journal} {Nat Rev Mater}\ }\textbf
  {\bibinfo {volume} {6}},\ \bibinfo {pages} {201–206} (\bibinfo {year}
  {2021})}\BibitemShut {NoStop}%
\bibitem [{\citenamefont {Triola}\ \emph {et~al.}(2020)\citenamefont {Triola},
  \citenamefont {Cayao},\ and\ \citenamefont {Black-Schaffer}}]{Triola2020}%
  \BibitemOpen
  \bibfield  {author} {\bibinfo {author} {\bibfnamefont {C.}~\bibnamefont
  {Triola}}, \bibinfo {author} {\bibfnamefont {J.}~\bibnamefont {Cayao}}, \
  and\ \bibinfo {author} {\bibfnamefont {A.~M.}\ \bibnamefont
  {Black-Schaffer}},\ }\href {\doibase https://doi.org/10.1002/andp.201900298}
  {\bibfield  {journal} {\bibinfo  {journal} {Annalen der Physik}\ }\textbf
  {\bibinfo {volume} {532}},\ \bibinfo {pages} {1900298} (\bibinfo {year}
  {2020})}\BibitemShut {NoStop}%
\bibitem [{\citenamefont {Eschrig}(2011)}]{Eschrig2011}%
  \BibitemOpen
  \bibfield  {author} {\bibinfo {author} {\bibfnamefont {M.}~\bibnamefont
  {Eschrig}},\ }\href {\doibase 10.1063/1.3541944} {\bibfield  {journal}
  {\bibinfo  {journal} {Physics Today}\ }\textbf {\bibinfo {volume} {64}},\
  \bibinfo {pages} {43} (\bibinfo {year} {2011})}\BibitemShut {NoStop}%
\bibitem [{\citenamefont {Linder}\ and\ \citenamefont
  {Robinson}(2015)}]{Linder2015}%
  \BibitemOpen
  \bibfield  {author} {\bibinfo {author} {\bibfnamefont {J.}~\bibnamefont
  {Linder}}\ and\ \bibinfo {author} {\bibfnamefont {J.}~\bibnamefont
  {Robinson}},\ }\href {\doibase 10.1038/nphys3242} {\bibfield  {journal}
  {\bibinfo  {journal} {Nature Phys}\ }\textbf {\bibinfo {volume} {11}},\
  \bibinfo {pages} {307–315} (\bibinfo {year} {2015})}\BibitemShut {NoStop}%
\bibitem [{\citenamefont {Kim}\ \emph {et~al.}(2018)\citenamefont {Kim},
  \citenamefont {Wang}, \citenamefont {Nakajima}, \citenamefont {Hu},
  \citenamefont {Ziemak}, \citenamefont {Syers}, \citenamefont {Wang},
  \citenamefont {Hodovanets}, \citenamefont {Denlinger}, \citenamefont
  {Brydon}, \citenamefont {Agterberg}, \citenamefont {Tanatar}, \citenamefont
  {Prozorov},\ and\ \citenamefont {Paglione}}]{Kim2018}%
  \BibitemOpen
  \bibfield  {author} {\bibinfo {author} {\bibfnamefont {H.}~\bibnamefont
  {Kim}}, \bibinfo {author} {\bibfnamefont {K.}~\bibnamefont {Wang}}, \bibinfo
  {author} {\bibfnamefont {Y.}~\bibnamefont {Nakajima}}, \bibinfo {author}
  {\bibfnamefont {R.}~\bibnamefont {Hu}}, \bibinfo {author} {\bibfnamefont
  {S.}~\bibnamefont {Ziemak}}, \bibinfo {author} {\bibfnamefont
  {P.}~\bibnamefont {Syers}}, \bibinfo {author} {\bibfnamefont
  {L.}~\bibnamefont {Wang}}, \bibinfo {author} {\bibfnamefont {H.}~\bibnamefont
  {Hodovanets}}, \bibinfo {author} {\bibfnamefont {J.~D.}\ \bibnamefont
  {Denlinger}}, \bibinfo {author} {\bibfnamefont {P.~M.~R.}\ \bibnamefont
  {Brydon}}, \bibinfo {author} {\bibfnamefont {D.~F.}\ \bibnamefont
  {Agterberg}}, \bibinfo {author} {\bibfnamefont {M.~A.}\ \bibnamefont
  {Tanatar}}, \bibinfo {author} {\bibfnamefont {R.}~\bibnamefont {Prozorov}}, \
  and\ \bibinfo {author} {\bibfnamefont {J.}~\bibnamefont {Paglione}},\ }\href
  {\doibase 10.1126/sciadv.aao4513} {\bibfield  {journal} {\bibinfo  {journal}
  {Science Advances}\ }\textbf {\bibinfo {volume} {4}},\ \bibinfo {pages}
  {eaao4513} (\bibinfo {year} {2018})}\BibitemShut {NoStop}%
\bibitem [{\citenamefont {Khim}\ \emph {et~al.}(2021)\citenamefont {Khim},
  \citenamefont {Landaeta}, \citenamefont {Banda}, \citenamefont {Bannor},
  \citenamefont {Brando}, \citenamefont {Brydon}, \citenamefont {Hafner},
  \citenamefont {K\"uchler}, \citenamefont {Cardoso-Gil}, \citenamefont
  {Stockert}, \citenamefont {Mackenzie}, \citenamefont {Agterberg},
  \citenamefont {Geibel},\ and\ \citenamefont {Hassinger}}]{Khim2021}%
  \BibitemOpen
  \bibfield  {author} {\bibinfo {author} {\bibfnamefont {S.}~\bibnamefont
  {Khim}}, \bibinfo {author} {\bibfnamefont {J.~F.}\ \bibnamefont {Landaeta}},
  \bibinfo {author} {\bibfnamefont {J.}~\bibnamefont {Banda}}, \bibinfo
  {author} {\bibfnamefont {N.}~\bibnamefont {Bannor}}, \bibinfo {author}
  {\bibfnamefont {M.}~\bibnamefont {Brando}}, \bibinfo {author} {\bibfnamefont
  {P.~M.~R.}\ \bibnamefont {Brydon}}, \bibinfo {author} {\bibfnamefont
  {D.}~\bibnamefont {Hafner}}, \bibinfo {author} {\bibfnamefont
  {R.}~\bibnamefont {K\"uchler}}, \bibinfo {author} {\bibfnamefont
  {R.}~\bibnamefont {Cardoso-Gil}}, \bibinfo {author} {\bibfnamefont
  {U.}~\bibnamefont {Stockert}}, \bibinfo {author} {\bibfnamefont {A.~P.}\
  \bibnamefont {Mackenzie}}, \bibinfo {author} {\bibfnamefont {D.~F.}\
  \bibnamefont {Agterberg}}, \bibinfo {author} {\bibfnamefont {C.}~\bibnamefont
  {Geibel}}, \ and\ \bibinfo {author} {\bibfnamefont {E.}~\bibnamefont
  {Hassinger}},\ }\href {\doibase 10.1126/science.abe7518} {\bibfield
  {journal} {\bibinfo  {journal} {Science}\ }\textbf {\bibinfo {volume}
  {373}},\ \bibinfo {pages} {1012} (\bibinfo {year} {2021})}\BibitemShut
  {NoStop}%
\bibitem [{\citenamefont {Ebert}\ \emph {et~al.}(2011)\citenamefont {Ebert},
  \citenamefont {K{\"o}dderitzsch},\ and\ \citenamefont
  {Min{\'a}r}}]{Ebert2011}%
  \BibitemOpen
  \bibfield  {author} {\bibinfo {author} {\bibfnamefont {H.}~\bibnamefont
  {Ebert}}, \bibinfo {author} {\bibfnamefont {D.}~\bibnamefont
  {K{\"o}dderitzsch}}, \ and\ \bibinfo {author} {\bibfnamefont
  {J.}~\bibnamefont {Min{\'a}r}},\ }\href {\doibase
  10.1088/0034-4885/74/9/096501} {\bibfield  {journal} {\bibinfo  {journal}
  {Rep. Prog. Phys.}\ }\textbf {\bibinfo {volume} {74}},\ \bibinfo {pages}
  {096501} (\bibinfo {year} {2011})}\BibitemShut {NoStop}%
\bibitem [{\citenamefont {Zabloudil}\ \emph {et~al.}(2005)\citenamefont
  {Zabloudil}, \citenamefont {Hammerling}, \citenamefont {Szunyogh},\ and\
  \citenamefont {Weinberger}}]{KKRbook}%
  \BibitemOpen
  \bibfield  {author} {\bibinfo {author} {\bibfnamefont {J.}~\bibnamefont
  {Zabloudil}}, \bibinfo {author} {\bibfnamefont {R.}~\bibnamefont
  {Hammerling}}, \bibinfo {author} {\bibfnamefont {L.}~\bibnamefont
  {Szunyogh}}, \ and\ \bibinfo {author} {\bibfnamefont {P.}~\bibnamefont
  {Weinberger}},\ }\href@noop {} {\emph {\bibinfo {title} {{Electron Scattering
  in Solid Matter: A Theoretical and Computational Treatise}}}},\ \bibinfo
  {series} {Springer Series in Solid-State Sciences}, Vol.\ \bibinfo {volume}
  {147}\ (\bibinfo  {publisher} {Springer, New York},\ \bibinfo {year}
  {2005})\BibitemShut {NoStop}%
\bibitem [{\citenamefont {Vosko}\ \emph {et~al.}(1980)\citenamefont {Vosko},
  \citenamefont {Wilk},\ and\ \citenamefont {Nusair}}]{Vosko1980}%
  \BibitemOpen
  \bibfield  {author} {\bibinfo {author} {\bibfnamefont {S.~H.}\ \bibnamefont
  {Vosko}}, \bibinfo {author} {\bibfnamefont {L.}~\bibnamefont {Wilk}}, \ and\
  \bibinfo {author} {\bibfnamefont {M.}~\bibnamefont {Nusair}},\ }\href
  {\doibase 10.1139/p80-159} {\bibfield  {journal} {\bibinfo  {journal} {Can.
  J. Phys.}\ }\textbf {\bibinfo {volume} {58}},\ \bibinfo {pages} {1200}
  (\bibinfo {year} {1980})}\BibitemShut {NoStop}%
\bibitem [{\citenamefont {Stefanou}\ \emph {et~al.}(1990)\citenamefont
  {Stefanou}, \citenamefont {Akai},\ and\ \citenamefont
  {Zeller}}]{Stefanou1990}%
  \BibitemOpen
  \bibfield  {author} {\bibinfo {author} {\bibfnamefont {N.}~\bibnamefont
  {Stefanou}}, \bibinfo {author} {\bibfnamefont {H.}~\bibnamefont {Akai}}, \
  and\ \bibinfo {author} {\bibfnamefont {R.}~\bibnamefont {Zeller}},\ }\href
  {\doibase 10.1016/0010-4655(90)90009-P} {\bibfield  {journal} {\bibinfo
  {journal} {Comput. Phys. Commun.}\ }\textbf {\bibinfo {volume} {60}},\
  \bibinfo {pages} {231} (\bibinfo {year} {1990})}\BibitemShut {NoStop}%
\bibitem [{\citenamefont {Stefanou}\ and\ \citenamefont
  {Zeller}(1991)}]{Stefanou1991}%
  \BibitemOpen
  \bibfield  {author} {\bibinfo {author} {\bibfnamefont {N.}~\bibnamefont
  {Stefanou}}\ and\ \bibinfo {author} {\bibfnamefont {R.}~\bibnamefont
  {Zeller}},\ }\href {\doibase 10.1088/0953-8984/3/39/006} {\bibfield
  {journal} {\bibinfo  {journal} {J. Phys.: Cond. Matter}\ }\textbf {\bibinfo
  {volume} {3}},\ \bibinfo {pages} {7599} (\bibinfo {year} {1991})}\BibitemShut
  {NoStop}%
\bibitem [{\citenamefont {R\"u{\ss}mann}\ \emph
  {et~al.}(2021{\natexlab{a}})\citenamefont {R\"u{\ss}mann}, \citenamefont
  {Bertoldo}, \citenamefont {Br\"oder}, \citenamefont {Wasmer}, \citenamefont
  {Mozumder}, \citenamefont {Chico},\ and\ \citenamefont
  {Bl\"ugel}}]{aiida-kkr-code}%
  \BibitemOpen
  \bibfield  {author} {\bibinfo {author} {\bibfnamefont {P.}~\bibnamefont
  {R\"u{\ss}mann}}, \bibinfo {author} {\bibfnamefont {F.}~\bibnamefont
  {Bertoldo}}, \bibinfo {author} {\bibfnamefont {J.}~\bibnamefont {Br\"oder}},
  \bibinfo {author} {\bibfnamefont {J.}~\bibnamefont {Wasmer}}, \bibinfo
  {author} {\bibfnamefont {R.}~\bibnamefont {Mozumder}}, \bibinfo {author}
  {\bibfnamefont {J.}~\bibnamefont {Chico}}, \ and\ \bibinfo {author}
  {\bibfnamefont {S.}~\bibnamefont {Bl\"ugel}},\ }\href {\doibase
  https://doi.org/10.5281/zenodo.3628250} {\bibfield  {journal} {\bibinfo
  {journal} {Zenodo}\ } (\bibinfo {year} {2021}{\natexlab{a}}),\
  https://doi.org/10.5281/zenodo.3628250}\BibitemShut {NoStop}%
\bibitem [{\citenamefont {R\"u{\ss}mann}\ \emph
  {et~al.}(2021{\natexlab{b}})\citenamefont {R\"u{\ss}mann}, \citenamefont
  {Bertoldo},\ and\ \citenamefont {Bl\"ugel}}]{aiida-kkr-paper}%
  \BibitemOpen
  \bibfield  {author} {\bibinfo {author} {\bibfnamefont {P.}~\bibnamefont
  {R\"u{\ss}mann}}, \bibinfo {author} {\bibfnamefont {F.}~\bibnamefont
  {Bertoldo}}, \ and\ \bibinfo {author} {\bibfnamefont {S.}~\bibnamefont
  {Bl\"ugel}},\ }\href {\doibase 10.1038/s41524-020-00482-5} {\bibfield
  {journal} {\bibinfo  {journal} {npj Comput. Mater.}\ }\textbf {\bibinfo
  {volume} {7}},\ \bibinfo {pages} {13} (\bibinfo {year}
  {2021}{\natexlab{b}})}\BibitemShut {NoStop}%
\bibitem [{\citenamefont {Huber}\ \emph {et~al.}(2020)\citenamefont {Huber},
  \citenamefont {Zoupanos}, \citenamefont {Uhrin}, \citenamefont {Talirz},
  \citenamefont {Kahle}, \citenamefont {Häuselmann}, \citenamefont {Gresch},
  \citenamefont {M\"uller}, \citenamefont {Yakutovich}, \citenamefont
  {Andersen}, \citenamefont {Ramirez}, \citenamefont {Adorf}, \citenamefont
  {Gargiulo}, \citenamefont {Kumbhar}, \citenamefont {Passaro}, \citenamefont
  {Johnston}, \citenamefont {Merkys}, \citenamefont {Cepellotti}, \citenamefont
  {Mounet}, \citenamefont {Marzari}, \citenamefont {Kozinsky},\ and\
  \citenamefont {Pizzi}}]{aiida}%
  \BibitemOpen
  \bibfield  {author} {\bibinfo {author} {\bibfnamefont {S.~P.}\ \bibnamefont
  {Huber}}, \bibinfo {author} {\bibfnamefont {S.}~\bibnamefont {Zoupanos}},
  \bibinfo {author} {\bibfnamefont {M.}~\bibnamefont {Uhrin}}, \bibinfo
  {author} {\bibfnamefont {L.}~\bibnamefont {Talirz}}, \bibinfo {author}
  {\bibfnamefont {L.}~\bibnamefont {Kahle}}, \bibinfo {author} {\bibfnamefont
  {R.}~\bibnamefont {Häuselmann}}, \bibinfo {author} {\bibfnamefont
  {D.}~\bibnamefont {Gresch}}, \bibinfo {author} {\bibfnamefont
  {T.}~\bibnamefont {M\"uller}}, \bibinfo {author} {\bibfnamefont {A.~V.}\
  \bibnamefont {Yakutovich}}, \bibinfo {author} {\bibfnamefont {C.~W.}\
  \bibnamefont {Andersen}}, \bibinfo {author} {\bibfnamefont {F.~F.}\
  \bibnamefont {Ramirez}}, \bibinfo {author} {\bibfnamefont {C.~S.}\
  \bibnamefont {Adorf}}, \bibinfo {author} {\bibfnamefont {F.}~\bibnamefont
  {Gargiulo}}, \bibinfo {author} {\bibfnamefont {S.}~\bibnamefont {Kumbhar}},
  \bibinfo {author} {\bibfnamefont {E.}~\bibnamefont {Passaro}}, \bibinfo
  {author} {\bibfnamefont {C.}~\bibnamefont {Johnston}}, \bibinfo {author}
  {\bibfnamefont {A.}~\bibnamefont {Merkys}}, \bibinfo {author} {\bibfnamefont
  {A.}~\bibnamefont {Cepellotti}}, \bibinfo {author} {\bibfnamefont
  {N.}~\bibnamefont {Mounet}}, \bibinfo {author} {\bibfnamefont
  {N.}~\bibnamefont {Marzari}}, \bibinfo {author} {\bibfnamefont
  {B.}~\bibnamefont {Kozinsky}}, \ and\ \bibinfo {author} {\bibfnamefont
  {G.}~\bibnamefont {Pizzi}},\ }\href {\doibase 10.1038/s41597-020-00638-4}
  {\bibfield  {journal} {\bibinfo  {journal} {Sci. Data}\ }\textbf {\bibinfo
  {volume} {7}},\ \bibinfo {pages} {300} (\bibinfo {year} {2020})}\BibitemShut
  {NoStop}%
\bibitem [{\citenamefont {Wilkinson}\ \emph {et~al.}(2016)\citenamefont
  {Wilkinson}, \citenamefont {Dumontier}, \citenamefont {Aalbersberg},
  \citenamefont {Appleton}, \citenamefont {Axton}, \citenamefont {Baak},
  \citenamefont {Blomberg}, \citenamefont {Boiten}, \citenamefont {{da Silva
  Santos}}, \citenamefont {Bourne}, \citenamefont {Bouwman}, \citenamefont
  {Brookes}, \citenamefont {Clark}, \citenamefont {Crosas}, \citenamefont
  {Dillo}, \citenamefont {Dumon}, \citenamefont {Edmunds}, \citenamefont
  {Evelo}, \citenamefont {Finkers}, \citenamefont {Gonzalez-Beltran},
  \citenamefont {Gray}, \citenamefont {Groth}, \citenamefont {Goble},
  \citenamefont {Grethe}, \citenamefont {Heringa}, \citenamefont {{'t Hoen}},
  \citenamefont {Hooft}, \citenamefont {Kuhn}, \citenamefont {Kok},
  \citenamefont {Kok}, \citenamefont {Lusher}, \citenamefont {Martone},
  \citenamefont {Mons}, \citenamefont {Packer}, \citenamefont {Persson},
  \citenamefont {Rocca-Serra}, \citenamefont {Roos}, \citenamefont {van
  Schaik}, \citenamefont {Sansone}, \citenamefont {Schultes}, \citenamefont
  {Sengstag}, \citenamefont {Slater}, \citenamefont {Strawn}, \citenamefont
  {Swertz}, \citenamefont {Thompson}, \citenamefont {van~der Lei},
  \citenamefont {van Mulligen}, \citenamefont {Velterop}, \citenamefont
  {Waagmeester}, \citenamefont {Wittenburg}, \citenamefont {Wolstencroft},
  \citenamefont {Zhao},\ and\ \citenamefont {Mons}}]{Wilkinson2016}%
  \BibitemOpen
  \bibfield  {author} {\bibinfo {author} {\bibfnamefont {M.~D.}\ \bibnamefont
  {Wilkinson}}, \bibinfo {author} {\bibfnamefont {M.}~\bibnamefont
  {Dumontier}}, \bibinfo {author} {\bibfnamefont {I.~J.}\ \bibnamefont
  {Aalbersberg}}, \bibinfo {author} {\bibfnamefont {G.}~\bibnamefont
  {Appleton}}, \bibinfo {author} {\bibfnamefont {M.}~\bibnamefont {Axton}},
  \bibinfo {author} {\bibfnamefont {A.}~\bibnamefont {Baak}}, \bibinfo {author}
  {\bibfnamefont {N.}~\bibnamefont {Blomberg}}, \bibinfo {author}
  {\bibfnamefont {J.-W.}\ \bibnamefont {Boiten}}, \bibinfo {author}
  {\bibfnamefont {L.~B.}\ \bibnamefont {{da Silva Santos}}}, \bibinfo {author}
  {\bibfnamefont {P.~E.}\ \bibnamefont {Bourne}}, \bibinfo {author}
  {\bibfnamefont {J.}~\bibnamefont {Bouwman}}, \bibinfo {author} {\bibfnamefont
  {A.~J.}\ \bibnamefont {Brookes}}, \bibinfo {author} {\bibfnamefont
  {T.}~\bibnamefont {Clark}}, \bibinfo {author} {\bibfnamefont
  {M.}~\bibnamefont {Crosas}}, \bibinfo {author} {\bibfnamefont
  {I.}~\bibnamefont {Dillo}}, \bibinfo {author} {\bibfnamefont
  {O.}~\bibnamefont {Dumon}}, \bibinfo {author} {\bibfnamefont
  {S.}~\bibnamefont {Edmunds}}, \bibinfo {author} {\bibfnamefont {C.~T.}\
  \bibnamefont {Evelo}}, \bibinfo {author} {\bibfnamefont {R.}~\bibnamefont
  {Finkers}}, \bibinfo {author} {\bibfnamefont {A.}~\bibnamefont
  {Gonzalez-Beltran}}, \bibinfo {author} {\bibfnamefont {A.~J.~G.}\
  \bibnamefont {Gray}}, \bibinfo {author} {\bibfnamefont {P.}~\bibnamefont
  {Groth}}, \bibinfo {author} {\bibfnamefont {C.}~\bibnamefont {Goble}},
  \bibinfo {author} {\bibfnamefont {J.~S.}\ \bibnamefont {Grethe}}, \bibinfo
  {author} {\bibfnamefont {J.}~\bibnamefont {Heringa}}, \bibinfo {author}
  {\bibfnamefont {P.~A.~C.}\ \bibnamefont {{'t Hoen}}}, \bibinfo {author}
  {\bibfnamefont {R.}~\bibnamefont {Hooft}}, \bibinfo {author} {\bibfnamefont
  {T.}~\bibnamefont {Kuhn}}, \bibinfo {author} {\bibfnamefont {R.}~\bibnamefont
  {Kok}}, \bibinfo {author} {\bibfnamefont {J.}~\bibnamefont {Kok}}, \bibinfo
  {author} {\bibfnamefont {S.~J.}\ \bibnamefont {Lusher}}, \bibinfo {author}
  {\bibfnamefont {M.~E.}\ \bibnamefont {Martone}}, \bibinfo {author}
  {\bibfnamefont {A.}~\bibnamefont {Mons}}, \bibinfo {author} {\bibfnamefont
  {A.~L.}\ \bibnamefont {Packer}}, \bibinfo {author} {\bibfnamefont
  {B.}~\bibnamefont {Persson}}, \bibinfo {author} {\bibfnamefont
  {P.}~\bibnamefont {Rocca-Serra}}, \bibinfo {author} {\bibfnamefont
  {M.}~\bibnamefont {Roos}}, \bibinfo {author} {\bibfnamefont {R.}~\bibnamefont
  {van Schaik}}, \bibinfo {author} {\bibfnamefont {S.-A.}\ \bibnamefont
  {Sansone}}, \bibinfo {author} {\bibfnamefont {E.}~\bibnamefont {Schultes}},
  \bibinfo {author} {\bibfnamefont {T.}~\bibnamefont {Sengstag}}, \bibinfo
  {author} {\bibfnamefont {T.}~\bibnamefont {Slater}}, \bibinfo {author}
  {\bibfnamefont {G.}~\bibnamefont {Strawn}}, \bibinfo {author} {\bibfnamefont
  {M.~A.}\ \bibnamefont {Swertz}}, \bibinfo {author} {\bibfnamefont
  {M.}~\bibnamefont {Thompson}}, \bibinfo {author} {\bibfnamefont
  {J.}~\bibnamefont {van~der Lei}}, \bibinfo {author} {\bibfnamefont
  {E.}~\bibnamefont {van Mulligen}}, \bibinfo {author} {\bibfnamefont
  {J.}~\bibnamefont {Velterop}}, \bibinfo {author} {\bibfnamefont
  {A.}~\bibnamefont {Waagmeester}}, \bibinfo {author} {\bibfnamefont
  {P.}~\bibnamefont {Wittenburg}}, \bibinfo {author} {\bibfnamefont
  {K.}~\bibnamefont {Wolstencroft}}, \bibinfo {author} {\bibfnamefont
  {J.}~\bibnamefont {Zhao}}, \ and\ \bibinfo {author} {\bibfnamefont
  {B.}~\bibnamefont {Mons}},\ }\href {\doibase 10.1038/sdata.2016.18}
  {\bibfield  {journal} {\bibinfo  {journal} {Sci. Data}\ }\textbf {\bibinfo
  {volume} {3}},\ \bibinfo {pages} {160018} (\bibinfo {year}
  {2016})}\BibitemShut {NoStop}%
\bibitem [{\citenamefont {Talirz}\ \emph {et~al.}(2020)\citenamefont {Talirz},
  \citenamefont {Kumbhar}, \citenamefont {Passaro}, \citenamefont {Yakutovich},
  \citenamefont {Granata}, \citenamefont {Gargiulo}, \citenamefont {Borelli},
  \citenamefont {Uhrin}, \citenamefont {Huber}, \citenamefont {Zoupanos},
  \citenamefont {Adorf}, \citenamefont {Andersen}, \citenamefont {Sch\"utt},
  \citenamefont {Pignedoli}, \citenamefont {Passerone}, \citenamefont
  {VandeVondele}, \citenamefont {Thomas C.~Schulthess},\ and\ \citenamefont
  {Marzari}}]{Talirz2020}%
  \BibitemOpen
  \bibfield  {author} {\bibinfo {author} {\bibfnamefont {L.}~\bibnamefont
  {Talirz}}, \bibinfo {author} {\bibfnamefont {S.}~\bibnamefont {Kumbhar}},
  \bibinfo {author} {\bibfnamefont {E.}~\bibnamefont {Passaro}}, \bibinfo
  {author} {\bibfnamefont {A.~V.}\ \bibnamefont {Yakutovich}}, \bibinfo
  {author} {\bibfnamefont {V.}~\bibnamefont {Granata}}, \bibinfo {author}
  {\bibfnamefont {F.}~\bibnamefont {Gargiulo}}, \bibinfo {author}
  {\bibfnamefont {M.}~\bibnamefont {Borelli}}, \bibinfo {author} {\bibfnamefont
  {M.}~\bibnamefont {Uhrin}}, \bibinfo {author} {\bibfnamefont {S.~P.}\
  \bibnamefont {Huber}}, \bibinfo {author} {\bibfnamefont {S.}~\bibnamefont
  {Zoupanos}}, \bibinfo {author} {\bibfnamefont {C.~S.}\ \bibnamefont {Adorf}},
  \bibinfo {author} {\bibfnamefont {C.~W.}\ \bibnamefont {Andersen}}, \bibinfo
  {author} {\bibfnamefont {O.}~\bibnamefont {Sch\"utt}}, \bibinfo {author}
  {\bibfnamefont {C.~A.}\ \bibnamefont {Pignedoli}}, \bibinfo {author}
  {\bibfnamefont {D.}~\bibnamefont {Passerone}}, \bibinfo {author}
  {\bibfnamefont {J.}~\bibnamefont {VandeVondele}}, \bibinfo {author}
  {\bibfnamefont {G.~P.}\ \bibnamefont {Thomas C.~Schulthess}, \bibfnamefont
  {Berend~Smit}}, \ and\ \bibinfo {author} {\bibfnamefont {N.}~\bibnamefont
  {Marzari}},\ }\href {\doibase 10.1038/s41597-020-00637-5} {\bibfield
  {journal} {\bibinfo  {journal} {Sci. Data}\ }\textbf {\bibinfo {volume}
  {7}},\ \bibinfo {pages} {299} (\bibinfo {year} {2020})}\BibitemShut {NoStop}%
\bibitem [{\citenamefont {R\"u{\ss}mann}\ \emph {et~al.}(2023)\citenamefont
  {R\"u{\ss}mann}, \citenamefont {Bahari}, \citenamefont {Bl\"ugel},\ and\
  \citenamefont {Trauzettel}}]{doi-dataset}%
  \BibitemOpen
  \bibfield  {author} {\bibinfo {author} {\bibfnamefont {P.}~\bibnamefont
  {R\"u{\ss}mann}}, \bibinfo {author} {\bibfnamefont {M.}~\bibnamefont
  {Bahari}}, \bibinfo {author} {\bibfnamefont {S.}~\bibnamefont {Bl\"ugel}}, \
  and\ \bibinfo {author} {\bibfnamefont {B.}~\bibnamefont {Trauzettel}},\
  }\href {\doibase doi: 10.24435/materialscloud:20-9z} {\bibfield  {journal}
  {\bibinfo  {journal} {Materials Cloud Archive}\ }\textbf {\bibinfo {volume}
  {2023.X}} (\bibinfo {year} {2023}),\ doi:
  10.24435/materialscloud:20-9z}\BibitemShut {NoStop}%
\end{thebibliography}%


\appendix
\renewcommand{\thefigure}{A\arabic{figure}}
\setcounter{figure}{0}
\renewcommand{\thetable}{A\,\Roman{table}}
\setcounter{table}{0}

\section{Computational details of the DFT simulations} \label{app:ComputationDetails}

Our density functional theory calculations rely on multiple-scattering theory and employ the relativistic Korringa-Kohn-Rostoker Green function method (KKR)~\cite{Ebert2011, KKRbook} as implemented in the JuKKR code~\cite{jukkr}.
We use the local density approximation (LDA) to parameterize the normal state exchange correlation functional~\cite{Vosko1980} and an $\ell_\mathrm{max}=2$ cutoff in the angular momentum expansion of the space filling Voronoi cells around the atomic centers where we make use of the exact (i.e.\ full-potential) description of the atomic shapes~\cite{Stefanou1990,Stefanou1991}. {We use a two-dimensional geometry where periodicity is assumed in the plane, but a finite layer thickness is used in the direction along the heterostructure.}

The series of DFT calculations in this study are orchestrated with the help of the AiiDA-KKR plugin~\cite{aiida-kkr-code, aiida-kkr-paper} to the AiiDA infrastructure~\cite{aiida}. This has the advantage that the data provenance is automatically stored in compliance to the FAIR principles of open data~\cite{Wilkinson2016}. The complete data set that includes the full provenance of the calculations is made publicly available in the materials cloud {repository}~\cite{Talirz2020, doi-dataset}.

The source codes of the AiiDA-KKR plugin~\cite{aiida-kkr-paper, aiida-kkr-code} and the JuKKR code~\cite{jukkr} are published as open source software under the MIT license at \url{https://github.com/JuDFTteam/aiida-kkr} and \url{https://iffgit.fz-juelich.de/kkr/jukkr}, respectively.

\section{Additional details of the normal state electronic structure from DFT}\label{app:DFTnormal}

\begin{figure}
	\centering
	\includegraphics[width=\linewidth]{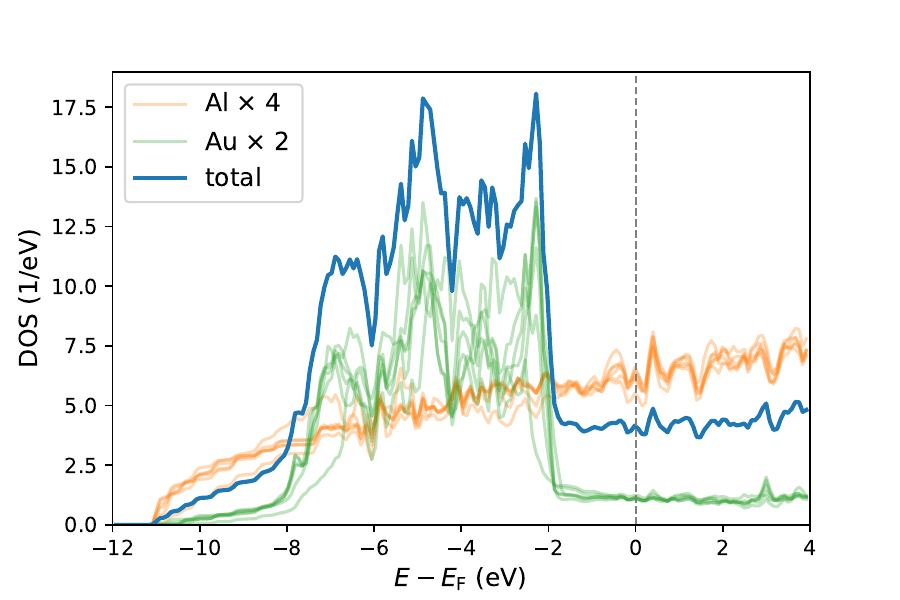}
	\caption{Density of states of the Al/Au heterostructure. The faint orange and green lines indicate the contributions of the individual Al and Au layers to the total DOS (blue line).  
	}
	\label{fig:suppDOS}
\end{figure}

Figure~\ref{fig:suppDOS} shows the total and layer-resolved density of states {(DOS)} of the Al/Au heterostructure as computed from DFT. The fully occupied $d$-shell of Au can be seen between $E-E_F=-8$\,eV and $E-E_F=-2$\,eV and we {confirm} the well-known fact that the DOS at $E_F$ has predominantly $s-p_z$ character. {The electrons in aluminum show an almost, i.e.\ except for small van Hove singularities due to lifting of degeneracies in $sp$-bands by crystal fields, free-electron nature which can be recognized in the typical square-root shape of the DOS.}

\begin{figure}
	\centering
	\includegraphics[width=\linewidth]{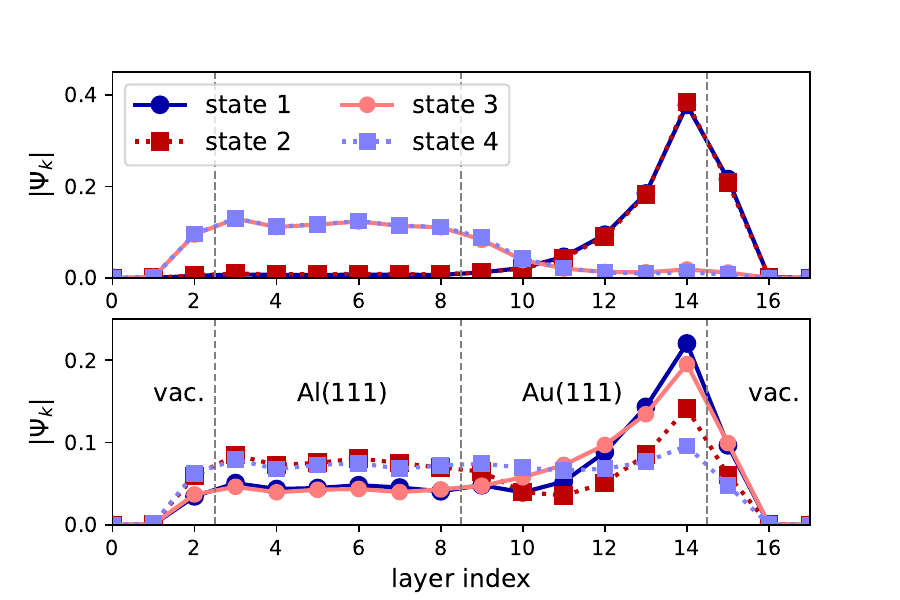}
	\caption{Wave function localization for the states 1--4 at the values of (top) $k_x=0.1\,$\AA$^{-1}$ and (bottom) $k_x=0.3\,$\AA$^{-1}$, which highlights the significant hybridization of the Au and Al-derived states at larger momenta.
	}
	\label{fig:suppLocalization}
\end{figure}

Figure~\ref{fig:suppLocalization} shows the wavefunction localization throughout the Al/Au heterostructure for the states labeled 1-4 in Fig.~\ref{Fig1.pdf}(b) of the main text.
The wavefunction localization is shown for two values of the momentum; the top panel illustrates the situation close to $\Gamma$ at $k_x=0.1\,$\AA$^{-1}$ where the Al and Au-derived states are clearly distinct and the bottom panel shows the situation at larger momentum of $k_x=0.3\,$\AA$^{-1}$ where the four bands interact. At smaller momentum, the Rashba surface states (1,2) are exponentially localized at the Au-vacuum interface and have a negligible overlap with the states 3 and 4 which are delocalized throughout the Al film. 
In contrast, at larger momenta the four states hybridize which can be seen in the significant weight of all states in both the layers of Al \emph{and} Au. This is particularly visible in comparing the top and bottom panels of Fig.~\ref{fig:suppLocalization} around layer 5 (in Al) and layer 14 (Au surface).

\section{Band basis representation}\label{Masoud Appendix: Band basis}

Using the Gram–Schmidt method, we derive orthonormal eigenvectors for the normal state Hamiltonian given in Eq.~\eqref{eq:modelNormalState} as
\begin{align}
  \!\!\!\!\!\!\!\!\hat{\mathcal{V}}(\mathbf{k})&\!\!=\!\!\!\left(\!\!\!\begin{array}{cccc}
\frac{-i\sqrt{2}\gamma}{2\sqrt{\gamma^{2}+4F_{0}^{2}}} & \frac{i\sqrt{2}P}{2\sqrt{P^{2}+4F_{0}^{2}}} & \frac{i\sqrt{2}\upsilon}{\sqrt{\Psi^{2}+4F_{0}^{2}}} & \frac{-i\sqrt{2}\Omega}{2\sqrt{Q^{2}+4F_{0}^{2}}}\\
\frac{\sqrt{\gamma^{2}+4F_{0}^{2}}}{\sqrt{8}\Theta^{-}} & \frac{\sqrt{P^{2}+4F_{0}^{2}}}{\sqrt{8}\Theta^{+}} & \frac{-\sqrt{2}\upsilon}{\sqrt{\Psi^{2}+4F_{0}^{2}}} & \frac{-\sqrt{Q^{2}+4F_{0}^{2}}}{\sqrt{8}\Theta^{+}}\\
\frac{-i\sqrt{2}F_{0}}{\sqrt{\gamma^{2}+4F_{0}^{2}}} & \frac{i\sqrt{2}F_{0}}{\sqrt{P^{2}+4F_{0}^{2}}} & \frac{-i\sqrt{8}F_{0}}{\sqrt{\Psi^{2}+4F_{0}^{2}}} & \frac{i\sqrt{2}F_{0}}{\sqrt{Q^{2}+4F_{0}^{2}}}\\
\frac{\sqrt{2}F_{0}}{\sqrt{\gamma^{2}+4F_{0}^{2}}} & \frac{\sqrt{2}F_{0}}{\sqrt{P^{2}+4F_{0}^{2}}} & \frac{\sqrt{8}F_{0}}{\sqrt{\Psi^{2}+4F_{0}^{2}}} & \frac{\sqrt{2}F_{0}}{\sqrt{Q^{2}+4F_{0}^{2}}}
    \end{array}\!\!\!\right)\!,\label{Eigenvectors}\!\!\!\!
\end{align}
where $[\hat{\mathcal{V}}(\mathbf{k})]^{\dagger}\hat{\mathcal{V}}(\mathbf{k})=1$ and we have defined
\begin{align}
\Theta^{+} & =\sqrt{(\Upsilon_{1}^{+})^{2}+4F_{0}^{2}}, & \Theta^{-} & =\sqrt{(\Upsilon_{1}^{\pm})^{2}+4F_{0}^{2}},\\
\Upsilon_{1}^{\pm} & =\mathcal{E}_{\text{Al}}-\mathcal{E}_{\text{Au}}^{\pm}, & \Upsilon_{2}^{\pm} & =\mathcal{E}_{\text{Al}}+\mathcal{E}_{\text{Au}}^{\pm},\\
\gamma & =\Theta^{-}+\Upsilon_{1}^{-}, & P & =\Theta^{+}+\Upsilon_{2}^{-},\\
\upsilon & =\Theta^{-}-\Upsilon_{1}^{-}, & \Psi & =\Theta^{+}+\Upsilon_{1}^{+},\\
\Omega & =\Theta^{+}-\Upsilon_{1}^{+}, & Q & =\Theta^{+}+\Upsilon_{1}^{+}.
\end{align}

The band basis representation for the BdG Hamiltonian can be obtained as follows. The superconducting Hamiltonian is defined by $H=\sum_{\mathbf{k}}\hat{\psi}_{\mathbf{k}}^{\dagger}\hat{\mathcal{H}}(\mathbf{k})\hat{\psi}_{\mathbf{k}}$
where the Nambu basis is $\hat{\psi}_{\mathbf{k}}=(\hat{\varphi}_{\mathbf{k}}^{\dagger},\hat{\varphi}_{-\mathbf{k}}^{\dagger T})^{T}$,
and the BdG Hamiltonian is given by
\begin{equation}
\hat{\mathcal{H}}(\mathbf{k})=\left(\begin{array}{cc}
\hat{H}(\mathbf{k}) & \text{diag}(\Delta i\hat{\sigma}_{y},0)\\{}
[\text{diag}(\Delta i\hat{\sigma}_{y},0)]^{\dagger} & -\hat{H}^{T}(-\mathbf{k})
\end{array}\right).
\end{equation}
where $\hat{H}(\mathbf{k})$ is the $4\times4$ matrix form of the normal state Hamiltonian. The band basis representation of the superconducting Hamiltonian can be obtained through the similarity transformation $\hat{H}_{\text{BdG}}(\mathbf{k}) =\hat{U}^{\dagger}\hat{\mathcal{H}}(\mathbf{k})\hat{U}$ with $\hat{U}=\hat{\mathcal{V}}(\mathbf{k})\oplus[\hat{\mathcal{V}}(-\mathbf{k})]^{\dagger T}$.


\section{Downfolding method}\label{APP:FoldingDown}

In this section, we explain how to employ the downfolding method to
obtain a $4\times4$ effective Hamiltonian to describe spectral properties
of the system for a specified energy window. Our starting point is
a general superconducting Hamiltonian, represented in the eigenspace
of the normal state given by the $8\times8$ matrix
\begin{equation}
    \hat{H}=\left(\begin{array}{cc|cc}
    \hat{N}_{1} & 0 & \hat{\varDelta}_{1} & \hat{\varDelta}_{2}\\
    0 & \hat{N}_{2} & \hat{\varDelta}_{3} & \hat{\varDelta}_{4}\\
    \hline \hat{\varDelta}_{1}^{\dagger} & \hat{\varDelta}_{3}^{\dagger} & \hat{h}_{1} & 0\\
    \hat{\varDelta}_{2}^{\dagger} & \hat{\varDelta}_{4}^{\dagger} & 0 & \hat{h}_{2}
    \end{array}\right),
\end{equation}
where $\hat{N}_{1(2)}$ ($\hat{h}_{1(2)}$) is a $2\times2$ diagonal
sub-block matrix containing a pair of energy bands in the normal state,
and $\hat{\varDelta}_{1(4)}$ and $\hat{\varDelta}_{2(3)}$ are the
pairing matrices projected onto the intra-(inter-)bands. Note that $\hat{\varDelta}_{1(4)}$ and $\hat{\varDelta}_{2(3)}$ induce full and partial pairing gaps at Fermi energy and finite excitation energies, respectively. Without loss
of generality, we change the basis of $\hat{H}$ with the unitary transformation
$\hat{U}$ to let the diagonal sub-blocks contain the electron-hole
components with a pairing potential. This can be done with $\hat{H}^{\prime}=\hat{U}^{\dagger}\hat{H}\hat{U}$
where $\hat{U}$ is given by
\begin{eqnarray}
    \hat{U}^{-1} & = & \hat{U}^{\dagger}=\left(\begin{array}{cccc}
    1 & 0 & 0 & 0\\
    0 & 0 & 0 & 1\\
    0 & 1 & 0 & 0\\
    0 & 0 & 1 & 0
    \end{array}\right),
\end{eqnarray}
and $\hat{H}^{\prime}$ becomes
\begin{equation}
    \hat{H}^{\prime}=\left(\begin{array}{cc|cc}
    \hat{N}_{1} & \hat{\varDelta}_{2} & 0 & \hat{\varDelta}_{1}\\
    \hat{\varDelta}_{2}^{\dagger} & \hat{h}_{2} & \hat{\varDelta}_{4}^{\dagger} & 0\\
    \hline 0 & \hat{\varDelta}_{4} & \hat{N}_{2} & \hat{\varDelta}_{3}\\
    \hat{\varDelta}_{1}^{\dagger} & 0 & \hat{\varDelta}_{3}^{\dagger} & \hat{h}_{1}
    \end{array}\right)\equiv\!\left(\!\!\begin{array}{c|c}
    \hat{M}_{11} & \hat{M}_{12}\\
    \hline \hat{M}_{21} & \hat{M}_{22}
    \end{array}\!\!\right)\label{App: General band basis Hamil}
\end{equation}
In multiband systems, the downfolding method paves the way to obtain
the spectral properties of a desired sub-block, e.g., $\hat{M}_{11}$,
taking into account the perturbative effects of other blocks. To understand
the method, we consider the eigenvalue problem for $H^{\prime}$ given
by
\begin{align}
    \left(\!\!\begin{array}{cc}
    \hat{M}_{11} & \hat{M}_{12}\\
    \hat{M}_{21} & \hat{M}_{22}
    \end{array}\!\!\right)\left(\begin{array}{c}
    \hat{\psi}_{A}\\
    \hat{\psi}_{B}
    \end{array}\right) & =E\left(\begin{array}{c}
    \hat{\psi}_{A}\\
    \hat{\psi}_{B}
    \end{array}\right),\label{App: Eigenvalue problem}
\end{align}
where $(\hat{\psi}_{A},\hat{\psi}_{B})^{T}$ is the eigenvector associated
to eigenenergy $E$. Equation~\eqref{App: Eigenvalue problem} is a coupled
equation which can be written as
\begin{align}
    \hat{M}_{11}\hat{\psi}_{A}+\hat{M}_{12}\hat{\psi}_{B} & =E\hat{\psi}_{A},\label{App: Eg1}\\
    \hat{M}_{21}\hat{\psi}_{A}+\hat{M}_{22}\hat{\psi}_{B} & =E\hat{\psi}_{B}.\label{APP: Eg2}
\end{align}
Inserting $\hat{\psi}_{B}=\left(E-\hat{M}_{22}\right)^{-1}\hat{M}_{21}\hat{\psi}_{A}$ into Eq.~$\text{\eqref{App: Eg1}}$
results in $\hat{M}_{11}^{\text{eff}}\hat{\psi}_{A}=E\hat{\psi}_{A}$
with
\begin{equation}
    \hat{M}_{11}^{\text{eff}}=\hat{M}_{11}+\hat{M}_{12}\hat{\Lambda}^{-1}\hat{M}_{21},
\end{equation}
where $\hat{\Lambda}=(\omega\hat{I}_{4}-\hat{M}_{22})^{-1}$, $\hat{I}_{4}$
is the $4\times4$ identity matrix, and $\omega$ denotes a constant
close to the energy range where the desired pairing happens. Thus,
eigenvalues $\hat{M}_{11}^{\text{eff}}$ describes the spectral properties
of $\hat{M}_{11}$ taking into account the effects of other
sub-blocks. We now aim to find an expression for
$\hat{\Lambda}^{-1}$. To do so, we define $\hat{\Lambda}\equiv\hat{\varepsilon}+\hat{\tilde{\Delta}}$
where $\hat{\varepsilon}$ ($\hat{\tilde{\Delta}}$) is the normal
state (pairing) part given by
\begin{align}
    \hat{\varepsilon} & =\left(\begin{array}{cc}
    \omega\hat{I}_{2}-\hat{N}_{2} & 0\\
    0 & \omega\hat{I}_{2}-\hat{h}_{1}
    \end{array}\right),\\
    \hat{\tilde{\Delta}} & =\left(\begin{array}{cc}
    0 & -\hat{\varDelta}_{3}\\
    -\hat{\varDelta}_{3}^{\dagger} & 0
    \end{array}\right).
\end{align}
We can find the inverse of $\hat{\Lambda}$ using the Neumann series
expansion up to second order given by
\begin{align}
    \!\hat{\Lambda}^{-1} & \!=\hat{\varepsilon}^{-1}(\hat{I}+\hat{\tilde{\Delta}}\hat{\varepsilon}^{-1})^{-1}=\hat{\varepsilon}^{-1}\sum_{n=0}^{\infty}(-\hat{\tilde{\Delta}}\hat{\varepsilon}^{-1})^{n}\label{App:Neumann-0}\\
     & \!\approx\hat{\varepsilon}^{-1}(\hat{I}-\hat{\tilde{\Delta}}\hat{\varepsilon}^{-1})=\hat{\varepsilon}^{-1}-\hat{\varepsilon}^{-1}\hat{\tilde{\Delta}}\hat{\varepsilon}^{-1}\label{App:Neumann-1}\\
     & \!=\!\left(\!\!\begin{array}{cc}
    \!\frac{1}{\omega\hat{\sigma}_{0}-\hat{N}_{2}} & \!\frac{1}{\omega\hat{\sigma}_{0}-\hat{N}_{2}}\hat{\varDelta}_{3}\frac{1}{\omega\hat{\sigma}_{0}-\hat{h}_{1}}\\
    \!\frac{1}{\omega\hat{\sigma}_{0}-\hat{h}_{1}}\big[\hat{\varDelta}_{3}\big]^{\dagger}\frac{1}{\omega\hat{\sigma}_{0}-\hat{N}_{2}} & \!\frac{1}{\omega\hat{\sigma}_{0}-\hat{h}_{1}}
    \end{array}\!\!\right)\!.\!\!
\end{align}
Note that Eq.~\eqref{App:Neumann-0} converges when the norm of $\hat{\tilde{\Delta}}\hat{\varepsilon}^{-1}$
is smaller than unity which can be fulfilled in the weak pairing limit.
After some algebra, we arrive at an explicit relation for $\hat{M}_{11}^{\text{eff}}$
that is
\begin{equation}
    \hat{M}_{11}^{\text{eff}}=\left(\begin{array}{cc}
    \hat{N}_{1}+\hat{\xi}_{1} & \hat{\varDelta}_{2}^{\text{eff}}\\{}
    [\hat{\varDelta}_{2}^{\text{eff}}]^{\dagger} & \hat{h}_{2}+\hat{\xi}_{2}
    \end{array}\right),\label{APP: general effective matrix}
\end{equation}
where the energy shifts arising from the multiband nature as
\begin{align}
    \hat{\xi}_{1} & =\hat{\varDelta}_{1}\frac{1}{\omega\hat{\sigma}_{0}-\hat{h}_{1}}\text{\ensuremath{\hat{\varDelta}_{1}^{\dagger}}},\\
    \hat{\xi}_{2} & =\hat{\varDelta}_{4}^{\dagger}\frac{1}{\omega\hat{\sigma}_{0}-\hat{N}_{2}}\hat{\varDelta}_{4}.
\end{align}
Additionally, the effective $2\times2$ pairing matrix takes the form
\begin{equation}
    \hat{\varDelta}_{2}^{\text{eff}}=\hat{\varDelta}_{2}+\hat{\varDelta}_{1}\frac{1}{\omega\hat{\sigma}_{0}-\hat{h}_{1}}\hat{\varDelta}_{3}^{\dagger}\frac{1}{\omega\hat{\sigma}_{0}-\hat{N}_{2}}\hat{\varDelta}_{4}.
\end{equation}
According to Eqs. (16) and (24), the projected pairing matrices $\hat{\varDelta}_{1,2,3,4}$
have only nonvanishing off-diagonal elements, and also $\hat{N}_{1}+\hat{\xi}_{1}\equiv\text{diag}(\zeta_{1},\zeta_{2})$
and $\hat{h}_{2}+\hat{\xi}_{2}\equiv\text{diag}(\zeta_{3},\zeta_{4})$
are diagonal matrices due to time-reversal symmetry. In this case,
$\hat{\varDelta}_{2}^{\text{eff}}$ becomes an off-diagonal matrix
explicitly given by
\begin{widetext}
    \begin{equation}
        \hat{\varDelta}_{2}^{\text{eff}}=\left(\begin{array}{cc}
        0 & (\hat{\varDelta}_{2})_{12}+\frac{(\hat{\varDelta}_{1})_{12}[(\hat{\varDelta}_{3})_{12}]^{*}(\hat{\varDelta}_{4})_{12}}{[\omega-(\hat{h}_{1})_{22}][\omega-(\hat{N}_{2})_{11}]}\\
        (\hat{\varDelta}_{2})_{21}+\frac{(\hat{\varDelta}_{1})_{21}[(\hat{\varDelta}_{3})_{21}]^{*}(\hat{\varDelta}_{4})_{21}}{[\omega-(\hat{h}_{1})_{11}][\omega-(\hat{N}_{2})_{22}]} & 0
        \end{array}\right),
    \end{equation}
\end{widetext}
where $()_{ij}$ indicates the matrix element of the given matrix.
To find the pairing symmetry for $\hat{\varDelta}_{2}^{\text{eff}}$,
we can multiply it with the inverse of the Cooper pair symmetrization factor,
i.e., $(i\hat{\sigma}_{y})^{-1}$ leading to the effective pseudo-spin-singlet ($\varphi$) and pseudo-spin-triplet ($\mathbf{d}$-vector) components,
i.e., $\hat{\varDelta}_{2}^{\text{eff}}(i\hat{\sigma}_{y})^{-1}=\varphi\hat{\sigma}_{0}+\mathbf{d}\cdot\hat{\mathbf{\sigma}}$.
In our model, $\hat{\varDelta}_{2}^{\text{eff}}$ is off-diagonal
reflecting an effective mixed pairing state having nonvanishing pseudo-spin-singlet and triplet components of the $\mathbf{d}$-vector  explicitly given by
\begin{widetext}
    \begin{align}
        \varphi & =\frac{1}{2}\left((\hat{\varDelta}_{2})_{12}-(\hat{\varDelta}_{2})_{21}+\frac{(\hat{\varDelta}_{1})_{12}[(\hat{\varDelta}_{3})_{12}]^{*}(\hat{\varDelta}_{4})_{12}}{[\omega-(\hat{h}_{1})_{22}][\omega-(\hat{N}_{2})_{11}]}-\frac{(\hat{\varDelta}_{1})_{21}[(\hat{\varDelta}_{3})_{21}]^{*}(\hat{\varDelta}_{4})_{21}}{[\omega-(\hat{h}_{1})_{11}][\omega-(\hat{N}_{2})_{22}]}\right),\label{APP: general singlet}\\
        d_{z} & =\frac{1}{2}\left((\hat{\varDelta}_{2})_{12}+(\hat{\varDelta}_{2})_{21}+\frac{(\hat{\varDelta}_{1})_{12}[(\hat{\varDelta}_{3})_{12}]^{*}(\hat{\varDelta}_{4})_{12}}{[\omega-(\hat{h}_{1})_{22}][\omega-(\hat{N}_{2})_{11}]}+\frac{(\hat{\varDelta}_{1})_{21}[(\hat{\varDelta}_{3})_{21}]^{*}(\hat{\varDelta}_{4})_{21}}{[\omega-(\hat{h}_{1})_{11}][\omega-(\hat{N}_{2})_{22}]}\right).\label{APP: general triplet}
    \end{align}
\end{widetext}
Considering Eqs.~$\text{\eqref{APP: general singlet}}$ and $\text{\eqref{APP: general triplet}}$,
$\hat{M}_{11}^{\text{eff}}$ becomes
\begin{equation}
    \hat{M}_{11}^{\text{eff}}=\left(\begin{array}{cccc}
    \zeta_{1} & 0 & 0 & d_{z}+\varphi\\
    0 & \zeta_{2} & d_{z}-\varphi & 0\\
    0 & d_{z}^{*}-\varphi^{*} & \zeta_{3} & 0\\
    d_{z}^{*}+\varphi^{*} & 0 & 0 & \zeta_{4}
    \end{array}\right).
\end{equation}
Interestingly, $\hat{M}_{11}^{\text{eff}}$ can preserve pseudo-spin
rotational symmetry. The matrix form for such an operator is defined
by
\begin{equation}
    \hat{\Delta}_{\mathbf{n}}(\theta)\equiv\text{diag}\left(e^{i\theta/2(\mathbf{n}\cdot\hat{\mathbf{\sigma}})},e^{-i\theta/2(\mathbf{n}\cdot\hat{\mathbf{\sigma}})^{*}}\right),
\end{equation}
where $\theta$ is the angle of rotation in the pseudo-spin space.
Note that $\hat{M}_{11}^{\text{eff}}$ preserves pseudo-spin-$\pi$
rotational symmetry along the $z$-axis, i.e., $[\hat{M}_{11}^{\text{eff}},\hat{\mathcal{D}}_{n_{z}}(\pi)]=0,$
with
\begin{equation}
    \hat{\Delta}_{n_{z}}(\pi)=\left(\begin{array}{cccc}
    i & 0 & 0 & 0\\
    0 & -i & 0 & 0\\
    0 & 0 & -i & 0\\
    0 & 0 & 0 & i
    \end{array}\right).
\end{equation}
Representing $\hat{M}_{11}^{\text{eff}}$ in the eigenspace of $\hat{\mathcal{D}}_{n_{z}}(\pi)$
denoted by $\hat{\mathcal{U}}$, we decouple the effective Hamiltonian
into two sectors given by
\begin{align}
    \hat{\mathcal{M}}_{11}^{\text{eff}} & =\hat{\mathcal{U}}^{-1}\hat{M}_{11}^{\text{eff}}\hat{\mathcal{U}}=\text{diag}(\hat{\mathcal{H}}_{1},\hat{\mathcal{H}}_{2}),
\end{align}
where
\begin{align}
    \hat{\mathcal{H}}_{1} & =\left(\begin{array}{cc}
    \zeta_{4} & d_{z}^{*}+\varphi^{*}\\
    d_{z}+\varphi & \zeta_{1}
    \end{array}\right),\\
    \hat{\mathcal{H}}_{2} & =\left(\begin{array}{cc}
    \zeta_{3} & d_{z}^{*}-\varphi^{*}\\
    d_{z}-\varphi & \zeta_{4}
    \end{array}\right).
\end{align}
Therefore, the effective superconducting spectra become
\begin{align}
    \mathrm{E} & =\frac{1}{2}\left(\zeta_{1}+\zeta_{4}\pm\sqrt{(\zeta_{1}-\zeta_{4})^{2}+4\delta^{2}}\right).
\end{align}
where the magnitude of the avoided crossing is
\begin{align}
    \delta & =\sqrt{|\varphi|^{2}+|d_{z}|^{2}\pm(\varphi^{*}d_{z}+\varphi d_{z}^{*})}.
\end{align}


\section{Effective low-energy theory}

In this section, we employ the general formalism of the downfolding
method, described in App.~\ref{APP:FoldingDown}, to obtain an effective
low(finite)-energy intra(inter)-band superconducting Hamiltonian for
the Al/Au model. We first derive the low-energy
formalism while the finite-energy pairing is formulated subsequently.

\subsection{Band basis label with $\nu=\nu^{\prime}=+$}

Consider the BdG Hamiltonian represented in the eigenspace of the
normal state model given by
\begin{align}
    \hat{\mathscr{H}}_{\mathbf{k}} & =\left(\begin{array}{cccc}
    \hat{N}_{\mathbf{k}}^{++} & 0 & \hat{\varDelta}_{\mathbf{k}}^{++} & \hat{\varDelta}_{\mathbf{k}}^{+-}\\
    0 & \hat{N}_{\mathbf{k}}^{--} & \hat{\varDelta}_{\mathbf{k}}^{-+} & \hat{\varDelta}_{\mathbf{k}}^{--}\\{}
    [\hat{\varDelta}_{\mathbf{k}}^{++}]^{\dagger} & [\hat{\varDelta}_{\mathbf{k}}^{-+}]^{\dagger} & -\hat{N}_{-\mathbf{k}}^{++} & 0\\{}
    [\hat{\varDelta}_{\mathbf{k}}^{+-}]^{\dagger} & [\hat{\varDelta}_{\mathbf{k}}^{--}]^{\dagger} & 0 & -\hat{N}_{-\mathbf{k}}^{--}
    \end{array}\right).
\end{align}
To derive the effective superconducting Hamiltonian at the Fermi energy,
we change the basis to an intra-band formalism through a unitary transformation
$\hat{\mathscr{H}}_{\mathbf{k}}^{\prime}=\hat{U}^{\prime\dagger}\hat{\mathscr{H}}_{\mathbf{k}}\hat{U}^{\prime}$
where $\hat{U}^{\prime}$ is given by
\begin{eqnarray}
    (\hat{U}^{\prime})^{-1} & = & \hat{U}^{\prime\dagger}=\left(\begin{array}{cccc}
    1 & 0 & 0 & 0\\
    0 & 0 & 1 & 0\\
    0 & 1 & 0 & 0\\
    0 & 0 & 0 & 1
    \end{array}\right),
\end{eqnarray}
and $\hat{\mathscr{H}}_{\mathbf{k}}^{\prime}$ becomes
\begin{align}
    \hat{\mathscr{H}}_{\mathbf{k}}^{\prime}\! & =\!\left(\!\!\begin{array}{cc|cc}
    \hat{N}_{\mathbf{k}}^{++} & \hat{\varDelta}_{\mathbf{k}}^{++} & 0 & \!\!\hat{\varDelta}_{\mathbf{k}}^{+-}\\
    \big[\hat{\varDelta}_{\mathbf{k}}^{++}\big]^{\dagger} & -\hat{N}_{-\mathbf{k}}^{++} & [\hat{\varDelta}_{\mathbf{k}}^{-+}]^{\dagger} & \!\!0\\
    \hline 0 & \hat{\varDelta}_{\mathbf{k}}^{-+} & \hat{N}_{\mathbf{k}}^{--} & \!\!\hat{\varDelta}_{\mathbf{k}}^{--}\\
    \big[\hat{\varDelta}_{\mathbf{k}}^{+-}\big]^{\dagger} & 0 & \big[\hat{\varDelta}_{\mathbf{k}}^{--}\big]^{\dagger} & \!\!-\hat{N}_{-\mathbf{k}}^{--}
    \end{array}\!\!\right).\!\label{APP: Intrabandpairing}
\end{align}
The first $4\times4$ block describes the
superconducting sector with predominant aluminum orbital character
in the normal state for small momenta. Comparing Eq.~\eqref{APP: Intrabandpairing}
with Eq.~\eqref{App: General band basis Hamil}, we deduce that
\begin{align}
    \hat{N}_{1} & =\hat{N}_{\mathbf{k}}^{++},\ \ \hat{h}_{2}=-\hat{N}_{-\mathbf{k}}^{++},\\
    \hat{N}_{2} & =\hat{N}_{\mathbf{k}}^{--},\ \ \hat{h}_{1}=-\hat{N}_{-\mathbf{k}}^{--}\\
    \hat{\varDelta}_{1} & =\hat{\varDelta}_{\mathbf{k}}^{+-},\ \ \hat{\varDelta}_{2}=\hat{\varDelta}_{\mathbf{k}}^{++},\\
    \hat{\varDelta}_{3} & =\hat{\varDelta}_{\mathbf{k}}^{--},\ \ \hat{\varDelta}_{4}=\hat{\varDelta}_{\mathbf{k}}^{-+}.
\end{align}
Substituting the above results into Eq.~$\text{\eqref{APP: general effective matrix}}$
and setting $\omega=0$, we obtain the effective Hamiltonian describing
superconducting spectral properties of energy bands with predominant
aluminum orbital character in the normal state at the Fermi energy
\begin{equation}
    \hat{H}_{\mathbf{k},\text{eff}}^{++}=\left(\begin{array}{cc}
    \hat{N}_{\mathbf{k}}^{++}+\hat{\xi}_{1} & \hat{\varDelta}_{\mathbf{k},\text{eff}}^{++}\\{}
    [\hat{\varDelta}_{\mathbf{k},\text{eff}}^{++}]^{\dagger} & -\hat{N}_{-\mathbf{k}}^{++}+\hat{\xi}_{2}
    \end{array}\right),
\end{equation}
where the energy shifts, arising from the inter-orbital pairing, are 
\begin{align}
    \hat{\xi}_{1} & =+\hat{\varDelta}_{\mathbf{k}}^{+-}\frac{1}{\hat{N}_{-\mathbf{k}}^{--}}[\hat{\varDelta}_{\mathbf{k}}^{+-}]^{\dagger},\\
    \hat{\xi}_{2} & =-\big[\hat{\varDelta}_{\mathbf{k}}^{-+}\big]^{\dagger}\frac{1}{\hat{N}_{\mathbf{k}}^{--}}\hat{\varDelta}_{\mathbf{k}}^{-+}.
\end{align}
The effective low-energy pairing potential for the predominant
aluminum bands takes the form
\begin{equation}
    \hat{\varDelta}_{\mathbf{k},\text{eff}}^{++}=\hat{\varDelta}_{\mathbf{k}}^{++}+\hat{\varDelta}_{\mathbf{k}}^{+-}\frac{1}{\hat{N}_{-\mathbf{k}}^{--}}[\hat{\varDelta}_{\mathbf{k}}^{--}]^{\dagger}\frac{1}{-\hat{N}_{\mathbf{k}}^{--}}\hat{\varDelta}_{\mathbf{k}}^{-+}.\label{APP: Effective pairing Aluminum}
\end{equation}
Note that the second term is arising from the interplay between inter-orbital
pairing with pairing of energy bands with predominant Au character.

\subsection{Band basis label with $\nu=\nu^{\prime}=-$}

The effective superconducting Hamiltonian for the predominant Au
sector can be derived through a unitary transformation $\hat{\mathscr{H}}_{\mathbf{k}}^{\prime\prime}=\hat{U}^{\prime\dagger}\hat{\mathscr{H}}_{\mathbf{k}}\hat{U}^{\prime}$
with
\begin{align}
    \hat{\mathscr{H}}_{\mathbf{k}}^{\prime\prime}\! & =\!\left(\!\!\begin{array}{cc|cc}
    \hat{N}_{\mathbf{k}}^{--} & \!\!\hat{\varDelta}_{\mathbf{k}}^{--} & 0 & \hat{\varDelta}_{\mathbf{k}}^{-+}\\
    \big[\hat{\varDelta}_{\mathbf{k}}^{--}\big]^{\dagger} & \!\!-\hat{N}_{-\mathbf{k}}^{--} & \big[\hat{\varDelta}_{\mathbf{k}}^{+-}\big]^{\dagger} & 0\\
    \hline 0 & \!\!\hat{\varDelta}_{\mathbf{k}}^{+-} & \hat{N}_{\mathbf{k}}^{++} & \hat{\varDelta}_{\mathbf{k}}^{++}\\
    \big[\hat{\varDelta}_{\mathbf{k}}^{-+}\big]^{\dagger} & \!\!0 & \big[\hat{\varDelta}_{\mathbf{k}}^{++}\big]^{\dagger} & -\hat{N}_{-\mathbf{k}}^{++}
    \end{array}\!\!\right),\label{APP: gold BdG Hamil}
\end{align}
where $\hat{U}^{\prime}$ is given by
\begin{equation}
    (\hat{U}^{\prime})^{-1}=(\hat{U}^{\prime})^{\dagger}=\left(\begin{array}{cccc}
    0 & 1 & 0 & 0\\
    0 & 0 & 0 & 1\\
    1 & 0 & 0 & 0\\
    0 & 0 & 1 & 0
    \end{array}\right).
\end{equation}
Comparing Eq.~\eqref{APP: gold BdG Hamil} with Eq.~\eqref{App: General band basis Hamil},
we obtain
\begin{align}
    \hat{N}_{1} & =\hat{N}_{\mathbf{k}}^{--},\ \ \hat{h}_{2}=-\hat{N}_{-\mathbf{k}}^{--},\\
    \hat{N}_{2} & =\hat{N}_{\mathbf{k}}^{++},\ \ \hat{h}_{1}=-\hat{N}_{-\mathbf{k}}^{++},\\
    \hat{\varDelta}_{1} & =\hat{\varDelta}_{\mathbf{k}}^{-+},\ \ \hat{\varDelta}_{2}=\hat{\varDelta}_{\mathbf{k}}^{--},\\
    \hat{\varDelta}_{3} & =\hat{\varDelta}_{\mathbf{k}}^{++},\ \ \hat{\varDelta}_{4}=\hat{\varDelta}_{\mathbf{k}}^{+-}.
\end{align}
At the Fermi energy $\omega=0$, inserting the above relations to Eq.~$\text{\eqref{APP: general effective matrix}}$, we explicitly derive
the effective superconducting Hamiltonian for the energy bands with
predominant Au orbital character in the normal state for small momenta
given by
\begin{equation}
    \hat{H}_{\mathbf{k},\text{eff}}^{--}=\left(\begin{array}{cc}
    \hat{N}_{\mathbf{k}}^{--}+\hat{\xi}_{1} & \hat{\varDelta}_{\mathbf{k},\text{eff}}^{--}\\{}
    [\hat{\varDelta}_{\mathbf{k},\text{eff}}^{--}]^{\dagger} & -\hat{N}_{-\mathbf{k}}^{--}+\hat{\xi}_{2}
    \end{array}\right),
\end{equation}
where the energy shifts induced by multiband effects are
\begin{align}
    \hat{\xi}_{1} & =+\hat{\varDelta}_{\mathbf{k}}^{-+}\frac{1}{\hat{N}_{-\mathbf{k}}^{++}}[\hat{\varDelta}_{\mathbf{k}}^{-+}]^{\dagger},\\
    \hat{\xi}_{2} & =-[\hat{\varDelta}_{\mathbf{k}}^{+-}]^{\dagger}\frac{1}{\hat{N}_{\mathbf{k}}^{++}}\hat{\varDelta}_{\mathbf{k}}^{+-}.
\end{align}
Moreover, the effective low-energy pairing potential for the predominant
Au bands in the normal state becomes
\begin{equation}
    \hat{\varDelta}_{\mathbf{k},\text{eff}}^{--}=\hat{\varDelta}_{\mathbf{k}}^{--}+\hat{\varDelta}_{\mathbf{k}}^{-+}\frac{1}{\hat{N}_{-\mathbf{k}}^{++}}[\hat{\varDelta}_{\mathbf{k}}^{++}]^{\dagger}\frac{1}{-\hat{N}_{\mathbf{k}}^{++}}\hat{\varDelta}_{\mathbf{k}}^{+-}.\label{APP: Effective pairing gold}
\end{equation}

\subsection{Inter-orbital sector}

To obtain an effective inter-orbital superconducting Hamiltonian and
study the BdG spectra at finite excitation energies, it is helpful to
represent the BdG Hamiltonian in the inter-band basis. This can be done
by the unitary transformation $\hat{\mathscr{H}}_{\mathbf{k}}^{\prime\prime\prime}=\hat{U}^{\prime\prime\dagger}\hat{\mathscr{H}}_{\mathbf{k}}\hat{U}^{\prime\prime}$
with
\begin{align}
    \hat{\mathscr{H}}_{\mathbf{k}}^{\prime\prime\prime}\! & =\!\left(\!\!\begin{array}{cc|cc}
    \hat{N}_{\mathbf{k}}^{++} & \hat{\varDelta}_{\mathbf{k}}^{+-} & 0 & \hat{\varDelta}_{\mathbf{k}}^{++}\\{}
    [\hat{\varDelta}_{\mathbf{k}}^{+-}]^{\dagger} & -\hat{N}_{-\mathbf{k}}^{--} & [\hat{\varDelta}_{\mathbf{k}}^{--}]^{\dagger} & 0\\
    \hline 0 & \hat{\varDelta}_{\mathbf{k}}^{--} & \hat{N}_{\mathbf{k}}^{--} & \hat{\varDelta}_{\mathbf{k}}^{-+}\\{}
    [\hat{\varDelta}_{\mathbf{k}}^{++}]^{\dagger} & 0 & [\hat{\varDelta}_{\mathbf{k}}^{-+}]^{\dagger} & -\hat{N}_{-\mathbf{k}}^{++}
    \end{array}\!\!\right),\label{APP: Interband BdG}
\end{align}
where the unitary matrix $\hat{U}^{\prime\prime}$ is given by
\begin{equation}
    (\hat{U}^{\prime\prime})^{-1}=(\hat{U}^{\prime\prime})^{\dagger}=\left(\begin{array}{cccc}
    1 & 0 & 0 & 0\\
    0 & 0 & 1 & 0\\
    0 & 0 & 0 & 1\\
    0 & 1 & 0 & 0
    \end{array}\right).
\end{equation}
In this case,  comparing Eq.~\eqref{APP: Interband BdG} with Eq.~\eqref{App: General band basis Hamil}, we arrive at
\begin{align}
    \hat{N}_{1} & =\hat{N}_{\mathbf{k}}^{++},\ \ \hat{h}_{2}=-\hat{N}_{-\mathbf{k}}^{--},\\
    \hat{N}_{2} & =\hat{N}_{\mathbf{k}}^{--},\ \ \hat{h}_{1}=-\hat{N}_{-\mathbf{k}}^{++},\\
    \hat{\varDelta}_{1} & =\hat{\varDelta}_{\mathbf{k}}^{++},\ \ \hat{\varDelta}_{2}=\hat{\varDelta}_{\mathbf{k}}^{+-},\\
    \hat{\varDelta}_{3} & =\hat{\varDelta}_{\mathbf{k}}^{-+},\ \ \hat{\varDelta}_{4}=\hat{\varDelta}_{\mathbf{k}}^{--}.
\end{align}
Since the inter-orbital pairing happens at finite excitation energy,
$\omega$ is no longer vanishing and becomes finite. Substituting
the above relations into Eq.~\eqref{APP: general effective matrix},
we find an explicit form for the effective inter-orbital superconducting
Hamiltonian
\begin{equation}
    \hat{H}_{\mathbf{k},\text{eff}}^{\text{IO}}=\left(\begin{array}{cc}
    \hat{N}_{\mathbf{k}}^{++}+\hat{\xi}_{1} & \hat{\varDelta}_{\mathbf{k},\text{eff}}^{+-}\\{}
    [\hat{\varDelta}_{\mathbf{k},\text{eff}}^{+-}]^{\dagger} & -\hat{N}_{-\mathbf{k}}^{--}+\hat{\xi}_{2}
    \end{array}\right),\label{APP: Inter-orbital pairing}
\end{equation}
where the energy shifts, induced by the intra-band effects, are given
by
\begin{align}
    \hat{\xi}_{1} & =\hat{\varDelta}_{\mathbf{k}}^{++}\frac{1}{\omega\hat{\sigma}_{0}+\hat{N}_{-\mathbf{k}}^{++}}[\hat{\varDelta}_{\mathbf{k}}^{++}]^{\dagger},\\
    \hat{\xi}_{2} & =[\hat{\varDelta}_{\mathbf{k}}^{--}]^{\dagger}\frac{1}{\omega\hat{\sigma}_{0}-\hat{N}_{\mathbf{k}}^{--}}\hat{\varDelta}_{\mathbf{k}}^{--}.
\end{align}
Note that $\hat{H}_{\mathbf{k},\text{eff}}^{\text{IO}}$ breaks particle-hole
symmetry due to the different diagonal entries arising from two different
energy bands. Finally, the effective $2\times2$ finite-energy pairing
matrix becomes
\begin{equation}
    \hat{\varDelta}_{\mathbf{k},\text{eff}}^{\text{IO}}=\hat{\varDelta}_{\mathbf{k}}^{+-}+\hat{\varDelta}_{\mathbf{k}}^{++}\frac{1}{\omega\hat{\sigma}_{0}+\hat{N}_{-\mathbf{k}}^{++}}[\hat{\varDelta}_{\mathbf{k}}^{-+}]^{\dagger}\frac{1}{\hat{\sigma}_{0}\omega-\hat{N}_{\mathbf{k}}^{--}}\hat{\varDelta}_{\mathbf{k}}^{--}.\label{APP: Effective pairing inter-orbital}
\end{equation}
Note that the second term originates from the interplay between
low-energy bands and their corresponding pairings with finite
energy pairing.

\section{Experimental detection of finite-energy avoided crossings in Au/Al} \label{APP:experimental detection}

As discussed in the main text, finite-energy Cooper pairing in Au/Al are of size $30-50\,\mu\mathrm{eV}$ to $100-200\,\mu\mathrm{eV}$. In order to resolve this, an energy resolution of $\Delta_E\approx10-20\,\mu\mathrm{eV}$ should suffice. If we assuming a thermal broadening of $3.5k_BT$, we can estimate that experiments need to be performed at $T = \Delta_E / (3.5k_B) \approx 30-60\,\mathrm{mK}$ in order to be able to resolve $\Delta E$. With state-of-the-art STM and transport experiments in dilution refrigerators, energy resolutions below $10\,\mu\mathrm{eV}$ at operating temperatures of 10\,mK are indeed possible~\cite{Schwenk2020}. Thus, we believe that our predicted finite-energy features in the superconducting electronic structure of Au/Al are observable.

\begin{table}[b]
    \centering
    \caption{Superconducting energy gap, critical temperature and critical magnetic field of Al of different film thicknesses.}
    \begin{tabular}{c|c|c|c}
        $\delta$ ($\mu\mathrm{eV}$) & $T_c$ (K) & $H_c$ (T) & References \\ \hline
        208--307 & -- &  -- & Ref.~\onlinecite{Court2008} \\
        -- & 1.2-2.8 & 0.01-5 & Ref.~\onlinecite{Meservey1971}
    \end{tabular}
    \label{tab:expDeltaTcHc}
\end{table}
\begin{table}[h]
    \centering
    \caption{Finite-energy avoided crossings ($\delta$ in $\mu\mathrm{eV}$) from DFT. The numbers are scaled values using a value {of the superconducting gap of Al of} $\delta_\mathrm{Al}=300\,\mu\mathrm{eV}$ for very thin films~\cite{Court2008}.}
    \begin{tabular}{c|ccccccc}
         &
        $\delta_\mathrm{Au}^-$ &
        $\delta_\mathrm{Au}^+$ &
        $\delta_\mathrm{IOP}^-$ &
        $\delta_\mathrm{IOP}^+$ &
        $\delta_\mathrm{IOP}^\mathrm{Au}$ &
        $\delta_\mathrm{IOP}^\mathrm{Au}$ &
        $\delta_\mathrm{IOP}^\mathrm{Au}$ \\\hline
        $\delta$ &
        114 &
        153 &
        180 &
        132 &
        0 &
        31 &
        52 \\
        $H/H_c$ & 
        0  &
        0  &
        0  &
        0 & 
        0 &
        $\sim0.3$ &
        $\sim0.4$
    \end{tabular}
    \label{tab:DFTDeltaHc}
\end{table}

\end{document}